\documentclass[prd,twocolumn,aps, floatfix,preprintnumbers,showpacs,nofootinbib]{revtex4-1}
\usepackage{amsthm}   
\usepackage{amsmath}  
\usepackage{amssymb}  
\usepackage{mathrsfs}
\usepackage{color}
\usepackage[all]{xy}
\usepackage[pdftex]{graphicx}
\usepackage{indentfirst}
\usepackage[pdftex]{hyperref}
\usepackage{subfigure}

\newcommand{\aap}{Astron.Astrophys.}

\newcommand{\mnras}{Mon. Not. Roy. Astron. Soc.}

\newcommand{\be}{\begin{equation}}
\newcommand{\ee}{\end{equation}}
\newcommand{\ba}{\begin{eqnarray}}
\newcommand{\ea}{\end{eqnarray}}

\newcommand{\en}{\nonumber\\}

\newcommand{\de}{\delta}

\newcommand{\dd}[1]{\dot{#1}}

\begin{document}
%
\title{The Cosmology of Atomic Dark Matter}
\author{Francis-Yan Cyr-Racine}\email{francis@phas.ubc.ca}
\affiliation{Department of Physics and Astronomy, University of British Columbia, Vancouver, BC, V6T 1Z1, Canada}
\author{Kris Sigurdson}\email{krs@phas.ubc.ca}
\affiliation{Department of Physics and Astronomy, University of British Columbia, Vancouver, BC, V6T 1Z1, Canada}
\date{\today}
\begin{abstract}
While, to ensure successful cosmology, dark matter (DM) must kinematically decouple from the standard model plasma very early in the history of the Universe,  it can remain  coupled to a bath of ``dark radiation'' until  a relatively  late epoch. One minimal theory that realizes such a scenario is the Atomic Dark Matter model, in which two fermions oppositely charged under a new U(1) dark force are initially coupled to a thermal bath of ``dark photons'' but 
eventually recombine into neutral atom-like bound states 
and begin forming gravitationally-bound structures.  As dark atoms have (dark) atom-sized geometric cross sections, this model also provides an example of self-interacting DM with a velocity-dependent cross section. 
Delayed kinetic decoupling in this scenario predicts novel DM properties on small scales but retains the success of cold DM on larger scales.  We calculate the atomic physics necessary to capture the thermal history of this dark sector and show significant improvements over the standard atomic hydrogen calculation are needed. We solve the Boltzmann equations that govern the evolution of cosmological fluctuations in this model and find in detail the impact of the atomic DM scenario on the matter power spectrum and the cosmic microwave background (CMB).  This scenario imprints a new length scale, the Dark-Acoustic-Oscillation (DAO) scale, on the matter density field. This DAO scale shapes the small-scale matter power spectrum and determines the minimal DM halo mass at late times which may be many orders of magnitude larger than in a typical WIMP scenario. This model necessarily includes an extra dark radiation component, which may be favoured by current CMB  experiments, and we quantify CMB signatures that distinguish an atomic DM scenario from a standard $\Lambda$CDM model containing extra free-streaming particles. We finally discuss the impacts of atomic DM on galactic dynamics and show that these provide the strongest constraints on the model. 
\end{abstract}
\maketitle

\section{Introduction}
Although dark matter (DM) has been known to exist for several decades \cite{1933AcHPh...6..110Z,1937ApJ....86..217Z,1980ApJ...238..471R}, its physical nature remains one of the deepest mysteries of modern science. Observations show that DM is mostly cold, collisionless and interacts very weakly (if at all) with standard-model (SM) particles. Many models that fit this picture very well have been proposed through the years, including the prominent weakly-interacting-massive-particle (WIMP) models. Distinguishing between these different scenarios is crucial if we have any hope to pinpoint the nature of DM. In this respect, models incorporating new physics that predict novel observational signatures have a clear advantage in disentangling the DM puzzle. 

While the cold dark-matter (CDM) paradigm \cite{Blumenthal:1984bp,Davis:1985rj} has been extremely successful at describing observations from cosmological scales to galactic scales, recent observations of small nearby galaxies seem to be in tension with the CDM scenario. In addition to the so-called ``missing-satellite problem" \cite{Klypin:1999uc,Moore:1999nt,2010AdAst2010E...8K} which refers to the apparent under-abundance of light Milky-Way satellites, observations of the inner mass profiles of Dwarf Spheroidal (dSph) galaxies indicate that they are consistent with a core while CDM simulations favor a cuspier profile \cite{2004MNRAS.351..903G,deBlok:2009sp,Walker:2011zu,deNaray:2011hy,2012MNRAS.420.2034S}. Further, it has been pointed out recently that the most massive galactic subhalos from CDM simulations are too dense to host the brightest Milky Way satellites \cite{BoylanKolchin:2011de,BoylanKolchin:2011dk}. While it is plausible that these problems could be alleviated by including the appropriate baryonic physics in CDM simulations \cite{Haiman:1996rc,Thoul:1996by,MacLow:1998wv,Barkana:1999me,Somerville:2001km,Shapiro:2003gxa,Tassis:2006zt,Pfrommer:2011bg,2012MNRAS.421.3464P,2012MNRAS.421.3522H,Uhlig:2012rt,VeraCiro:2012na}, these observations may be pointing toward physics beyond the vanilla CDM paradigm.    

We adopt here the point of view that the above tensions between the $\Lambda$CDM paradigm and the astrophysical observations may be resolved by modifying the microphysics of the DM sector. Various scenarios have been proposed along those lines in the literature, most of which could either be classified as interacting DM models  \cite{Spergel:1999mh}, warm DM scenarios \cite{Bode:2000gq,Dalcanton:2000hn,Zentner:2003yd,Smith:2011ev}, or even hot DM models (see e.g.~\cite{Sigurdson:2009uz}). The former is tightly constrained by the observed ellipticity of DM halos \cite{MiraldaEscude:2000qt,Yoshida:2000bx} and by the apparent survivability of DM halos to evaporation in clusters \cite{Gnedin:2000ea}, while the latter is unlikely to be able to address all of the CDM issues discussed above \cite{Maccio:2012qf}.   It was realized recently \cite{Loeb:2010gj, 2012MNRAS.tmp.3127V,Aarssen:2012fx} that interacting DM with a velocity-dependent cross section could avoid the ellipticity and evaporation constraints while alleviating the tension between the simulations and the Milky-Way satellites.

In this paper, we investigate in detail the cosmology of a dark sector made of hidden hydrogen-like bound states \cite{Goldberg:1986nk,Kaplan:2009de,Kaplan:2011yj,Cline:2012is}. This so-called ``Atomic Dark Matter" model retains the success of CDM on large cosmological scales but modifies the DM dynamics on sub-galactic scales. Similar to conventional atoms, atomic DM is kinematically coupled to a thermal bath of ``dark" radiation (DR) until late times, which significantly delays the growth of matter perturbations on small scales. During the decoupling epoch, diffusion and acoustic damping substancially reduce the amplitude of sub-horizon perturbations, effectively wiping out structures on scales below this damping horizon, hence mimicking the effect of free-streaming. This model thus naturally provides a way to suppress the faint end of the galaxy luminosity function. After kinematic decoupling, the acoustic oscillations of the dark plasma remain imprinted on the small-scale matter power spectrum. Note that these physical processes also take place in a canonical WIMP scenario \cite{Loeb:2005pm,Bertschinger:2006nq,kris}. The crucial difference here is that the kinematic decoupling of atomic DM typically happens much later than that of a WIMP and therefore can have an impact on much larger, if not observable, scales. 

Beyond its effect on the matter power spectrum, the atomic DM scenario also impacts the cosmic microwave background (CMB) through the effects of the DR on the amplitude and phases of cosmological perturbations. We identify a key CMB signature that distinguishes the atomic DM scenario from a simple $\Lambda$CDM model incorporating extra relativistic degrees of freedom. Indeed, while models incorporating extra radiation always assume it to free-stream like neutrinos, the DR in the atomic DM model can only start free-streaming after it decouples from the DM. Cosmological modes entering the Hubble horizon before the onset of the free-streaming display a slightly different behavior than Fourier modes crossing the causal horizon after the decoupling of the DR.

Since dark atoms have a much larger geometric cross section than point particles, this scenario falls into the category of interacting DM models. Interestingly, the atomic physics naturally gives rise to a velocity-dependent interaction cross section. It is therefore possible that atomic DM could address some of the dSph-galaxy problems while evading the ellipticity and evaporation constraints. In the following, we focus on the ellipticity constraint and show that while it strongly constrains the parameter space of the model, there are plenty of parameter values for which the galactic dynamics is unaffected. 

For clarity and completeness, it is important to mention that the term ``dark atoms" has been used in various contexts in the literature. For example, dark atoms naturally appears in Mirror DM models \cite{Blinnikov:1982eh,Blinnikov:1983gh,Khlopov:1989fj,Foot:2004pa,Ciarcelluti:2010zz}.  In \cite{Khlopov:2010pq,Khlopov:2010ik}, the term ``dark atom" refers to a bound state between a new stable particle charged under ordinary electromagnetism and a helium nucleus. The authors of \cite{Behbahani:2010xa} explored a model in which dark matter is made of supersymmetric dark atoms. In the present work, the phrase ``dark atom" is exclusively used to describe the bound state of two dark fermions (i.e. neutral under the SM gauge group) oppositely charged under a new gauge $U(1)_D$ interaction.    

This paper is organized as follow. In section \ref{model}, we introduce the atomic DM model and discuss the parameters necessary to describe the theory. In section \ref{thermal}, we analyze the thermal history of atomic DM from its initial hot plasma state to its cold and collisionless stage at late times. We include a thorough discussion of the thermal and kinetic decoupling epoch and present an in-depth analysis of the dark-recombination process.  In section \ref{perturbations}, we solve the modified cosmological perturbation equations and discuss the different regimes that a DM fluctuation encounters. In section \ref{cosmology}, we present an analysis of the new features in the matter power spectrum and in the CMB due to the dark atomic physics. We also discuss the Lyman-$\alpha$ (Ly-$\alpha$) forest bounds on the parameter space of atomic DM. In section \ref{astro}, we consider the stringent astrophysical constraints on the model that are obtained by imposing that DM is effectively collision-less inside galactic halos. We revisit in section \ref{direct-detection} the direct-detection signatures of atomic DM proposed in the literature in light of our new analysis. We finally discuss our results in section \ref{discussion} and point out possible new avenues of research.

\section{The Model}\label{model}
The atomic DM model \cite{Goldberg:1986nk,Kaplan:2009de,Kaplan:2011yj} is composed of two oppositely-charged massive fermions interacting through a dark massless (or nearly massless) $U(1)_D$ gauge boson\footnote{We note that dark atoms do not have to be made of two spin-1/2 fermions. For instance, they could arise from a bound state of two scalar particles or from a bound state of a spin-1/2 fermion with a scalar particle.}. In analogy with the regular baryonic sector, we call the lighter fermion ``dark electron", the heavier fermion ``dark proton" and the massless gauge boson ``dark photon". In this work, we assume that the DM relic abundance is set by some UV physics in the early Universe. For example, the dark sector could contain an asymmetry that fixes the DM relic density (see e.g.~\cite{Kaplan:2009ag}). Since we are mostly interested in the late-time cosmological and astrophysical impacts of atomic DM, we do not expect the details of this high-energy completion to play any role in our results. See \cite{Kaplan:2011yj,Davoudiasl:2010am,Davoudiasl:2011fj,Blinov:2012hq,Petraki:2011mv} for examples of DM production mechanism that can lead to atom-like DM particles. 

Four parameters are necessary to fully characterize the physics of the dark sector in this model. These are the DM mass (i.e.~the mass of the bound state) $m_{D}$, the dark fine-structure constant $\alpha_D$, the binding energy of the bound state $B_D$ and the present-day ratio of the DR temperature to the CMB temperature $\xi\equiv (T_D/T_{\text{CMB}})|_{z=0}$. Other combinations of parameters are possible but this particular set is physically transparent since $B_D/\xi$ fixes the redshift of dark recombination, $m_{D}$ fixes the number density of DM particles and $\alpha_D$ governs the microscopic interactions between the dark sector constituents. These parameters are subject to the consistency constraint
\be\label{consistency_c}
\frac{m_{D}}{B_D}\geq\frac{8}{\alpha_D^2}-1,
\ee
which ensures that $m_{\bf e,p}\leq (m_D+B_D)/2$. This bound is saturated when the two fermions have equal masses. Here, $m_{\bf e}$ and $m_{\bf p}$ stand for the dark-electron mass and the dark-proton mass, respectively. We give their values in terms of $m_D$, $B_D$, and $\alpha_D$ in appendix \ref{m_ep}. 

On a more general level, the atomic DM scenario can be considered as a toy model of a more complete theory involving a hidden dark plasma in the early Universe. Indeed, atomic DM contains the all key ingredients of a dark-plasma theory (dark radiation, multiple particles, kinetic and thermal decoupling, modified growth of DM fluctuations, long-range and short-range interactions, etc) with only minimal physical inputs. As such, the results presented in this paper should be understood in the broader context of a generalized dark-plasma theory ( see e.g.~\cite{Das:2010ts} for general cosmological constraints on this type of model).

Interestingly, the atomic DM scenario naturally englobes the hidden charged DM models discussed in \cite{Ackerman:2008gi,Feng:2009mn} as special cases. Moreover, in the limit of very large atomic binding energy and large dark fine-structure constant, dark atoms become basically undistinguishable from standard CDM particles. Therefore, the atomic DM scenario is a rather general testbed for physics beyond the vanilla WIMP/CDM paradigm.

\section{Thermal History}\label{thermal}
In the early Universe, the dark sector forms a tightly-coupled plasma much like the standard baryon-photon plasma. As the Universe cools down, three important transitions need to carefully be taken into account. First, once the dark sector temperature falls below $B_D$, it becomes energetically favorable for the dark fermions to recombine into neutral dark atoms. Second, once the momentum transfer rate between the DR and the dark fermions falls below the Hubble expansion rate (kinetic decoupling), the DM effectively ceases to be dragged along by the radiation and can start to clump and form structures. Finally, once the energy transfer rate between the DM and the DR falls below the Hubble rate (thermal decoupling), the DM temperature $T_{DM}$ ceases to track that of DR and start cooling adiabatically. Accurately capturing these transitions and computing their impact on cosmological observables is a major goal of this paper. 

We begin this section by determining the big-bang nucleosynthesis (BBN) bound on the dark-photon temperature. We then discuss the recombination of dark atoms and their thermal coupling to the DR, emphasizing the differences between dark atoms and regular atomic hydrogen. We finally present the solutions to the joint evolution of the  dark-atom ionized fraction and temperature.
\subsection{BBN Limit on Dark-Sector Temperature}
\begin{figure}[t]
\begin{centering}
\includegraphics[width=0.45\textwidth]{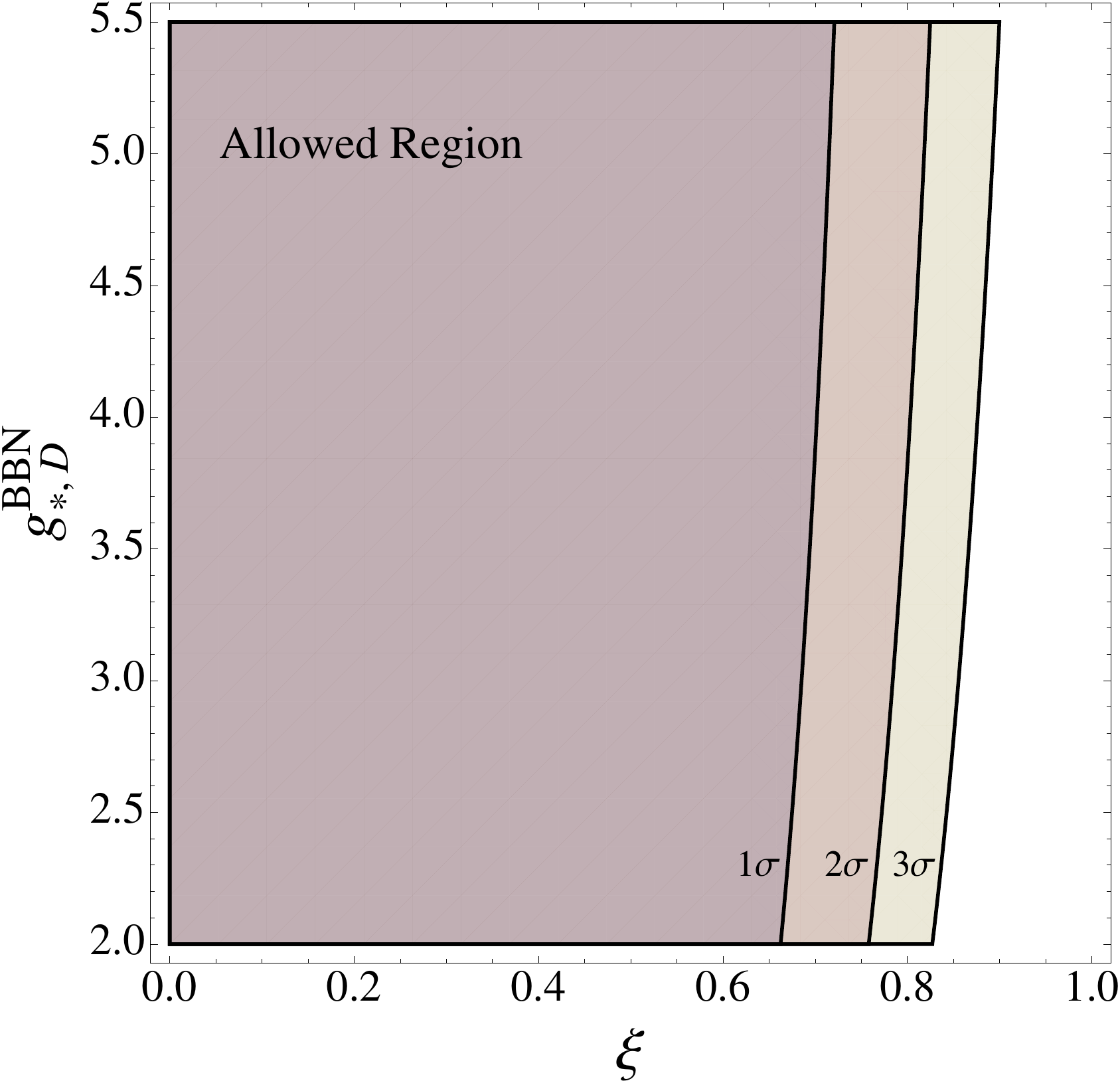}
\caption[Joint BBN constraints on the present-day dark sector temperature and on the effective number of dark sector relativistic degrees of freedom at the time of nucleosynthesis.]{Joint BBN constraints on the present-day dark sector temperature and on the effective number of dark sector relativistic degrees of freedom at the time of nucleosynthesis. As indicated, we display contours corresponding to 1-, 2-, and 3-$\sigma$ constraints.}
\label{bbn_plot}
\end{centering}
\end{figure}

Observations of the relative abundance of light elements put a bound on the possible number of relativistic degrees of freedom at the time of BBN. This limit is usually quoted in terms of the effective number of light neutrino species in thermal equilibrium at BBN; here we shall use the conservative estimate $N_{\nu}=3.24\pm1.2$ 
(95$\%$ confidence) derived in Ref.~\cite{Cyburt:2004yc}. More recent estimates \cite{Mangano:2011ar,Nollett:2011aa,Aver:2010wq,Izotov:2010ca,Simha:2008zj} of $N_{\nu}$ are statistically consistent with this value.  Assuming that the dark sector contributes $g_{*,D}^{\text{BBN}}$ relativistic degrees of freedom during BBN and further assuming three species of SM neutrinos, we obtain the bound
\be\label{Nnu_const}
g_{*,D}^{\text{BBN}}\xi_{\text{BBN}}^4\leq2.52,
\ee
where $\xi_{\text{BBN}}$ is the ratio of the dark sector and visible temperatures at the time of nucleosynthesis. In the minimal atomic DM scenario considered in this work, DM is totally decoupled from SM particles and therefore the entropy of the dark sector and the visible sector are separately conserved
\be\label{entropy_conservation}
\frac{g_{*S,D}^{\text{BBN}}\xi_{\text{BBN}}^3}{g_{*S,D}^{\text{today}}\xi^3} = \frac{g_{*S,\text{vis}}^{\text{BBN}}}{g_{*S,\text{vis}}^{\text{today}}},
\ee
\begin{figure}[t!]
\begin{centering}
\includegraphics[width=0.5\textwidth]{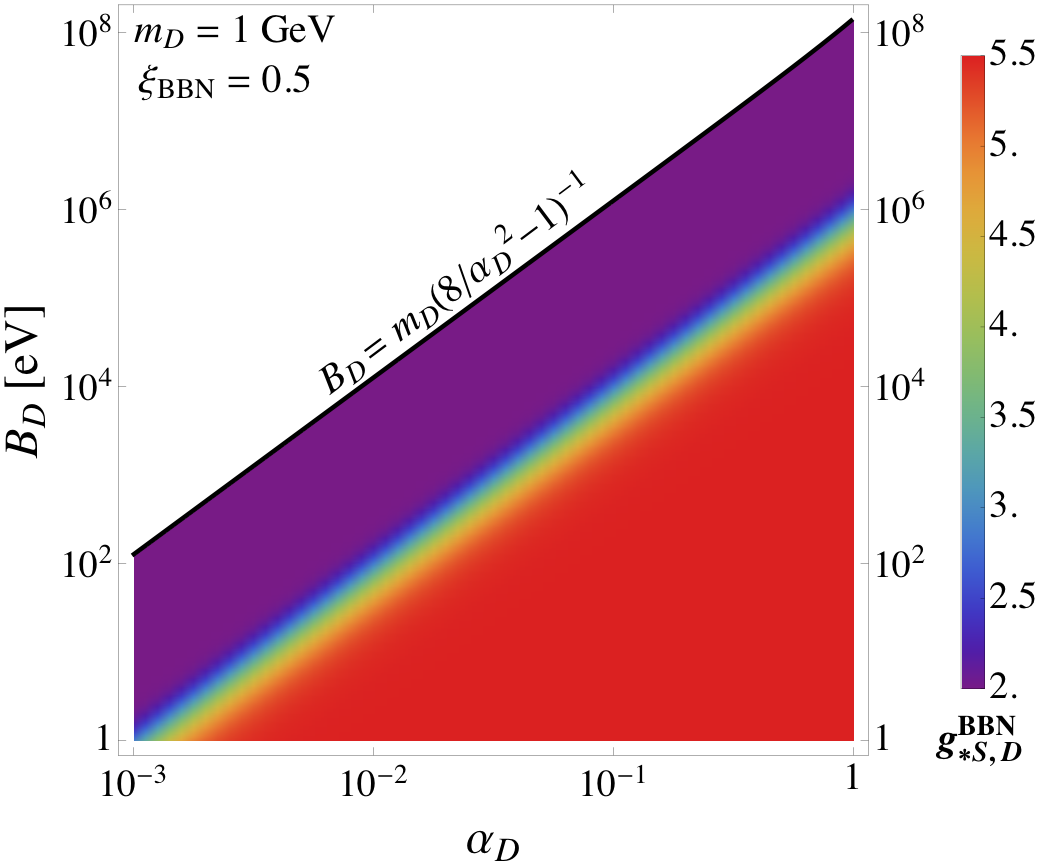}
\caption[Effective number of dark sector relativistic degrees of freedom at the time of nucleosynthesis as a function of $\alpha_D$ and $B_D$.]{Effective number of dark sector relativistic degrees of freedom at the time of nucleosynthesis as a function of $\alpha_D$ and $B_D$ for dark atoms with mass $m_D=1$ GeV. Here, we have fixed $\xi_{\text{BBN}}=0.5$. We also display the consistency constraint given by Eq.~\ref{consistency_c} above which dark atoms do not exist.}
\label{gD_plot}
\end{centering}
\end{figure}
where $g_{*S,D}$ is the present-day effective number of degrees of freedom contributing to the entropy of the dark sector, $s_ D\propto g_{*S,D}T_D^3$. For the simplest model of dark atoms considered here, we have $g_{*S,D}^{\text{today}}=2$ (i.e. only dark photons contribute). Similarly, $g_{*S,\text{vis}}$ is the effective number of degrees of freedom contributing to the entropy of the visible sector. For the particle content of the SM, we have $g_{*S,\text{vis}}^{\text{BBN}}=10.75$ and $g_{*S,\text{vis}}^{\text{today}}=3.94$. During BBN, both dark photons and dark electrons (together with their antiparticles) can contribute to $g_{*S,D}^{\text{BBN}}$  (we assume that the dark proton is massive enough to be non-relativistic at the time of BBN).
 These dark components are kept in thermal equilibrium through Compton scattering and we therefore always have 
$g_{*S,D}^{\text{BBN}} = g_{*,D}^{\text{BBN}}$. Figure \ref{gD_plot} shows the dependence of $g_{*S,D}^{\text{BBN}}$ on $\alpha_D$ and $B_D$ for dark atoms with mass $m_D=1$ GeV and for $\xi_{\text{BBN}}=0.5$. We see that there is a large parameter space for which dark electrons are relativistic at BBN, leading to $g_{*S,D}^{\text{BBN}} = 11/2$ for these models.  At late times, the ratio of the dark sector and visible-sector temperatures is given by 
\be\label{xi_vs_xi_BBN}
\xi = \left(\frac{g_{*S,\text{vis}}^{\text{today}}g_{*S,D}^{\text{BBN}}}{g_{*S,\text{vis}}^{\text{BBN}}g_{*S,D}^{\text{today}}}\right)^{1/3}\xi_{\text{BBN}}.
\ee
Note that if $g_{*S,D}^{\text{BBN}} = 11/2$, that is, if the dark electrons and anti-electrons annihilate after BBN, then $\xi\approx\xi_{\text{BBN}}$ since both the visible and dark sector are reheated by the same amount in this case (assuming the dark electron is a Dirac fermion). Substituting Eq.~\ref{xi_vs_xi_BBN} into Eq.~\ref{Nnu_const}, we obtain the constraint
\be\label{BBN_bound}
\frac{\xi}{(g_{*,D}^{\text{BBN}})^{1/12}}\lesssim0.71\quad(95\%\,\text{confidence}).
\ee
We display this constraint in Fig.~\ref{bbn_plot} where we observe that $\xi\geq1$ is disfavored by at least 4 standard deviations if we consider BBN alone. Note that atomic DM models generally predict a different number of effective relativistic degrees of freedom at BBN than at the time of hydrogen recombination. Given a choice of dark parameters $(\alpha_D,B_D,m_D,\xi)$, we can compute $g_{*,D}(T_D)$ using Eq.~\ref{exact_g_*D}. The evolution of the dark-radiation temperature is then given by the implicit equation
\be\label{real_TD_evol_w g}
T_D(z)=T_D^{\text{today}}(1+z)\left(\frac{g^{\text{today}}_{*,D}}{g_{*,D}(T_D)}\right)^{1/3}.
\ee
This equation can easily be solved iteratively by substituting the zeroth order solution $T_D(z)=T_D^{\text{today}}(1+z)$ into $g_{*,D}(T_D)$. In practice however, the annihilation of dark electron and dark positron has very little effect on the cosmological evolution and to a very good approximation, we can take $\xi$ to be constant.

\subsection{Dark Recombination}\label{dark_recombination}
Once $T_D\ll B_D$, it becomes energetically favorable for the dark plasma to recombine into neutral dark atoms. Letting $n_{\bf e}$ and $n_{D}$ denote the number densities of \emph{dark electrons} and of \emph{dark protons} (both free and those bound in dark atoms)  respectively, we can define the ionization fraction of the dark plasma as $x_D\equiv n_{\bf e}/n_{D}$. 

The phenomenology of dark-atom recombination can significantly differ from that of standard atomic hydrogen. In the latter case, atoms cannot directly recombine to their ground state (Case A) as this results in the emission of a Lyman-continuum photon that can immediately ionize a neutral atom \cite{Peebles:1968ja,1968ZhETF..55..278Z}. In the case of dark atoms, it can be shown that for all but the most weakly-coupled models, the dark sector is always optically thick to dark Lyman-continuum photons and dark atoms cannot effectively recombine directly to their ground state. Dark recombination thus needs to proceed through excited states (Case B) for most models. For the very weakly-coupled models, the dark sector can become optically thin to the Lyman-continuum dark photons, hence allowing dark recombination directly to the ground state. However, the very small coupling between the different dark sector constituents implies that very little recombination actually takes place for these models. We therefore conclude that whenever dark recombination is at all appreciable, it proceeds through case B. Hence, we neglect ground-state recombination in the following.

We can classify dark-atom models in three broad categories according to their recombination phenomenology. The first category encompasses models for which the majority of the recombination process takes place in thermodynamic equilibrium. This category includes models having relatively large values of the dark fine-structure constant ($\alpha_D\gtrsim0.1$) and low-mass models for which the number density is very large during dark recombination. For these scenarios, the recombination history is very well captured by the Saha ionization equation until the recombination timescale becomes comparable to the Hubble expansion rate and the ionized fraction freezes out. In this case, the recombination process is insensitive to the exact population of the excited states and to the details of the radiative transfer between the various atomic lines, making the determination of the ionization history quite simple.

The second category encompasses models for the recombination timescale is slightly shorter or comparable to the Hubble expansion rate. Standard atomic hydrogen falls into this category. For these models, the details of the transitions between the excited states and ground state are relevant. The basic picture is as follow. Once a dark electron has reached an excited state, it rapidly cascades down to the $n=2$ state. Dark electrons can then reach the ground state via the $2s$-$1s$ two-photon transition or by emitting a Lyman-$\alpha$ dark photon which redshifts out of the line wing. The recombination process is thus governed by the rate of escape of the dark Lyman-$\alpha$ photons out of their resonance line and by the rate of two-photon decay \cite{Peebles:1968ja,1968ZhETF..55..278Z,Seager:1999km}. While these two rates are comparable for regular atomic hydrogen, they may differ significantly in the case of dark atoms. In addition, it has been shown \cite{2006A&A...446...39C,2006AstL...32..795K,2008PhRvD..78b3001H,2007A&A...475..109C,2010PhRvD..81h3004K,2009A&A...496..619C,2008A&A...480..629C,2010A&A...512A..53C,2008AstL...34..439G,2009A&A...503..345C,2009PhRvD..80b3001H,2010PhRvD..82l3502A} that the details of the radiative transfer process can affect recombination at the few percent level for this type of models. While we do not take these corrections into account in this work, we expect them to have a negligible impact on our results.

The last category englobes models for which the recombination timescale is longer than the Hubble rate. Atomic DM scenarios having very small values of the dark fine-structure constant or very large masses fall into this category. Not surprisingly, the dark sector remains mostly ionized for these models. The small amount of neutral dark atoms that do form in this case is mostly controlled by the small value of the effective recombination rate and is, to a good approximation, insensitive to the radiative transfer between the various atomic lines.

For all models, we solve for the ionization history of the dark sector using the Effective Multi-Level Atom (EMLA) method presented in Refs.~\cite{AliHaimoud:2010ab,2011PhRvD..83d3513A}. For our purpose, we consider a four-level effective dark atom consisting of the two interface states $2s$ and $2p$, the ground state ($1s$), and a continuum. The recombination process is then governed by a set of rate equations of the form
\be\label{xD_eq}
\frac{dx_D}{dz}=\frac{x_D^2n_{D}(\mathcal{A}_D^{2s}+\mathcal{A}_D^{2p})-\mathcal{B}_D^{2s}x_{2s}- \mathcal{B}_D^{2p}x_{2p} }{H(z)(1+z)},
\ee
\ba\label{x_2s_eq}
\frac{dx_{2s}}{dz}&=&\frac{\mathcal{B}_D^{2s}x_{2s}-x_D^2n_{D}\mathcal{A}_D^{2s}+\mathcal{R}_{2p,2s} (3 x_{2s}-x_{2p})}{H(z)(1+z)} \en
&&+\frac{\Lambda_{2s,1s}(x_{2s}-(1-x_D)e^{-\frac{3B_D}{4T_D}})}{H(z)(1+z)},
\ea
\ba\label{x_2p_eq}
\frac{dx_{2p}}{dz}&=&\frac{\mathcal{B}_D^{2p}x_{2p}-x_D^2n_{D}\mathcal{A}_D^{2p}+\mathcal{R}_{2p,2s} (x_{2p}-3 x_{2s})}{H(z)(1+z)} \en
&&+\frac{R_{\text{Ly}\alpha}(x_{2p}-3(1-x_D)e^{-\frac{3B_D}{4T_D}})}{H(z)(1+z)},
\ea
where $\mathcal{A}_D^{nl}=\mathcal{A}_D^{nl}(T_{DM},T_D)$ is the effective recombination coefficient to the interface state $nl$, $\mathcal{B}_D^{nl}=\mathcal{B}_D^{nl}(T_D)$ is the effective photoionization rate of the interface state $nl$, $H(z)$ is the Hubble expansion rate and $z$ is the redshift which we use here as a time variable.   $R_{\text{Ly}\alpha}$ is the rate at which dark Lyman-$\alpha$ photons escape the Ly$\alpha$ resonance due to the expansion of the Universe \cite{2010PhRvD..82l3502A}
\be\label{RLyalpha}
R_{\text{Ly}\alpha}=\frac{2^{9}\alpha_D^3B_D}{3^8}\frac{(1-e^{-\tau_{\rm S}})}{\tau_{\rm S}},
\ee
where $\tau_{\rm S}$ is the Sobolev optical depth for the Lyman-$\alpha$ transition. Here and in the remainder of this paper, we set $\hbar=c=k_{\rm B}=1$. The second factor in Eq.~\ref{RLyalpha} is the probability that a Lyman-$\alpha$ dark photon redshifts out of the line wing before being reabsorbed. The Lyman-$\alpha$ Sobolev optical depth is given by \cite{2010PhRvD..82l3502A}
\be
\tau_{\rm S}=\frac{2^9\alpha_D^3B_Dn_D(1-x_D)}{3^78\pi H\nu_{\rm{Ly}\alpha}^3}\left(1-\frac{x_{2p}}{3(1-x_D)}\right),
\ee
where $\nu_{\rm{Ly}\alpha}$ is the frequency of the Lyman-$\alpha$ dark photons. $\Lambda_{2s,1s}$ denotes the rate of the forbidden two-photon transition between the $2s$ state and the ground state. It is given by \cite{1951ApJ...114..407S,Goldman:1981zz,Goldman:1989zz}
\be
\Lambda_{2s-1s}=\left(\frac{\alpha_D}{\alpha_{\rm em}}\right)^6\left(\frac{B_D}{B_{\rm H}}\right)8.22458\,\,\text{s}^{-1}.
\ee
Here $\alpha_{\rm em}$ is the SM fine-structure constant and $B_{\rm H}\simeq13.6$ eV is the binding energy of regular hydrogen. $\mathcal{R}_{2p,2s}$ is the effective transition rate between the interface states $2s$ and $2p$ and is given by \cite{AliHaimoud:2010ab}
\be
\mathcal{R}_{2p,2s} \equiv \sum_{nl}R_{2p\rightarrow nl}P_{nl\rightarrow 2s},
\ee
where $R_{2p\rightarrow nl}$ is the bound-bound transition rate between the $2p$ state and the ``interior" state $nl$ ($n > 2$) and $P_{nl\rightarrow 2s}$ stands for the probability that a dark atom in a state $nl$ will decay to the $2s$ state. Details of the computation of $R_{2p\rightarrow nl}$ and $P_{nl\rightarrow 2s}$ can be found in \cite{AliHaimoud:2010ab}.

\begin{figure}[t!]
\begin{centering}
\includegraphics[width=0.5\textwidth]{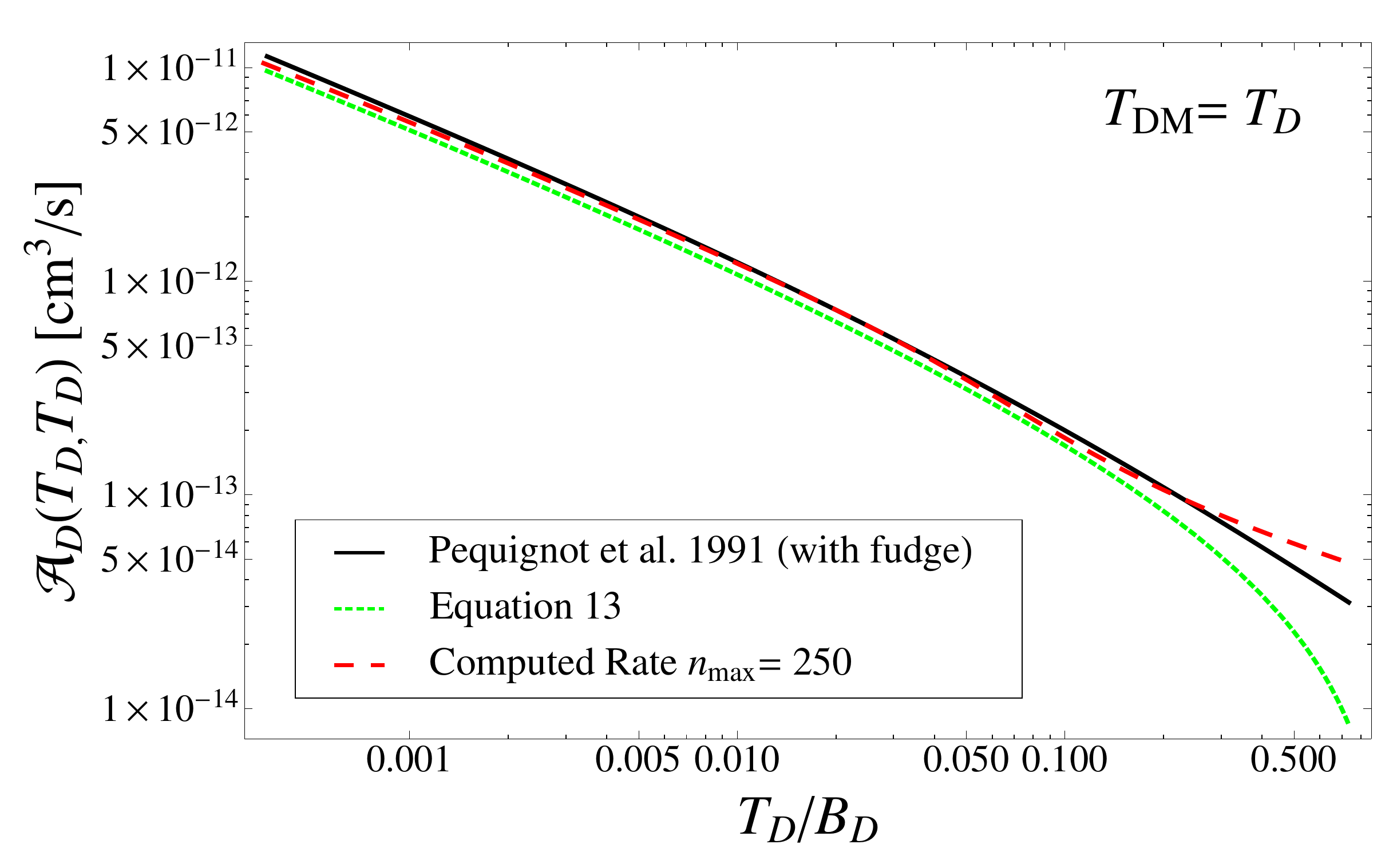}
\includegraphics[width=0.5\textwidth]{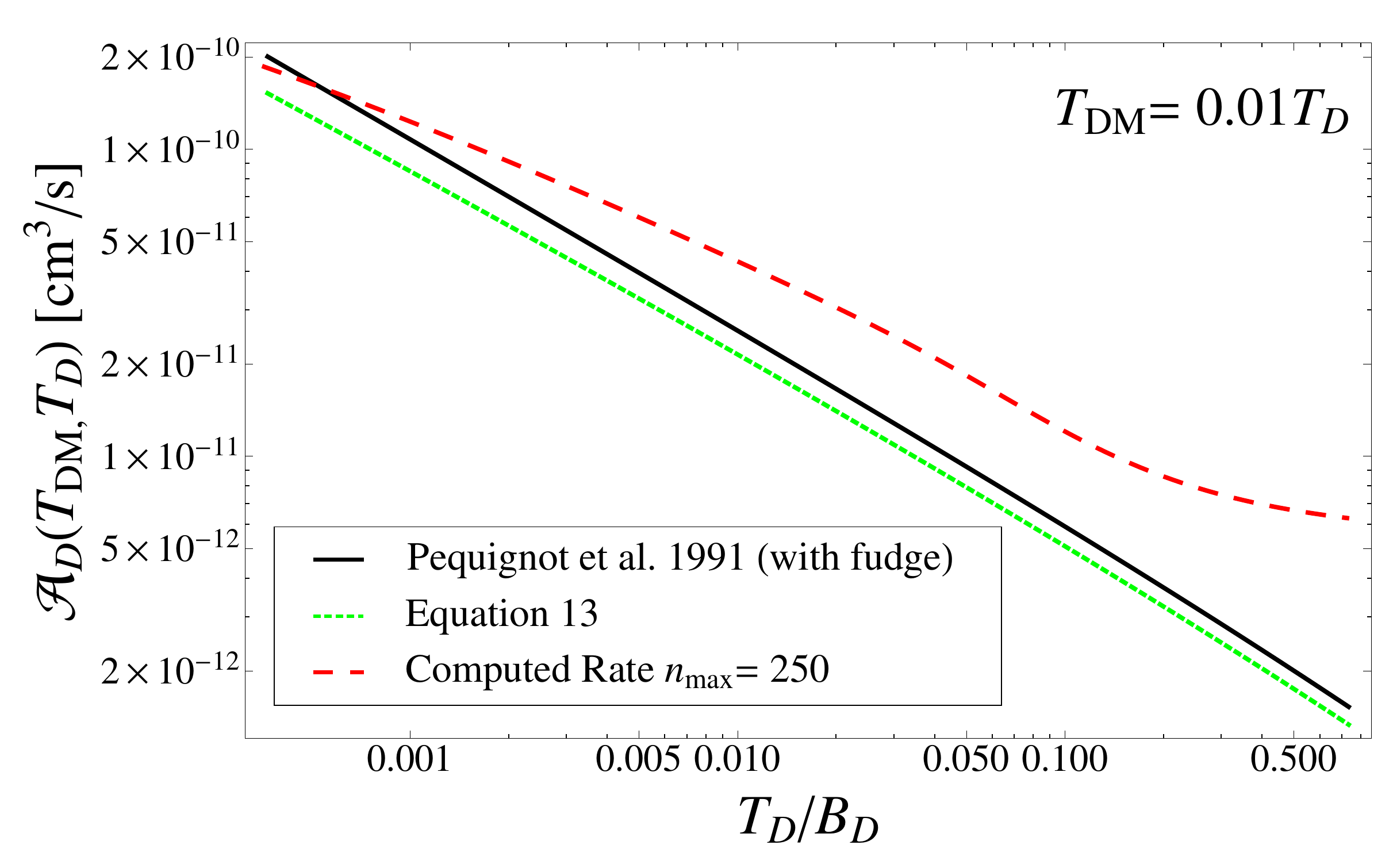}
\caption[Comparison between recombination rates.]{Comparison between recombination rates. We have chosen the dark sector parameters such that they match those of regular atomic hydrogen. We plot the approximate recombination rate given by Eq.~(\ref{rate_spitzer}) (green short-dashed line) as well as our rate computed according to Eq.~(\ref{exact_rec_rate}) including all shells up to $n_{\rm max}=250$ (red long-dashed line). For comparison, we also show the recombination rate given in Ref.~\cite{1991A&A...251..680P} corrected by a fudge factor of 1.14 as used in \texttt{Recfast} \cite{Seager:1999bc} (black solid line). \emph{Top Panel:} We compare the rates when the DM and DR are in thermal equilibrium such that $T_{DM}=T_D$. \emph{Lower Panel:} Similar to the top panel but with $T_{DM}=0.01T_D$.  }
\label{rec_rate_plot1}
\end{centering}
\end{figure}

Previous studies of atomic DM have used an approximate recombination coefficient given by \cite{spitzer_book,Ma:1995ey}
\be\label{rate_spitzer}
\mathcal{A}_D(T_{DM})=0.448\frac{64\pi}{\sqrt{27\pi}}\frac{\alpha_D^2}{\mu_D^2}\left(\frac{B_D}{T_{DM}}\right)^{1/2}\ln{\left(\frac{B_D}{T_{DM}}\right)}. 
\ee
This rate is problematic for three reasons. First, since it is only a function of the DM temperature, it fails to take into account the contribution from stimulated recombination which explicitly depends on $T_D$. Second, it does not take into account the effect of transitions among the high-$n$ atomic states and assume equilibrium among the angular momentum substates (an approximately $14\%$ correction). Finally, it is inaccurate for $T_{DM} \gtrsim B_D$ and $T_{DM} \ll B_D$. To improve this situation and obtain an accurate picture of dark recombination, we compute a new recombination coefficient using the method outlined in \cite{AliHaimoud:2010ab}. Our recombination coefficient is given by
\be\label{a_Dnl}
\mathcal{A}_D^{nl}(T_{DM},T_D)\equiv\alpha_{nl}+\sum_{n'l'}\alpha_{n'l'}P_{n'l'\rightarrow nl}.
\ee
Here, $P_{n'l'\rightarrow nl}$ stands for the probability that a dark atom in a state $n'l'$ will decay to a state $nl$. The volumetric recombination rate to an atomic state $nl$ is
\ba\label{alpha_rec}
\alpha_{nl}(T_{DM},T_D)&=&\frac{(2\pi)^{3/2}}{(\mu_DT_{DM})^{3/2}}\\
&&\times\int_0^{\infty}e^{-B_D\kappa^2/T_{DM}}\gamma_{nl}(\kappa)\en
&&\times\left[1+f_{\rm BB}(B_D(\kappa^2+n^{-2}),T_D)\right]d(\kappa^2).\nonumber
\ea
Here, $\kappa$ denotes the momentum of the incoming dark electron in units of $\alpha_D/2B_D$ and $f_{\rm BB}(E,T)\equiv(e^{E/T}-1)^{-1}$ is the dark-photon distribution function at energy $E$ for a blackbody spectrum at temperature $T$. The details of the atomic physics are encoded in the $\gamma_{nl}$ factor \cite{AliHaimoud:2010ab}
\ba\label{gamma_nl}
\gamma_{nl}(\kappa)&\equiv&\frac{1}{3\pi n^2}\alpha_D^3B_D(1+n^2\kappa^2)^3\en
&&\quad\times\sum_{l'=l\pm1}\text{max}(l,l')|g(n,l,\kappa,l')|^2,
\ea
where $g(n,l,\kappa,l')$ denotes the bound-free radial matrix elements \cite{1965MmRAS..69....1B}. Numerically computing the momentum integral in Eq.~(\ref{alpha_rec}) for each state $nl$ and performing the sums in Eq.~(\ref{a_Dnl}) up to $n_{\text{\rm max} }= 250$ yield a recombination coefficient of the form
\be\label{exact_rec_rate}
\mathcal{A}_D^{nl}(T_{DM},T_D)=\frac{2\sqrt{2\pi}\alpha_D^3}{3\mu_D^{3/2}\sqrt{T_{DM}}}G^{nl}_{250}(\frac{T_D}{B_D},\frac{T_{DM}}{T_D}),
\ee
where $G^{nl}_{250}$ are universal dimensionless functions encoding the details of the atomic physics and its interaction with the radiation field. These functions are independent of the model parameters ($\alpha_D$, $m_{D}$, $B_D$, $\xi$) and therefore need to be computed only once. For the purpose of numerical computation, we tabulate $G^{2s}_{250}$ and $G^{2p}_{250}$ on a grid of $T_D/B_D$ and $T_{DM}/T_D$ values and use an interpolation scheme to obtain accurate values of the effective recombination coefficients\footnote{The code to compute these coefficients was kindly provided by Yacine Ali-Ha{\"i}moud.}. The  photoionization rates are related to the above photorecombination coefficients through detailed balance
\be
\mathcal{B}^{nl}_D(T_D)=\left(\frac{\mu_DT_D}{2\pi}\right)^{3/2}e^{-\frac{B_D}{n^2T_D}}\mathcal{A}^{nl}_D(T_{DM}=T_D,T_D).
\ee
In terms of the universal dimensionless functions, this reads 
\be
\mathcal{B}^{nl}_D(T_D)=\frac{\alpha_D^3T_{D}}{3\pi}e^{-\frac{B_D}{n^2T_D}}G^{nl}_{250}(\frac{T_D}{B_D},1).
\ee
In Fig.~\ref{rec_rate_plot1}, we compare the effective total recombination rate $\mathcal{A}_D(T_{DM},T_D)\equiv\mathcal{A}_D^{2s}(T_{DM},T_D)+\mathcal{A}_D^{2p}(T_{DM},T_D)$ with the approximate rate given by Eq. (\ref{rate_spitzer}). The top panel compares the recombination rate when the DM and the DR are in thermal equilibrium with $T_{DM}=T_D$. We see that the approximate rate (\ref{rate_spitzer}) performs reasonably well over the temperature range where most of the recombination is happening ($0.007\lesssim T_D/B_D\lesssim0.02$). Most of the difference between Eq.~(\ref{rate_spitzer}) and our exact computation can be traced to the fact that the former neglects the transitions between the high-$n$ atomic levels and assumes that the angular momentum substates are in equilibrium. As shown in Ref.~\cite{Seager:1999bc}, this could in principle be taken into account by multiplying Eq.~(\ref{rate_spitzer}) by a fudge factor $\sim1.14$. We also show the recombination rate derived in Pequignot et al.~\cite{1991A&A...251..680P} corrected by this fudge factor as used in \texttt{Recfast} \cite{Seager:1999bc}. Not surprisingly, this last rate is an excellent fit to the exact rate over most of the important temperature range. However, we see that both the Pequignot et al.~and the rate given in Eq.~(\ref{rate_spitzer}) fails at high temperature and to a lesser extent, at low temperature. For regular atomic hydrogen, these errors are inconsequent since most of the baryonic plasma is ionized until $T_{\rm CMB}/B_H\sim0.02$ and has mostly recombined before $T_{\rm CMB}/B_H\sim0.003$. For weakly-coupled dark atoms however, these errors can have a substantial effect on the late-time ionization fraction of the dark sector.

The lower panel of Fig.~\ref{rec_rate_plot1} compares the different recombination rates when the DM temperature differs significantly from that of the DR. This can happen for example for a weakly-coupled dark sector which thermally decouples from the DR before the onset of or during recombination. In this case, both the canonical rate (\ref{rate_spitzer}) and that of Pequignot et al. fail to capture the correct temperature dependence over most of the important temperature range. This should not come as a surprise as these two rates are purely functions of $T_{DM}$ and cannot therefore capture the contribution from stimulated recombination which is a function on the DR temperature, $T_D$. For regular atomic hydrogen, this effect is unimportant since thermal decoupling of baryons happens well after recombination. Note that Eq.~(\ref{rate_spitzer}) systematically underestimates the recombination rate leading to an artificially large late-time ionization fraction for weakly-coupled dark atoms. Given the sensitivity of the recombination rate on the DM temperature, it is important to accurately capture its evolution through the stage of thermal decoupling which we discuss in the next section.   

Without solving any differential equation, it is possible to obtain an estimate for the late-time ionized fraction of the dark sector $\bar{x}_D$ by solving the condition 
\be
\bar{x}_Dn_D(\mathcal{A}_D^{2s}+\mathcal{A}_D^{2p})\simeq H,
\ee
which determines the ionized fraction when the recombination process goes out of equilibrium. Using the above expression for the recombination rates and evaluating them at $T_D/B_D\sim0.007$, we obtain
\be
\bar{x}_D\sim 2\times10^{-16}\frac{\xi}{\alpha_D^6}\left(\frac{\Omega_Dh^2}{0.11}\right)^{-1}\left(\frac{m_D}{\text{GeV}}\right)\left(\frac{B_D}{\text{keV}}\right).
\ee
While this expression is only accurate up to a factor as large as 10, it illustrates how the relic ionized fraction scales with the dark parameters.

In our above treatment of dark recombination, we have only considered the contribution from radiative processes. In principle, collisional processes could also contribute to the ionization, recombination, and bound-bound rates. To estimate whether we can safely neglect the contribution from collisional processes, we compare here the radiative rate for the $1s\rightarrow2p$ transition to its collisional counterpart. The radiative rate is given by \cite{AliHaimoud:2010ab}
\be
R_{1s\rightarrow2p}^{\rm rad}(T_D)=\frac{2^{10}\pi\alpha_D^3B_D}{3^7}e^{\frac{-3B_D}{4T_D}}.
\ee
The corresponding rate for a dark electron to collisionally excite a dark atom from its ground state to the $2p$ state is given by (see e.g. Ref.~\cite{PhysRevA.55.329})
\be
R_{1s\rightarrow2p}^{\rm coll}(T_{DM})\simeq\frac{9\sqrt{3}\alpha_D^2}{2^4\sqrt{\pi}B_D^2}n_{\bf e}\left(\frac{T_{DM}}{B_D}\right)^{-\frac{1}{2}}e^{\frac{-3B_D}{4T_{DM}}},
\ee
where we have assumed that the dark electrons have a Maxwellian velocity distribution with temperature $T_{DM}$. Demanding that the radiative rate dominates over the collisional rate at the onset of dark recombination leads to the condition
\be\label{rad_vs_coll}
\alpha_D\xi^3\left(\frac{m_D}{\rm GeV}\right)\left(\frac{\Omega_{ D}h^2}{0.11}\right)^{-1}>10^{-10},
\ee
where we have taken $T_D\sim T_{DM}\sim0.02B_D$ and $x_D\sim1$. We see that the radiative processes are expected to dominate over their collisional counterpart unless the dark sector is very cold ($\xi\ll1$), dark atoms are very light ($m_D\ll 1$ GeV) or very weakly coupled. As we will discuss in section \ref{cosmology}, all atomic DM models that are observationally relevant obey the condition given by Eq.~\ref{rad_vs_coll}. We therefore neglect any contribution from collisional processes in this work.
 
\subsection{Thermal Decoupling of Atomic Dark Matter}\label{thermal_decoupling}
In the early Universe, frequent interactions between the dark photons and the dark fermions keep the dark sector in thermal equilibrium at a single temperature. Dark photons Compton scatter off dark electrons, hence transferring energy to the DM gas. This energy is then redistributed among the dark sector fermions through Coulomb scattering between the dark electrons and the dark protons. The typical timescale for this process is  \cite{Kobayashi:2007wd}
\ba
\tau_{\text{\bf e-p}}&=&\frac{\sqrt{\mu_D}\,T_{DM}^{3/2}}{\sqrt{2\pi}\alpha_D^2n_{D}x_D\ln{\Lambda}}\\
&\simeq&\frac{3.8\times10^{-4}\xi^{\frac{3}{2}}}{\alpha_D^2x_D\ln\Lambda}\left(\frac{0.11}{\Omega_Dh^2}\right)\en
&&\quad\times\left(\frac{m_{D}}{\text{GeV}}\right)\left(\frac{\mu_D}{\text{MeV}}\right)^{\frac{1}{2}}\left(\frac{T}{\text{eV}}\right)^{-\frac{3}{2}}\text{s},\nonumber
\ea
where we have assumed $T_{DM}=T_D$ in going from the first line to the second equality. This timescale should therefore be considered as an upper limit since after thermal decoupling, we generally have $T_{DM} < T_D$. Here, $\ln{\Lambda}$ is the Coulomb logarithm and is approximately given by
\be
\ln \Lambda\simeq\ln\left[\frac{T_{DM}^{3/2}}{\sqrt{\pi n_{D}x_D}\alpha_D^{3/2}}\right], 
\ee
and has value $\ln\Lambda\sim30-60$ over the parameter space of interest. The Coulomb rate should be compared to the Hubble timescale during radiation domination 
\be\label{hubble_time_rad}
\tau_H\equiv\frac{1}{2} H^{-1}\simeq\frac{2.42\times10^{12}}{\sqrt{3.36+g_{*,D}\xi^4}}\left(\frac{T}{\text{eV}}\right)^{-2}\text{s} .
\ee
For all interesting atomic DM parameter space, we always have $\tau_{\text{\bf e-p}}\ll\tau_H$ and therefore, we can always assume that the dark fermions are in thermal equilibrium among themselves at a single temperature $T_{DM}$. The neutral dark atoms maintain thermal contact with the dark fermions through elastic collisions. To a good approximation, the cross section for a collision between a dark electron of energy $E_{\bf e}$ and a dark atom in its ground state is given by \cite{Stepanek200399}
\be  
\sigma_{\bf e-H}(E_{\bf e})\simeq\frac{320\alpha_D^2}{B_D^2}\frac{1}{\sqrt{1+\gamma(E_{\bf e}/B_D)^2}},
\ee 
which is valid for $E_{\bf e} < B_D$ and where $\gamma\simeq15.69$ is a best-fit parameter. Averaging this cross-section over the Maxwell-Boltzmann velocity distribution of the dark electrons and multiplying by the number density of neutral atoms, we obtain the elastic collision rate between dark atoms and dark electrons
\ba
\Gamma_{\bf e-H}&\simeq&\frac{5.3\times10^8\alpha_D^2(1-x_D)}{\left(1+15.07\left(\frac{T_{DM}}{B_D}\right)^{\frac{3}{2}}\right)^{0.576}}\left(\frac{T_{DM}}{m_{\bf e}}\right)^{\frac{1}{2}}\\
&&\quad\times\left(\frac{\Omega_Dh^2}{0.11}\right)\left(\frac{B_D}{\text{eV}}\right)^{-2}\left(\frac{m_D}{\text{GeV}}\right)^{-1}\left(\frac{T}{\text{eV}}\right)^{3}\text{s}^{-1},\nonumber
\ea
which is valid for $T_{DM}/B_D\lesssim1$. Again, the timescale associated with this collisional process $\tau_{\bf e-H}\equiv\Gamma_{\bf e-H}^{-1}$ is much shorter than the Hubble time over most of the dark atom parameter space. We can therefore safely assume that the whole DM sector (ions +  neutral dark atoms) is always in thermal equilibrium at a single temperature $T_{DM}$. For the remainder of this section, we thus focus our attention on the interaction between the dark fermions and the DR. 

Due to its steep dependence on the DR temperature, Compton heating is always the dominant energy-transfer mechanism between DM and DR at early times. The Compton heating rate is given by \cite{Seager:1999km}
\be
\Gamma_{\rm Compton}=\left[1+\left(\frac{m_{\bf e}}{m_{\bf p}}\right)^3\right]\frac{64\pi^3\alpha_D^2T_D^4}{135 m_{\bf e}^3}\frac{x_D}{1+x_D},
\ee
where $m_{\bf e}$ and $m_{\bf p}$ are the masses of the dark electron and dark proton, respectively. The prefactor in the square bracket accounts for Compton heating of both dark electrons and protons. For regular atomic hydrogen, the large photon-to-baryon ratio ensures that Compton heating alone efficiently maintains thermal contact between baryons and the photon bath well after the former recombine into neutral hydrogen and helium. For dark atoms however, there is a large parameter space for which Compton heating becomes inefficient at early times (i.e.~for $T_D\gg B_D$). In this case, one must consider other possible energy-exchange mechanisms between the DR and DM. Before dark recombination, photo-ionization heating and photo-recombination cooling are the most important mechanisms that can maintain thermal contact between DM and DR once Compton heating falls out of equilibrium. Free-free (Bremsstrahlung) cooling and free-free heating are also relevant energy-exchange mechanisms for dark atoms. Finally, dark photons can exchange energy with neutral dark atoms through Rayleigh scattering. The volumetric energy-exchange rates for these processes are (in energy per unit time per unit volume, see Appendix \ref{thermal_rates_app})
\be\label{p-i}
\Pi_{\rm p-i}(T_D)=\frac{\alpha_D^3T_D^2}{3\pi}x_{2s}n_{D}e^{-\frac{B_D}{4T_D}}F_{\rm p-i}(T_D/B_D),
\ee
\ba\label{p-r}
\Pi_{\rm p-r}=\frac{2\alpha_D^3\sqrt{2\pi T_{DM}}}{3\mu_D^{3/2}}x_D^2n_{D}^2F_{\rm p-r}(\frac{T_D}{B_D},\frac{T_{DM}}{T_D}),
\ea
\ba\label{ff}
\Pi_{\rm ff}&\simeq&\frac{16\alpha_D^3\bar{g}_{ff}\sqrt{2\pi T_{DM}}x_D^2n_{D}^2}{(3\mu_D)^{3/2}}\en
&&\times\left(\frac{\pi^2\epsilon(1+2\epsilon)-6\zeta(3)\epsilon^2}{6}\right),
\ea
\be\label{rayleigh}
\Pi_{\rm R} \simeq \frac{430080\zeta(9)\alpha_D^2n_{D}(1-x_D)T_D^9\epsilon}{\pi^2B_D^4m_{D}m_{\bf e}^2},
\ee
where $\Pi_{\rm p-i}$, $\Pi_{\rm p-r}$, $\Pi_{\rm ff}$ and $\Pi_{\rm R}$ are the photo-ionization heating, photo-recombination cooling, the net free-free heating rate, and the net Rayleigh heating rate, respectively. $F_{\rm p-i}$ and $F_{\rm p-r}$ are universal dimensionless functions parametrizing the details of the atomic physics and its interaction with the radiation field. These functions are independent of the model parameters ($\alpha_D,B_D,m_{D},\xi$) and therefore need to be computed only once. The fractional temperature difference is denoted by $\epsilon\equiv(T_D-T_{DM})/T_D$.
\begin{figure}[h!]
\begin{centering}
\includegraphics[width=0.5\textwidth]{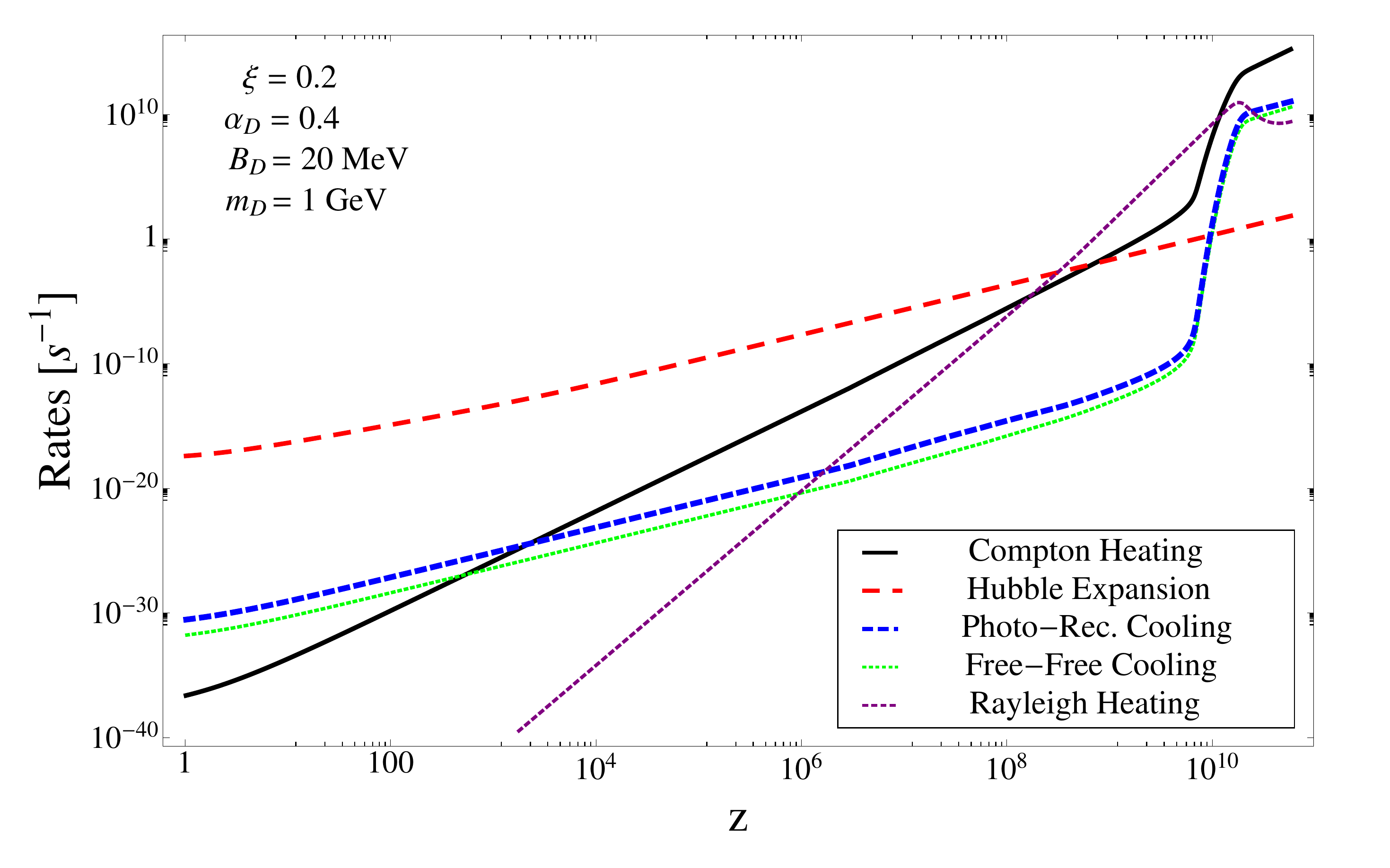}
\includegraphics[width=0.5\textwidth]{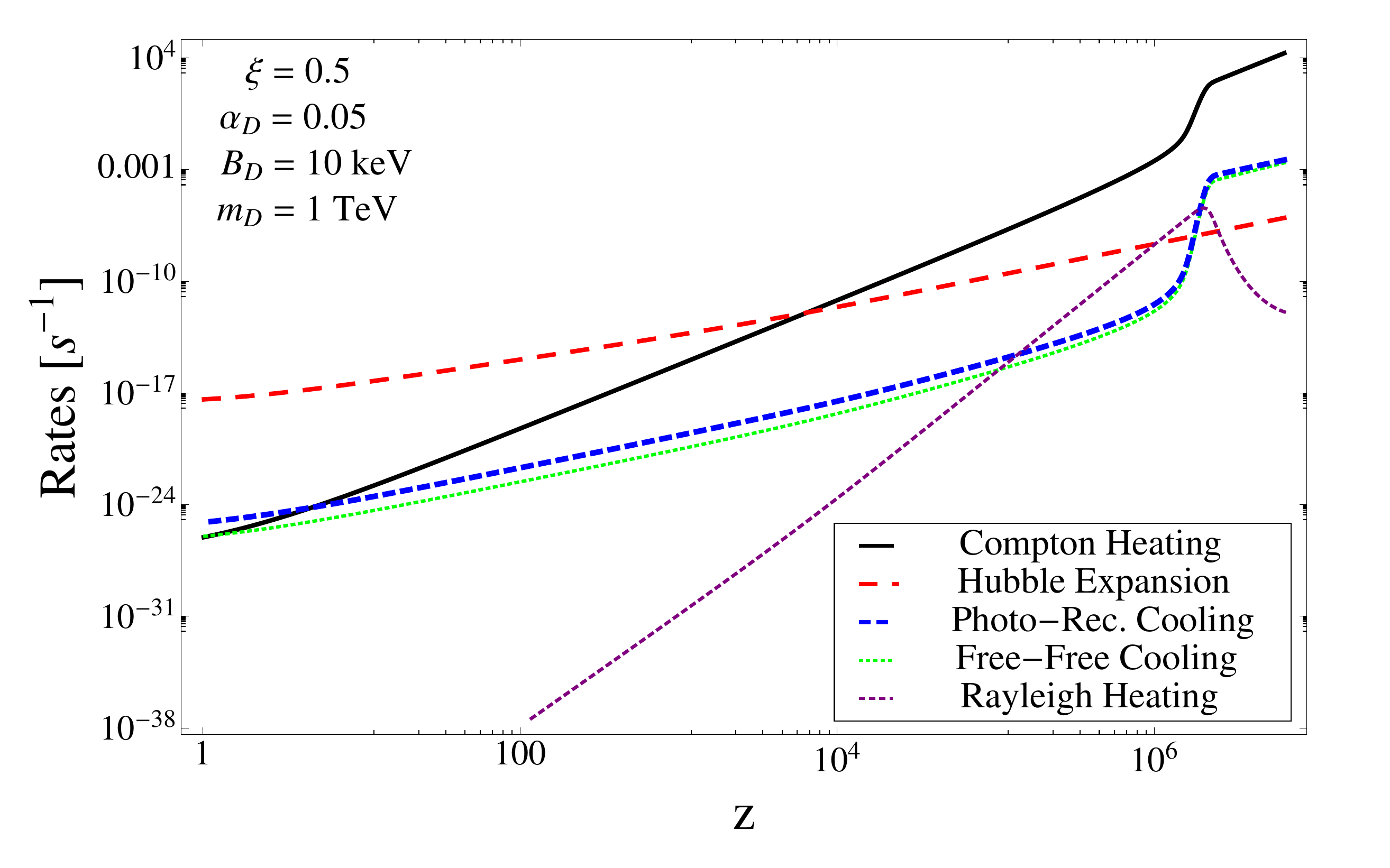}
\includegraphics[width=0.5\textwidth]{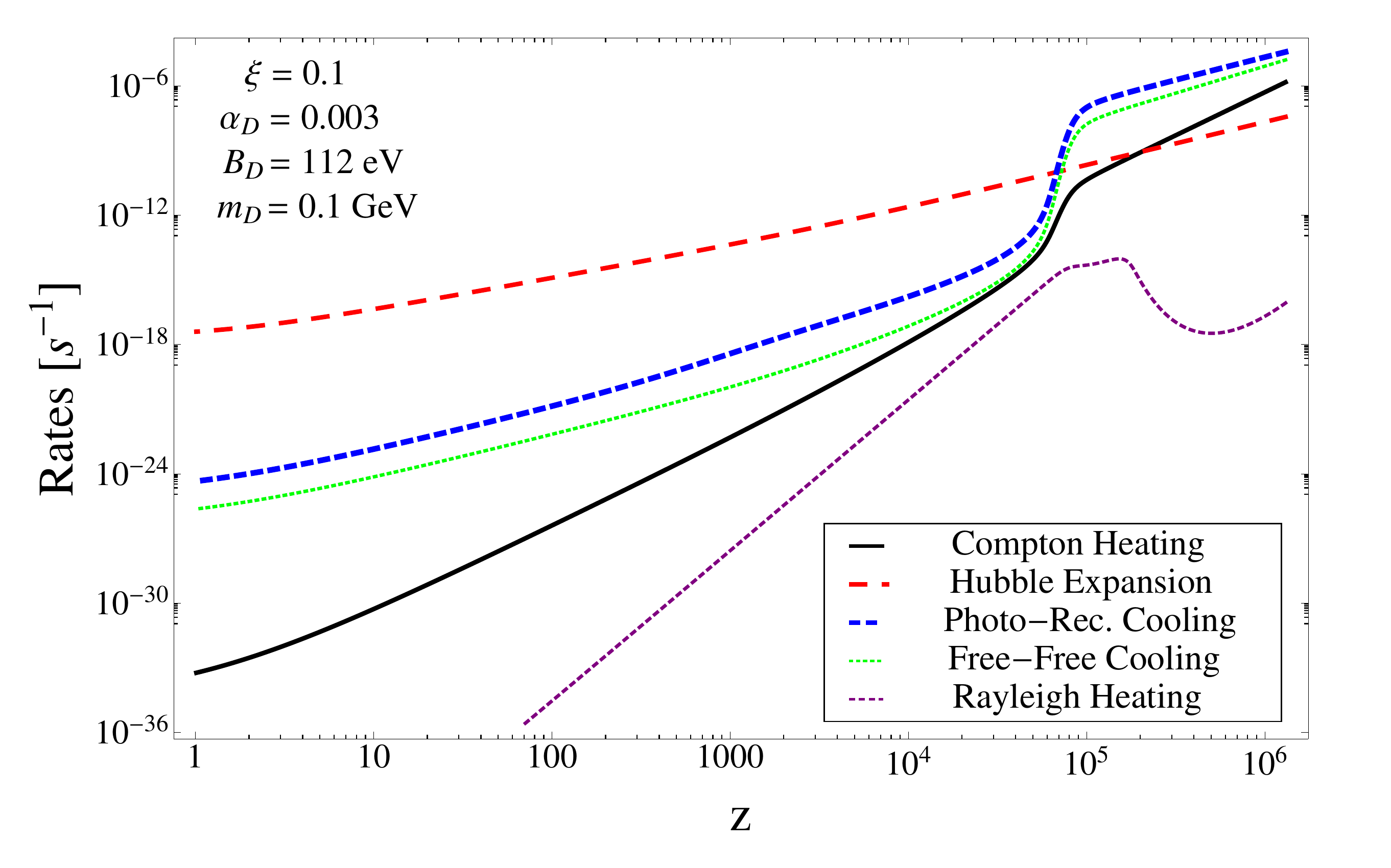}
\caption[Comparison between the rates of different energy-exchange mechanism.]{Comparison between the rates of different energy-exchange mechanism. We display the rates for Compton heating (solid, black), photo-recombination cooling (short-dashed, blue), and free-free cooling (dotted, green). We also show the Hubble expansion rate (long-dashed, red). The upper panel displays the evolution of the thermal rates for an atomic DM model with $\Upsilon_{BF}\sim700$ and $\Upsilon_R\sim6\times10^{-6}$, the middle panel has $\Upsilon_{BF}\sim280$ and $\Upsilon_R\sim400$, while the lower panel shows the evolution for a model with $\Upsilon_{BF}\sim5\times10^{-4}$ and $\Upsilon_R\sim4$. }
 \label{thermal_rates}
 \end{centering}
\end{figure}

It is instructive to compare the relative magnitude of the different energy-exchange mechanisms. The typical timescale required by the thermal processes to transfer an $\mathcal{O}(1)$ fraction of the kinetic energy of the DM is given by
\be
\tau_i\equiv\Gamma_i^{-1}=\left(\frac{2\Pi_i}{3T_{DM}n_{D}(1+x_D)}\right)^{-1},
\ee
where we have assumed the equipartition of energy among all dark constituents. Comparing Eqs.~(\ref{p-r}) and (\ref{ff}), we see that the free-free processes and the bound-free processes have  similar leading-order amplitudes. Explicitly taking ratio of the Compton heating rate to that of photo-recombination cooling, we obtain
\be
\frac{\Gamma_{\rm Compton}}{\Gamma_{\rm p-r}}\sim10^{-3}\frac{\alpha_D^2}{x_D}\left(\frac{T_{DM}}{B_D}\right)^{\frac{1}{2}}\left(\frac{m_{D}}{B_D}\right)\frac{\xi^4(1+z)}{F_{\rm p-r}}.
\ee
We first notice that it is always possible to find a high-enough redshift such that Compton heating dominates over the bound-free energy-exchange channels. As the Universe cools down, the photo-heating can become the dominant energy-exchange mechanism for low values of $\alpha_D$, for a cold dark sector ($\xi\ll1$), or for a light DM candidate. Generally, photo-recombination cooling, photo-ionization heating, and the free-free processes must be taken into account when
\ba
\Upsilon_{\rm BF}&\equiv&\frac{\Gamma_{\rm Compton}}{\Gamma_{\rm p-r}}\Big|_{z_{\rm drec}}\en
&\simeq&5.6\times10^5\alpha_D^2\xi^3\left(\frac{0.11}{\Omega_Dh^2}\right)\left(\frac{m_{D}}{\text{GeV}}\right)\lesssim1.
\ea
Here, $z_{\rm drec}$ is the redshift at which dark atoms recombine. For standard atomic hydrogen, we have $\Upsilon_{\rm BF}\sim10^2$ and can therefore neglect any contribution beyond Compton heating. 

For relatively large values of the coupling constant $\alpha_D$, recombination is generally very efficient,  resulting in a considerably low ionized fraction $x_D\ll1$ at late times. Consequently, thermal coupling mechanisms that depend on the presence of free ions to proceed such as Compton heating and free-free heating become relatively inefficient. When this happens, the only remaining mechanism that can maintain thermal equilibrium between the DR and DM for these models is Rayleigh heating.  Generally, Rayleigh heating takes over Compton heating as the dominant heat-exchange channel after dark recombination if the condition
\be
\Upsilon_{\rm R}\equiv\frac{\Gamma_{\rm Compton}}{\Gamma_{\rm R}}\Big|_{z_{\rm drec}}\simeq\frac{5\times10^4x_D}{1-x_D}\left(\frac{m_D}{m_{\bf e}}\right)\lesssim1,
\ee
is satisfied. This indicates that only models with very small left-over ionization fraction and moderate value of the ratio $m_D/m_{\bf e}$ can obtain a non-negligible contribution from Rayleigh heating. In Fig.~\ref{thermal_rates}, we compare the rates for the four dominant energy-exchange mechanisms to the Hubble rate. The upper panel shows a relatively strongly-coupled dark-atom model with $\Upsilon_{\rm R}\sim6\times10^{-6}$ and $\Upsilon_{\rm BF}\sim700$ where Rayleigh heating becomes the dominant thermal-coupling mechanism after the onset of dark recombination.  In the middle panel, we illustrate a model with $\Upsilon_{\rm BF}\sim280$ and $\Upsilon_{\rm R}\sim400$ where Compton heating is the only important mechanism until adiabatic cooling takes over at late times. The lower panel displays an alternate scenario with $\Upsilon_{\rm BF}\sim5\times10^{-4}$ and $\Upsilon_{\rm R}\sim4$ where photo-ionization heating and photo-recombination cooling dominate from early times until adiabatic cooling takes over after recombination.

We therefore see that the thermal evolution of atomic DM strongly depends on the specific choice of dark parameters. In contrast to regular atomic hydrogen whose thermal history can be accurately captured by only considering Compton heating, it is necessary in general to include complementary thermal coupling channels to precisely determine the thermal-decoupling temperature of dark atoms. Put differently, the thermal history of the baryon-photon plasma represents only one possibility among all the regimes that a plasma in an expanding universe can explore. In this respect, it is interesting to realize that the thermal evolution of the baryon-photon plasma is rather simple compared to what it could have been, had the parameters of SM been different. 

Putting the different pieces together, the Boltzmann equation governing the evolution of the DM temperature is then \cite{Seager:1999km}
\ba\label{TDM_eq}
(1+z)\frac{dT_{DM}}{dz}&=&2T_{DM}+\frac{2(\Pi_{\rm p-r}-\Pi_{\rm p-i}-\Pi_{\rm ff}+\Pi_{\rm R})}{3k_{\rm B}n_{D}(1+x_D)H(z)}\en
&&\hspace{-2.4cm}+\frac{64\pi^3\alpha_D^2T_D^4}{135 m_{\bf e}^3H(z)}\frac{x_D(T_{DM}-T_D)}{1+x_D}\left[1+\left(\frac{m_{\bf e}}{m_{\bf p}}\right)^3\right],
\ea
where the first term of the right-hand side corresponds to adiabatic cooling due to the expansion of the Universe, the second term takes into account the contribution from bound-free and free-free processes as well as Rayleigh scattering, and the last term describes the Compton heating of the DM gas.
\subsection{Joint-Evolution of DM Temperature and Ionization Fraction}
The thermal history of the dark sector sector is specified by simultaneously solving Eqs.~(\ref{xD_eq}), (\ref{x_2s_eq}), (\ref{x_2p_eq}), and (\ref{TDM_eq}) together with the initial conditions
\ba
x_D(z_{\rm i})=x_{D,{\rm Saha}}(z_{\rm i})&\qquad &x_{2s}(z_{\rm i}) = e^{-\frac{3B_D}{4T_D}}(1-x_D(z_{\rm i}))\en
x_{2p}(z_{\rm i}) = 3 x_{2s}(z_{\rm i}) &\qquad& T_{DM}(z_{\rm i})=T_D(z_{\rm i}),\nonumber
\ea
where $x_{D,{\rm Saha}}$ is the Saha equilibrium ionization fraction and $z_{\rm i}$ is the initial redshift. It is obtained by solving the Saha equation
\be
\frac{x_{D,{\rm Saha}}^2}{1-x_{D,{\rm Saha}}}=\frac{1}{n_{D}}\left[\left(\frac{m_{\bf e}m_{\bf p} T_D}{2\pi m_{D}}\right)^{3/2}e^{-B_D/T_D}\right].
\ee
These ionization and temperature evolution equations are extremely stiff at early times and therefore require the use of a stiff solver. We assume that adiabatic cooling dominates the evolution of the DR temperature such that
\be\label{approx_TD_evol_wo_g}
T_D(z)=T_{\rm CMB}^{(0)}\xi(1+z),
\ee
where $T_{\rm CMB}^{(0)}$ is the temperature of the CMB today. Note that dark recombination always happens after the dark electrons and positrons annihilate and we have therefore neglected the change in the number of relativistic species (see Eq.~\ref{real_TD_evol_w g}). Equation (\ref{approx_TD_evol_wo_g}) is valid as long as we can neglect the energy injected into the dark radiation bath in a Hubble time, that is
\be\label{derho_o_rho_def}
\frac{\de\rho_{\gamma,D}}{\rho_{\gamma,D}}\simeq\frac{\Pi_{\rm p-r}\tau_H}{\rho_{\gamma,D}}\ll1.
\ee
In the above, we have only included the contribution from photo-recombination. In practice, all mechanisms leading to a net energy transfer from the DM to the DR should be included. Before DM recombination, the right-hand side of Eq.~(\ref{derho_o_rho_def}) is almost a constant and is equal to
\be\label{derho_o_rho}
\frac{\de\rho_{\gamma,D}}{\rho_{\gamma,D}}\sim10^2\frac{\alpha_D^6}{\xi^4}\left(\frac{m_{D}}{\text{GeV}}\right)^{-2}\left(\frac{B_D}{\text{eV}}\right)^{-1}\ll1.
\ee
Models not respecting this bound are likely to require a more involved analysis since the DR field cannot be taken to be thermal in these cases. A full solution to the energy-transfer problem is out of the scope of this paper and we therefore focus on models obeying Eq.~(\ref{derho_o_rho}).

\begin{figure}
\includegraphics[width=0.5\textwidth]{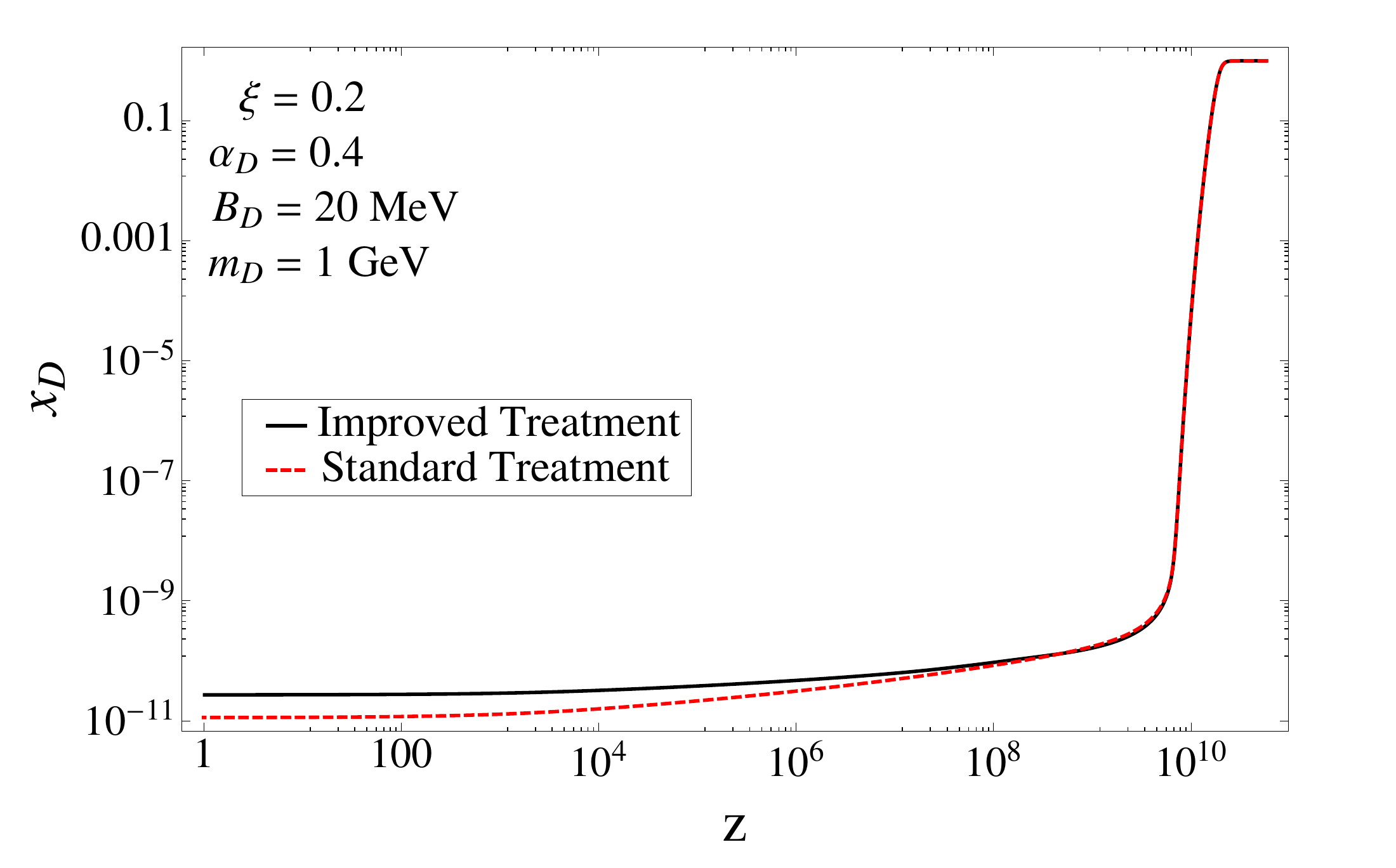}
\includegraphics[width=0.5\textwidth]{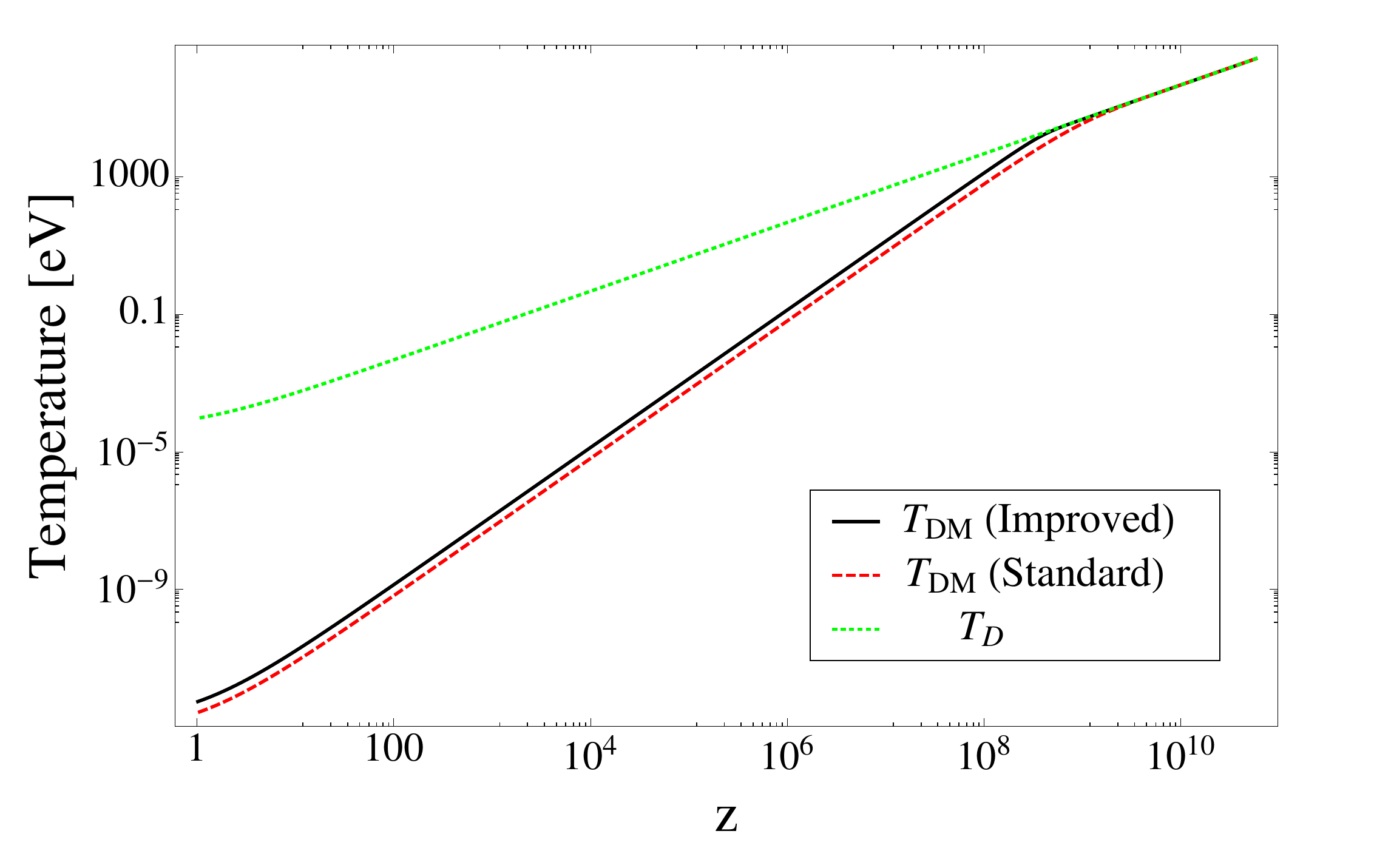}
\caption[Comparison between our improved treatment of dark recombination and the standard treatment for a strongly-coupled model.]{Comparison between our improved treatment of dark recombination and the standard treatment. We display results for a relatively strongly-coupled dark sector with $\Upsilon_{\rm R}\sim6\times10^{-6}$ and $\Upsilon_{\rm BF}\sim700$. The upper panel shows the evolution of the ionization fraction as a function of redshift while the lower panel shows the corresponding evolution of the DM and DR temperatures.}
\label{xe_plot_s}
\end{figure}

We can now compare our improved treatment of the dark-atom recombination and thermal decoupling to the ``standard treatment" described in Ref.~\cite{Kaplan:2009de}. The ``standard treatment" combines Eqs.~(\ref{xD_eq}), (\ref{x_2s_eq}), and (\ref{x_2p_eq}) into a single differential equation for $x_D$ and uses the approximate recombination rate given in Eq.~(\ref{rate_spitzer}). Its DM temperature evolution equation only includes Compton heating and adiabatic cooling. In Fig.~\ref{xe_plot_s}, we display the ionized-fraction and temperature evolution for a model with $\Upsilon_{\rm R}\sim6\times10^{-6}$ and $\Upsilon_{\rm BF}\sim700$. We see that the redshift evolution of our improved ionization calculation closely matches the standard treatment at early times. At late times however, our improved calculation predicts an ionization fraction that differs from the standard treatment by as much as $60\%$. The very good agreement at early times follows from the fact that the ionized fraction of the dark sector is initially mostly controlled by the Saha equilibrium condition, and therefore insensitive to the exact value of the recombination coefficient. At late times however, the Saha approximation breaks down and the ionized fraction becomes sensitive to the recombination rates. As can be seen in the lower panel of Fig.~\ref{xe_plot_s}, the inclusion of Rayleigh scattering postpones the thermal decoupling of DM, resulting in a DM gas that is hotter than one would expect by considering only Compton heating. Since the recombination coefficient is very sensitive to the DM temperature, this explains the somewhat large difference in $x_D$ at low redshift. 

\begin{figure}
\includegraphics[width=0.5\textwidth]{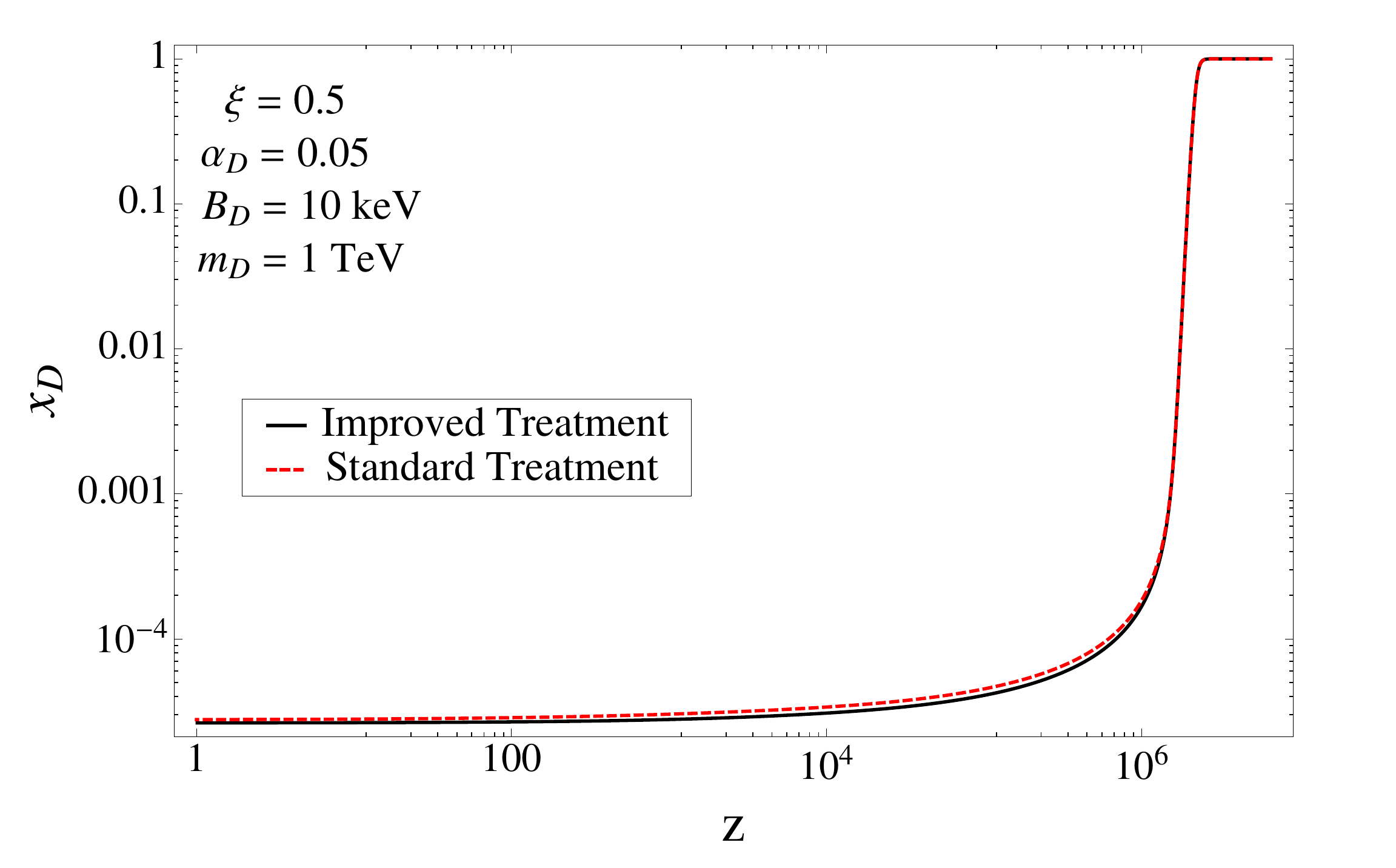}
\includegraphics[width=0.5\textwidth]{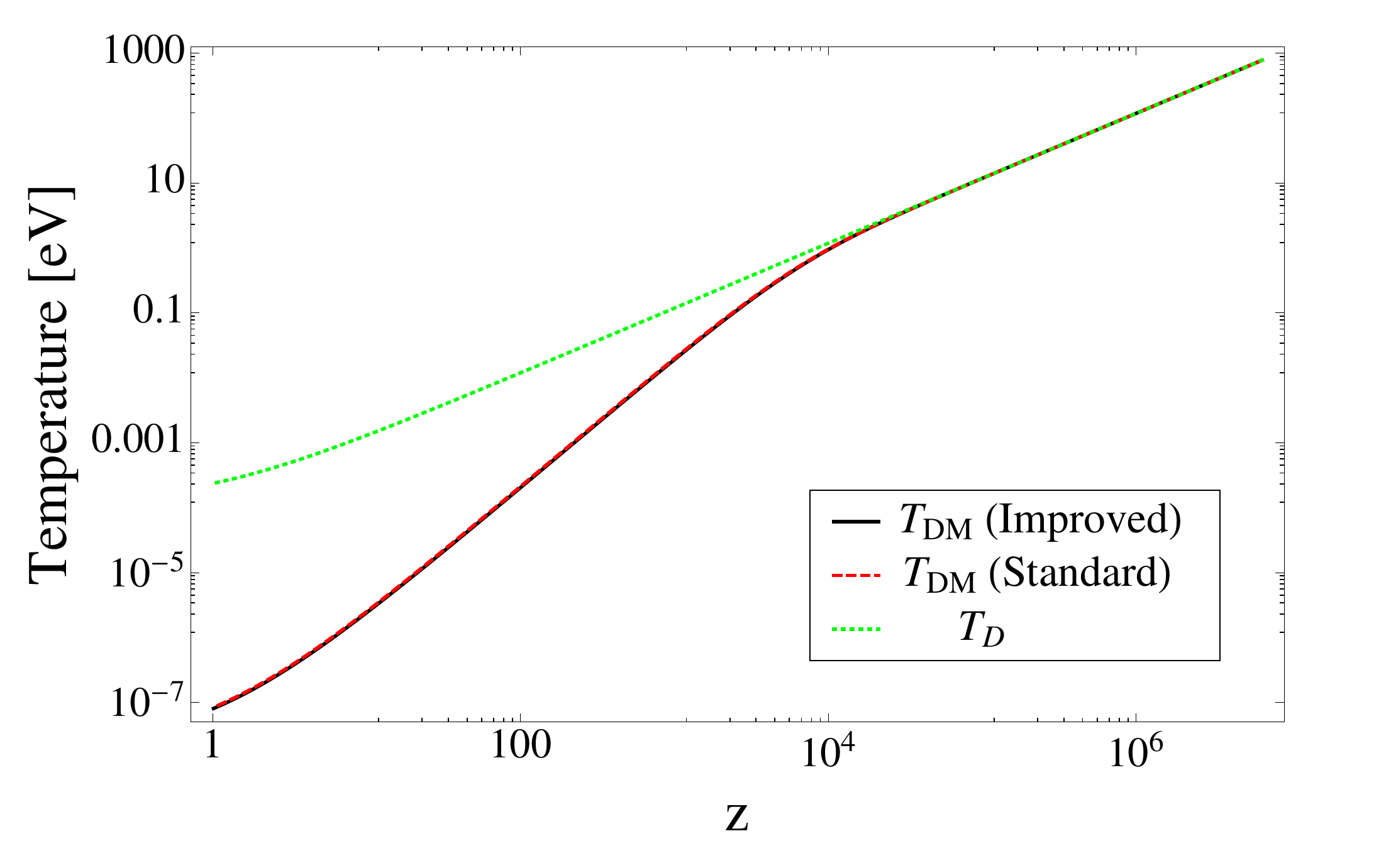}
\caption[Comparison between our improved treatment of dark recombination and the standard treatment for a Compton-heating-dominated dark sector.]{Comparison between our improved treatment of dark recombination and the standard treatment. We display results for a dark sector where Compton heating always dominates the thermal coupling between the DR and DM. Here, $\Upsilon_{\rm BF}\sim1.8\times10^5$ and $\Upsilon_{\rm R}\sim10^2$. The upper panel shows the evolution of the ionization fraction as a function of redshift while the lower panel shows the corresponding evolution of the DM and DR temperatures.}
\label{xe_plot_m}
\end{figure}

Figure \ref{xe_plot_m} displays the ionization and temperature evolution of a dark-atom model for which Compton heating is the dominant thermal coupling mechanism until late times ($\Upsilon_{\rm BF}\sim2\times10^5$ and $\Upsilon_{\rm R}\sim10^2$). In this case, the usual calculation accurately captures the behavior of both the ionization evolution and the DM temperature. This is a not a surprise, since the standard treatment was designed to capture this specific regime of the dark plasma. The small difference in the ionization fraction (up to $\sim12\%$) at late times is entirely due to our more accurate recombination coefficient which includes the effects of high-$n$ shells on the recombination process.

\begin{figure}
\includegraphics[width=0.5\textwidth]{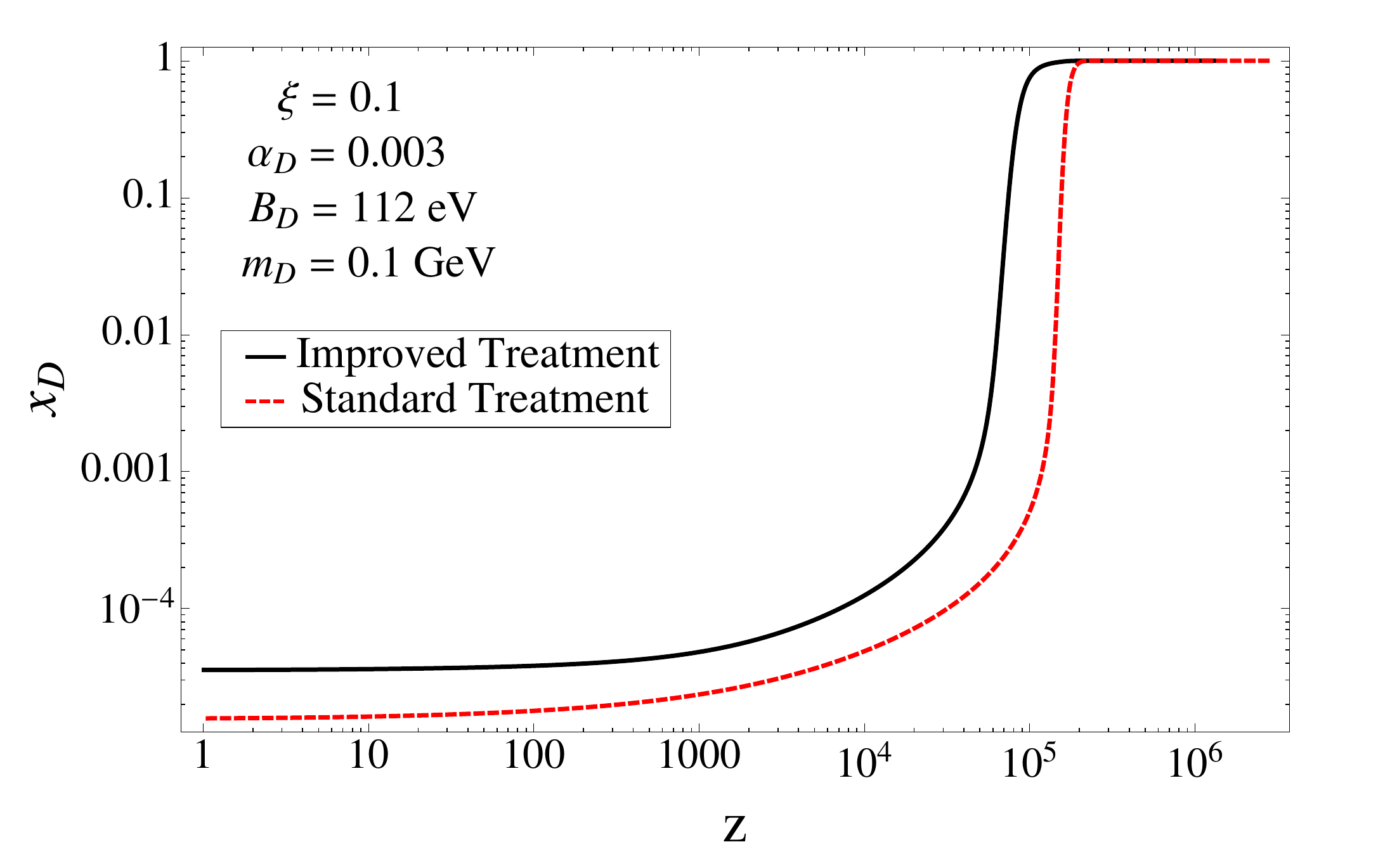}
\includegraphics[width=0.5\textwidth]{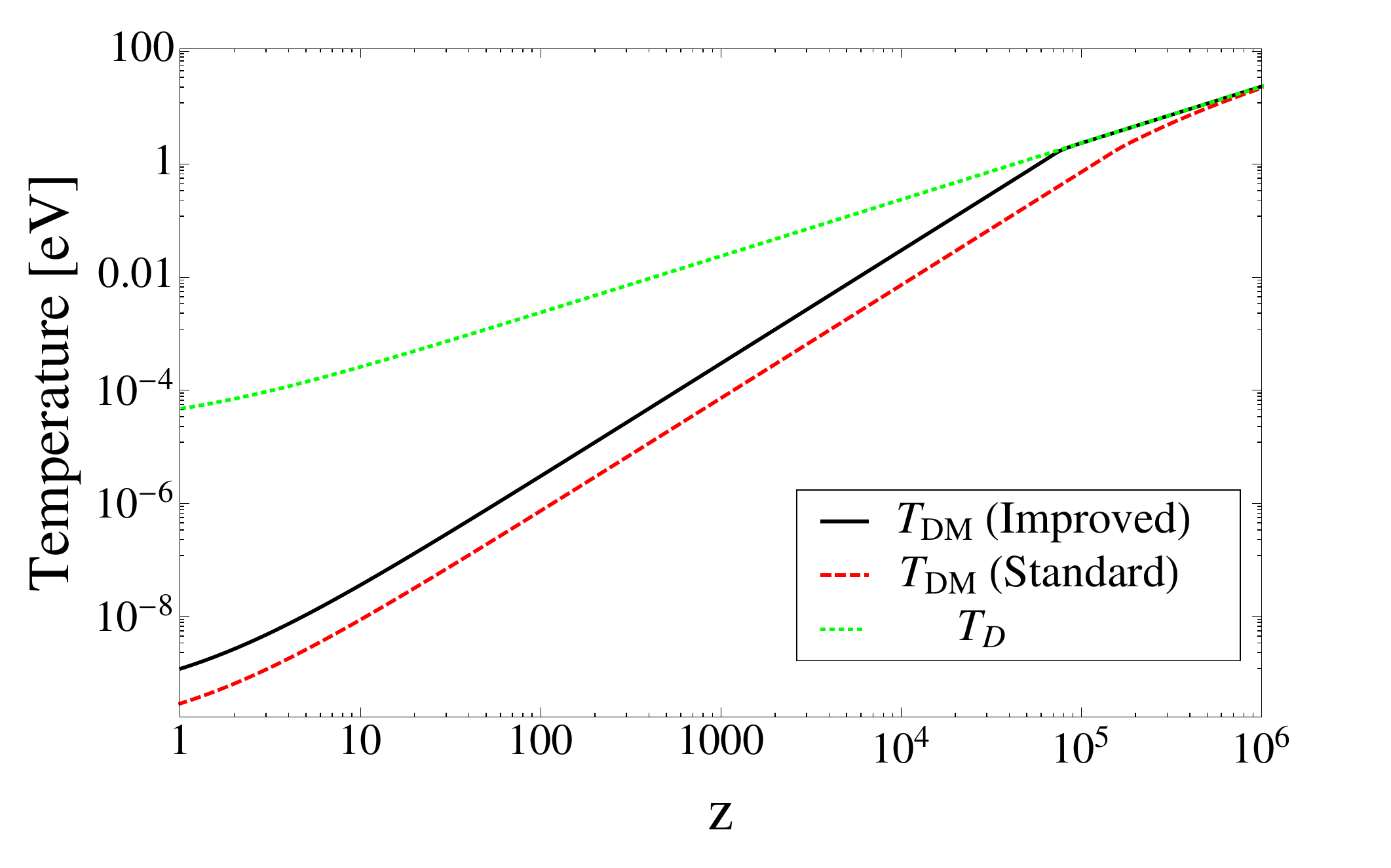}
\caption[Comparison between our improved treatment of dark recombination and the standard treatment for a weakly-coupled dark sector.]{Comparison between our improved treatment of dark recombination and the standard treatment. We display results for a weakly-coupled dark sector with $\Upsilon_{\rm BF}\sim5\times10^{-4}$ and $\Upsilon_{\rm R}\sim4$. The upper panel shows the evolution of the ionization fraction as a function of redshift while the lower panel shows the corresponding evolution of the DM and DR temperatures.}
\label{xe_plot_w}
\end{figure}

In Fig.~\ref{xe_plot_w}, we display the ionization history (upper panel) and temperature evolution (lower panel) for a model with $\Upsilon_{\rm BF}\sim5\times10^{-4}$ ($\Upsilon_{\rm R}\sim4$). Not only does the standard calculation fail to predict the right recombination redshift, it also underestimates the late-time ionization fraction by more than $50\%$. Since the Saha equilibrium does not hold for these weakly-coupled models, their ionization evolution is strongly determined by the value of the recombination coefficient. In the lower panel, we see that the inclusion of the bound-free and free-free processes delays the thermal decoupling of DM, hence postponing the onset of dark recombination. As the recombination coefficient is very sensitive to the DM temperature (see Fig.~\ref{rec_rate_plot1}), this delayed thermal decoupling acts to suppress the recombination rate, hence leaving a larger ionized fraction at late times.

In summary, the standard recombination treatment originally described in \cite{Peebles:1968ja} can only be accurately applied to dark-atom models for which both bound-free and Rayleigh heating are negligible. For these scenarios, the late-time difference between our improved calculation and the standard treatment is almost entirely due our more accurate recombination coefficient, which properly accounts for the effects of excited atomic states and for the difference between the DM and DR temperature. On the other hand, the standard treatment generally overestimates the thermal-decoupling temperature for atomic DM models with $\Upsilon_{\rm BF}\lesssim1$ or $\Upsilon_{\rm R}\lesssim1$. As a consequence, the standard treatment tends to underestimate the DM ionized fraction at late times for these types of models.   

\subsection{Existence of Dark Atoms}
\begin{figure}[t!]
\begin{centering}
\includegraphics[width=0.48\textwidth]{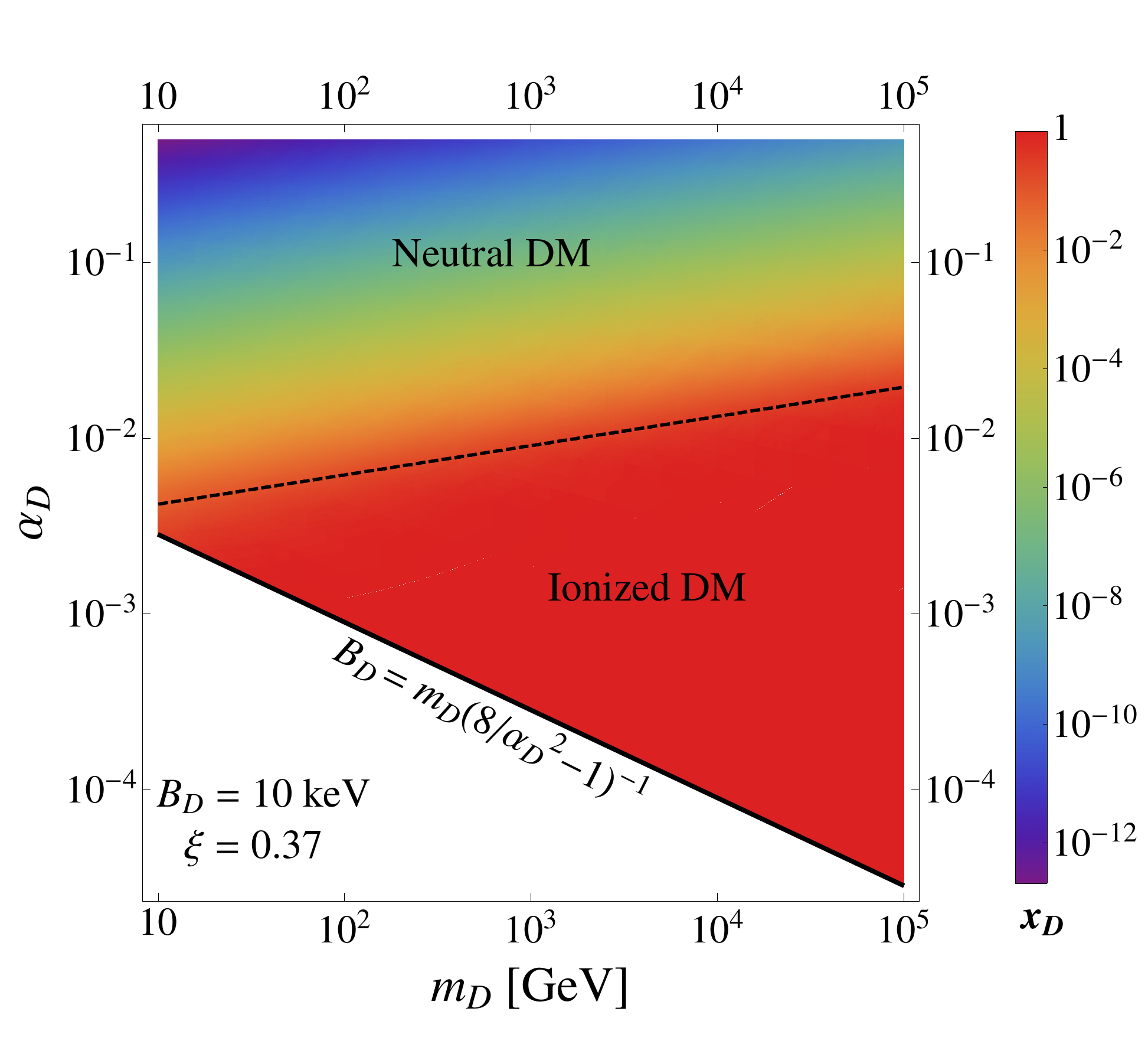}
\caption[Late-time ionized fraction as a function of $m_D$ and $\alpha_D$ for a dark-atom model with $B_D=10$ keV.]{Late-time ionized fraction as a function of $m_D$ and $\alpha_D$ for a dark-atom model with $B_D=10$ keV. The dashed line corresponds to the bound given in Eq.~(\ref{existence_of _DA}), delimiting the mostly ionized models (below the line) from the region of parameter space dominated by dark atoms at late times (above the line).}
\label{xD_10keV_plot}
\end{centering}
\end{figure}

As the dark fine-structure constant is decreased and the mass of the dark proton is increased, it becomes progressively more difficult for oppositely-charged dark fermions to find each other and form neutral bound states. There exist critical values of the masses and coupling constants beyond which dark atoms do not form and the dark plasma remains ionized even for $T_D\ll B_D$. Generally, this happens if the recombination rate is smaller than the Hubble expansion rate, when it becomes energetically favorable to form dark atoms. Using Eqs.~(\ref{exact_rec_rate}) and (\ref{hubble_time_rad}), dark atoms can form only if the following condition is satisfied
\be\label{existence_of _DA}
\frac{\alpha_D^6}{\xi}\left(\frac{\Omega_Dh^2}{0.11}\right)\left(\frac{m_D}{\text{GeV}}\right)^{-1}\left(\frac{B_D}{\text{keV}}\right)^{-1}\gtrsim1.5\times10^{-16},
\ee
where we have used $T_D\simeq T_{DM}\simeq0.02 B_D$, which corresponds to the usual dark sector temperature at the onset of dark recombination. Models violating this bound are effectively hidden charged DM models similarly to those discussed in Refs.~\cite{Ackerman:2008gi,Feng:2009mn}. We display the constraint (\ref{existence_of _DA}) in Fig.~\ref{xD_10keV_plot} (dashed line) together with the values of the late-time ionized fraction for a model with $B_D=10$ keV. We see that Eq.~(\ref{existence_of _DA}) delimits very well the region where the dark sector is mostly ionized at late times. 
%
\section{Evolution of Dark-Matter Perturbations}\label{perturbations}
Having determined the evolution of the background ionized fraction and temperatures, we now turn our attention to the evolution of cosmological perturbations in the atomic DM scenario. In this theory, the DM and DR form a tightly-coupled plasma in the early Universe, much like the baryon-photon fluid. Once modes enter the horizon, the pressure provided by the relativistic dark photons gives rise to a restoring force opposing the gravitational infall of DM, hence leading to dark acoustic oscillations (DAO) in the plasma.  Compared to a CDM model, the presence of these DAOs delays the onset of DM fluctuation growth until the epoch of kinetic decoupling. In addition, atomic DM fluctuations miss out on the kick due to the decaying gravitational potential when they cross inside the Hubble horizon \cite{Loeb:2005pm}. We therefore generically expect DM fluctuations to be suppressed on small scales in atomic DM models. 

We begin this section by giving the key equations describing the DM and DR perturbations, emphasizing the new collision term between these two constituents of the dark sector. We then discuss the different contributions to the opacity of the dark plasma and study their impact on the kinetic decoupling epoch. We then describe the various regimes that the perturbations in the dark plasma encounter as they evolve through the cosmic ages. We finally present numerical examples of these different regimes, both in Fourier space and in configuration space.

\subsection{Perturbation Equations}
The equations governing the evolution of atomic DM fluctuations are very similar to those describing the baryon-photon plasma. Special care must however be taken to include all the relevant contribution to the opacity of the DM to DR. The Boltzmann equations for DM are
\be\label{delta_D}
\dd{\de}_{D}+\theta_{D}-3\dd{\phi}=0,
\ee
\be\label{theta_D}
\dd{\theta}_{D}+\frac{\dd{a}}{a}\theta_D-c_D^2k^2\de_{D}-k^2\psi= \frac{R_D}{\tau_D}(\theta_{\tilde{\gamma}}-\theta_D),
\ee
where we closely followed the notation of Ref.~\cite{Ma:1995ey} in conformal Newtonian gauge. Here, $\de_D$ is the DM density constrast, $\theta_D$ and $\theta_{\tilde{\gamma}}$  are the divergence of the DM and DR velocity, respectively; $\phi$  and $\psi$ are the gravitational scalar potentials, $R_D\equiv4\rho_{\tilde{\gamma}}/3\rho_{D}$, $c_D$ is the sound speed of DM, $k$ is the wavenumber of the mode and $\tau_D^{-1}$ is the opacity of the dark plasma. Here, the subscript $\tilde{\gamma}$ always refers to the dark photons. The right-hand side of Eq.~(\ref{theta_D}) represents the collision term between the DM and the DR. At early times, we generally have $R_D\gg1$ and $\tau_D\ll\tau_H$, implying that the DM is effectively dragged along by the DR. 
The latter evolves according to the following Boltzmann equations:
\be\label{delta_gD}
\dd{\de}_{\tilde{\gamma}}+\frac{4}{3}\theta_{\tilde{\gamma}}-4\dd{\phi}=0;
\ee
\be\label{theta_gD}
\dd{\theta}_{\tilde{\gamma}}-k^2(\frac{1}{4}\de_{\tilde{\gamma}}-\frac{F_{\tilde{\gamma}2}}{2})-k^2\psi= \frac{1}{\tau_D}(\theta_D-\theta_{\tilde{\gamma}});
\ee
\be\label{FgD2}
\dd{F}_{\tilde{\gamma}2} = \frac{8}{15}\theta_{\tilde{\gamma}}-\frac{3}{5}kF_{\tilde{\gamma}3}-\frac{9}{10\tau_D}F_{\tilde{\gamma}2};
\ee
\be\label{FgDl}
\dd{F}_{\tilde{\gamma}l}=\frac{k}{2l+1}\left[l F_{\tilde{\gamma}(l-1)}-(l+1)F_{\tilde{\gamma}(l+1)}\right]-\frac{1}{\tau_D}F_{\tilde{\gamma}l}.
\ee
Eqs.~(\ref{delta_gD}) and (\ref{theta_gD}) describe the evolution of the dark-photon over-densities ($\de_{\tilde{\gamma}}$) and of the dark-photon velocity, respectively. It is also necessary to solve for the hierarchy of dark-photon multipoles (Eqs.~(\ref{FgD2}) and (\ref{FgDl})) to properly account for DR diffusion and its impact on DM perturbations. Since our focus is to describe the clustering of DM in this model, we do not solve for the DR polarization. 

During the radiation-dominated epoch, the energy density of the DR is generally subdominant compared to the contribution from regular photons and neutrinos (see Eq.~(\ref{BBN_bound})). Therefore, the time-dependence of the gravitational potentials $\phi$ and $\psi$ is very similar to the standard CDM case for which we have
\be
\phi\simeq-3\phi_{\rm p}\left[\frac{\sin{(k\tau/\sqrt{3})}-(k\tau/\sqrt{3})\cos{(k\tau/\sqrt{3})}}{(k\tau/\sqrt{3})^3}\right],
\ee
where $\tau$ stands for the conformal time and $\phi_{\rm p}$ is the primordial amplitude. Since the gravitational potential is an oscillatory function, Eqs.~(\ref{delta_D}) and (\ref{delta_gD}) essentially describe driven harmonic oscillators where the driving force is provided by the baryon-photon plasma. Indeed, taken as a whole, the equations describing the dark plasma and the baryon-photon plasma in the early Universe correspond to a system of coupled harmonic oscillators. Such a system is known to exhibit resonance phenomena whenever the driving frequency approaches the natural frequency of the oscillator. For the dark plasma however, we always have 
\be\label{R_condition}
R_D\ll R\equiv\frac{4\rho_{\gamma}}{3\rho_b},
\ee
where $\rho_{\gamma}$ and $\rho_b$ are the energy densities of regular photons and baryons, respectively. As the sound speed of a plasma is approximately given by $c_{\rm p}=1/\sqrt{3(1+R^{-1})}$, Eq.~(\ref{R_condition}) implies that the sound speed of the dark plasma is always smaller than that of the baryon-photon plasma. Thus, the dark plasma is never driven close to its resonance threshold\footnote{While resonant enhancement does not occur for the simple atomic DM scenario considered in this paper, it is nevertheless possible to construct a model where such resonance happens. This is an interesting possibility that we leave for future work}.
\subsection{Dark Opacity and Kinetic Decoupling}
The opacity of the dark plasma dictates the strength of the coupling between DM and DR. Heuristically, $\tau_D$ can be considered as the mean free path a dark photon travels between collisions with a dark ion or a dark atom. As such, $R_D/\tau_D$ is approximately the momentum-transfer rate between the DR and the DM. As in the case of the thermal coupling of DM to DR, many mechanisms contribute to the exchange of momentum between the two dark components. In addition to the usually-considered Compton-scattering term, we also include the contribution from Rayleigh scattering as well as the contribution from bound-free processes. The opacity is then
\be\label{opacity}
\tau_D^{-1}=\tau_{\rm Compton}^{-1}+\tau_{\rm R}^{-1}+\tau_{\rm p-i}^{-1}.
\ee
The Compton scattering term is given by
\be
\tau_{\rm Compton}^{-1}=an_{D}x_D\sigma_{{\rm T},D}\left[1+\left(\frac{m_{\bf e}}{m_{\bf p}}\right)^2\right],
\ee
where $\sigma_{{\rm T},D}\equiv 8\pi\alpha_D^2/(3 m_{\bf e}^2)$ is the dark Thomson cross section and $a$ is the scale factor. The factor in the bracket accounts for Compton scattering off dark protons. The contribution from Rayleigh scattering can be written as
\ba
\tau_{\rm R}^{-1}&=&an_{D}(1-x_D)\langle\sigma_{\rm R}\rangle\en
&\simeq&32\pi^4an_{D}(1-x_D)\sigma_{{\rm T},D}\left(\frac{T_D}{B_D}\right)^4,
\ea
which is valid for $T_D\ll B_D$. Here $\sigma_{\rm R}$ is the cross-section for Rayleigh scattering (see Eq.~(\ref{sigma_R})) and $\langle\ldots\rangle$ denotes a thermal average with respect to the Planck function describing the distribution of dark photons. Finally, the photoionization contribution is
\ba
\tau^{-1}_{\rm p-i}&=&a\sum_{n,l}n_{nl}\langle\sigma_{nl}\rangle\\
&\simeq&an_{D}x_{2s}\frac{\pi\alpha_D^3}{6 \zeta(3)T_D^2}e^{-B_D/(4T_D)}G_{250}^{2s}\left(\frac{T_D}{B_D},1\right),\nonumber
\ea
where $n_{nl}$ and $\sigma_{nl}$ are the number density and the photoionization cross-section for dark atoms in quantum state $nl$. 

For most atomic DM models, the Compton scattering contribution to the opacity dominates before dark recombination, with photoionization giving a subdominant contribution for models with $\Upsilon_{\rm BF} \lesssim 1$. Once a significant number of dark atoms has recombined, Rayleigh scattering can become the dominant source of opacity for models with $\Upsilon_{\rm R}\lesssim1$. 

The DR effectively begins free-streaming when its mean-free path becomes comparable to the size of the Hubble horizon, that is, $\tau_D^{-1}\simeq H$. In a model for which Compton scattering dominates the interactions between the DR and the DM, the onset of the free-streaming epoch is independent of the temperature and only depends of the fraction of ionized dark atoms,
\ba
x_D\big|^{\text{Compt}}_{\text{dec}}&\simeq&\min{\left[1,\frac{3.7\times10^{-10}}{\alpha_D^6}\right.}\\
&&\qquad\,\,
\times\left.\left(\frac{\Omega_D h^2}{0.11}\right)^{-1} \left(\frac{B_D}{\text{keV}}\right)^2\left(\frac{m_D}{\text{GeV}}\right)\right],\nonumber
\ea
where we have assumed that the DR decoupling happens in the radiation-dominated epoch and have neglected the contribution from dark protons. If the second argument in the $\min{[..,..]}$ function is larger than unity, it indicates that dark photons begin free-streaming at a very early epoch, making them difficult to distinguish from massless neutrinos.  For a strongly-coupled dark sector which have $\Upsilon_{\rm R}\lesssim1$, the decoupling of dark photons happens much later and is governed by Rayleigh scattering of dark photons off neutral dark atoms. This leads to a decoupling temperature
\be
\frac{T_D}{B_D}\Bigg|^{\text{R}}_{\text{dec}}\simeq\frac{7 \times10^{-4}}{\alpha_D^{3/2}}\left[\left(\frac{\Omega_D h^2}{0.11}\right)^{-1} \left(\frac{B_D}{\text{keV}}\right)^2\left(\frac{m_D}{\text{GeV}}\right)\right]^{\frac{1}{4}},
\ee
where we have taken the limit $x_D|_{\text{dec}}\ll1$. It is important to note that for most atomic DM scenarios, Compton scattering alone is responsible for the DM-DR coupling. 

The key quantity governing the size of the smallest DM structure in the Universe is the temperature at which DM kinematically decouples from the DR (i.e.~the dark drag epoch). This temperature  can approximately be determined by solving the condition $R_D/\tau_D\simeq H$. For models dominated by Compton scattering, this leads to a kinetic decoupling temperature given by
\be
\frac{T_{D}}{B_D}\Bigg|^{\text{Compt}}_{\text{drag}}\simeq\frac{5.8\times10^{-13}}{\alpha_D^6\xi^4x_D|_{\text{drag}}}\left(\frac{B_D}{\text{keV}}\right)\left(\frac{m_D}{\text{GeV}}\right).
\ee
This is effectively an implicit equation for the drag-epoch temperature since the right-hand side involves the ionized fraction of dark atoms evaluated at that epoch. Taking $x_D|_{\text{drag}}\sim1$ leads to a lower bound on the temperature at which DM ceases to be dragged along by the DR. If $\Upsilon_{\rm R}\lesssim1$, the kinetic decoupling occurs when Rayleigh scattering becomes ineffective. In this case, we obtain
\be
\frac{T_D}{B_D}\Bigg|^{\text{R}}_{\text{drag}}\simeq7 \times10^{-4}\left[\frac{1}{\alpha_D^6\xi^3}\left(\frac{B_D}{\text{keV}}\right)\left(\frac{m_D}{\text{GeV}}\right)\right]^{\frac{1}{5}},
\ee
where we have used $x_D|_{\text{drag}}\ll1$, which is a necessary condition for Rayleigh scattering to dominate over Compton scattering. 

DM fluctuations on subhorizon length scales at the drag epoch have a significantly different evolution than those of a standard $\Lambda$CDM model. In particular, the suppression of small-scale power due to the dark-photon pressure leads to a minimal halo mass at late times. We will discuss this effect further in section \ref{mps}.

\subsection{Regimes of the Dark Plasma}\label{DA-regime}
We now turn our attention to the formal solutions to Eqs.~(\ref{delta_D}) to (\ref{FgDl}). These admits different regimes depending on the relative values of the opacity, wavenumber, and Hubble expansion rate. These regimes are
\begin{enumerate}
\item{\bf Superhorizon Regime:} This regime occurs when the wavelength characterizing a mode is still larger than the Hubble horizon, $k<H$. As for the CDM case, cosmological perturbations do not significantly evolve in this regime. We therefore do not discuss this case any further since atomic DM is indistinguishable from standard CDM for these modes.
\item{\bf Dark-Acoustic-Oscillation (DAO) Regime:} Once modes cross into the Hubble horizon, the microphysics governing the interaction between the DM and the DR becomes effective. If the mean-free-path of dark photons between collisions with dark fermions is much smaller than the wavelength of a given mode ($k\tau_D\ll1$), then we can consider the DM and the DR to be tightly-coupled and to form an almost perfect fluid at this scale. In this case, the DR pressure can effectively counteract the pull from the gravitational potential, leading to the propagation of acoustic oscillations in the dark plasma. This DAO regime occurs for wavelengths larger than the diffusion length scale of the dark photons, that is,
\be\label{condition_DAO}
H<k<k_D
\ee
where $k_D$ is diffusion damping scale and is defined by \cite{1996ApJ...471..542H}
\be\label{k_D}
\hspace{1cm}\frac{1}{k_D^2(\tau)}=\int_0^\tau\frac{\tau_Dd\tau'}{6(1+R_D^{-1})}\left[\frac{1}{R_D^2+R_D}+\frac{8}{9}\right].
\ee
Heuristically, $1 /k_D$ corresponds approximately to the average distance travelled by a dark photon in a Hubble time. Note that Eq.~(\ref{k_D}) is the result of a first-order expansion in $k\tau_D$ of the dispersion relation of DAOs and therefore is only accurate in the limit $k\tau_D\ll1$.   In the DAO regime, DM fluctuations undergo constant-amplitude oscillations 
\be
\de_D\sim\text{exp}\left\{ik\int d\tau \tilde{c}_{\rm p}\right\},
\ee
where $\tilde{c}_{\rm p}=1/\sqrt{3(1+R_D^{-1})}$ is the sound speed of the dark plasma. Due to these oscillations, DM fluctuations entering the horizon in this regime miss out on both the logarithmic growth and the horizon kick. Thus, despite the absence of damping in the DAO regime (besides the small contribution from Hubble expansion), DM perturbations still generally display smaller amplitudes at these scales when compared to a standard CDM model. 
\item{\bf Diffusion-Damping Regime:}
Once the average distance travelled by a dark photon during a Hubble time becomes comparable with the wavelength of a given mode, DR can effectively diffuse out of over-densities at these scales. If the momentum transfer rate between the DR and the DM is larger than the plasma oscillation frequency,  
\be
k_D<k<\frac{1}{\tilde{c}_{\rm p}}\frac{R_D}{\tau_D},
\ee
then DM is dragged along by the escaping dark photons, effectively erasing DM fluctuations on these scales. The time-dependence of the DM perturbations is then given by
\be\label{exponential_supp}
\de_D\sim \text{exp}\left\{ik\int d\tau \tilde{c}_{\rm p}\right\}\text{exp}\left\{-\frac{k^2}{k_D^2}\right\},
\ee
leading to an exponential damping on fluctuations on scale smaller than the diffusion length. As before, we have assumed the dark plasma is tightly-coupled ($k\tau_D\ll1$) in deriving Eq.~(\ref{exponential_supp}). When this condition ceases to be satisfied, the damping of DM fluctuations is no longer exponential and Eq.~(\ref{exponential_supp}) breaks down. Note that the diffusion damping regime ends when $k\tilde{c}_{\rm p}\sim R_D/\tau_D$ since DM ceases to be dragged by the DR when this condition is satisfied.
\item{\bf Acoustic-Damping Regime}
When the mean-free-path of dark photons becomes comparable to the wavelength of a given mode ($k\tau_D\gtrsim1$), the dark plasma ceases to behave like a single perfect fluid. Due to the slow reaction of DM to the motion of dark photons in this regime, acoustic oscillations in the DM fluid start to significantly lag those propagating in the DR fluid. This in turns leads to viscous dissipation in the dark plasma resulting in the damping of DM fluctuations at these scales. Note however that this acoustic damping is weaker than the exponential damping characteristic of the tightly-coupled diffusion regime. Generally, acoustic damping occurs for modes satisfying the condition
\be
H<\frac{1}{\tau_D}<k<\frac{R_D}{\tilde{c}_{\rm p}\tau_D}.
\ee
The last inequality ensures that a significant amount of momentum is still transfered to the DM during acoustic damping. Physically, this regime is characterized by the development of anisotropic stress in the dark fluid associated with a significant dark-photon quadrupole moment. As such, it is therefore difficult to obtain an analytical solution to the evolution of DM fluctuations in this regime. 
\item{\bf Gravitationally-Dominated Regime:}
When the momentum transfer rate between the DR and atomic DM falls below the oscillation frequency of the dark photons, that is
\be\label{ADR}
H<\frac{R_D}{\tau_D}<k\tilde{c}_{\rm p},
\ee
the dark plasma ceases to behave like a single fluid and we must consider the DM and the DR fluctuations separately. In this regime, the evolution of the DM perturbations is determined by a competition between the gravitational potential dominated by the baryon-photon plasma and the remaining pressure of the DR. Since both of these contributions are oscillatory in nature, the evolution of the DM fluctuations in this regime can be rather complex. Essentially, we can view DM perturbations as forced oscillators driven by two distinct forces oscillating with difference frequencies. Explicitly, the equation governing the evolution of DM fluctuations is
\ba\label{2drivingterms}
\ddot{\de}_D+\left(H+\frac{R_D}{\tau_D}\right)\dot{\de}_D&=&S_G(k,\tau)
-\frac{R_D}{\tau_D}\theta_{\tilde{\gamma}},
\ea
where the gravitational source term is given by
\be
S_G(k,\tau)=3\ddot{\phi}-k^2\psi+3\dot{\phi}\left(H+\frac{R_D}{\tau_D}\right).
\ee
The dark-photon driving term can be obtained from rearranging Eqs.~(\ref{theta_gD}) and (\ref{delta_gD})
\be\label{theta_gDlate}
\hspace{1cm}\theta_{\tilde{\gamma}}=\theta_D+\frac{3\tau_D}{4}\left[\ddot{\de}_{\tilde{\gamma}}+\frac{k^2}{3}\de_{\tilde{\gamma}}+\frac{k^2}{6}F_{\tilde{\gamma}2}-4\dot{\phi}+\frac{4}{3}k^2\psi\right].
\ee
In the limit that a DM fluctuation enters this regime after being exponentially damped by the diffusion of dark photons, the evolution of DR perturbations is dominated by the gravitational potential. Neglecting the small quadrupole moment of the dark photons, this implies that the term in the square bracket of Eq. (\ref{theta_gDlate}) is very close to zero. We thus have $\theta_{\tilde{\gamma}}\simeq\theta_D$ in this limit and Eq.~(\ref{2drivingterms}) reduces to
\ba\label{log_growth}
\ddot{\de}_D+H\dot{\de}_D&=&3\ddot{\phi}-k^2\psi+3H\dot{\phi}.
\ea
Eq.~(\ref{log_growth}) is to the usual equation describing the growth of standard CDM fluctuations. It implies that DM fluctuations grow logarithmically with the scale factor during the radiation-dominated epoch. 

On the other hand, for modes that enter the gravitationally-dominated regime before being substantially damped by photon diffusion, the dark-photon driving term cannot be neglected in Eq.~(\ref{2drivingterms}). 
In this case, the dark-photon quadrupole moment is non-negligible in Eq.~(\ref{theta_gDlate}), leading to an oscillatory pressure term with
\be
\theta_{\tilde{\gamma}}\sim e^{\frac{ik\tau}{\sqrt{3}}}.
\ee
The gravitational potential contribution $S_G(k,\tau)$ also oscillates, but with a somewhat lower frequency $\sim k/\sqrt{3(1+R)}$. As these oscillators can randomly go in-phase and out-of phase, some damping or even some amplification can occur in this regime. However, since the momentum transfer rate $R_D/\tau_D$ is decaying more rapidly than $S_G(k,\tau)$, the gravitational potential rapidly becomes the dominant contribution to the evolution of DM fluctuations. Indeed, while $R_D/\tau_D\propto\tau^{-n}$ with $n\geq3$ (depending on which process dominates the calculation of the opacity),  $S_G(k,\tau)\propto\tau^{-2}$. Therefore, the DM fluctuations rapidly settle into the logarithmic growing mode.
\item{\bf Kinetically-Decoupled Regime:} As the momentum transfer rate between atomic DM and DR falls below the Hubble rate, $R_D/\tau_D<H$, DM essentially stops interacting with the dark photons and begin behaving like the standard CDM. Modes entering this regime during radiation domination begin growing logarithmically with the scale factor while modes entering during matter domination grow linearly with $a$. In all cases, modes that enter the Hubble horizon after kinetic decoupling are undistinguishable from those of standard CDM.
\end{enumerate}
\subsection{Numerical Solutions}
%
\begin{figure}[t!]
\includegraphics[width=0.5\textwidth]{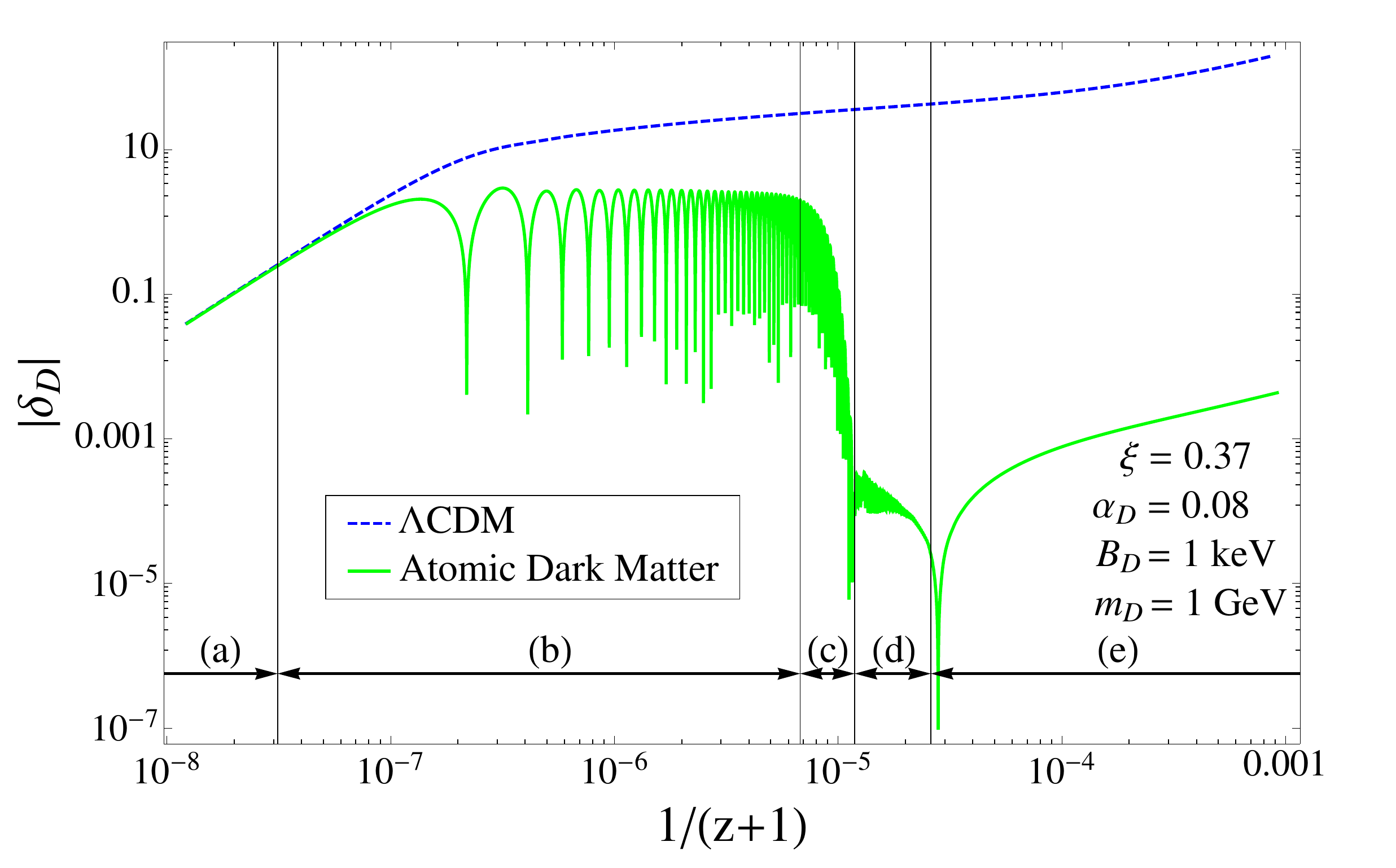}
\caption[Redshift evolution of an atomic DM fluctuation with $k=70$ Mpc$^{-1}$ in synchronous gauge.]{Redshift evolution of an atomic DM fluctuation with $k=70$ Mpc$^{-1}$ in synchronous gauge (green solid line). We identify on the figure the different regimes that the DM fluctuation encounters during its evolution. These are: (a) superhorizon; (b) DAO; (c) diffusion damping; (d) gravitationally-dominated; and (e) kinetically-decoupled. For the case displayed here, the fluctuation enters the horizon while the dark plasma is tightly-coupled and therefore undergoes acoustic oscillations. Once $k_D\sim k$, the fluctuation becomes exponentially damped by dark-photon diffusion. This damping ceases to be effective when the rate of momentum transfer between the DR and DM falls below the oscillation frequency of dark-photon fluctuations. When this happens, the fluctuation enters the gravitationally-dominated regime where the perturbation rapidly settles in tothe logarithmic growing mode. The sharp feature around $(z+1)^{-1}=3\times10^{-5}$ is an artifact of plotting the absolute value of the fluctuation. For comparison, we also show the behavior of a CDM fluctuation with the same wavenumber.}
\label{evolution_adm_s}
\end{figure}
\begin{figure}[t]
\includegraphics[width=0.5\textwidth]{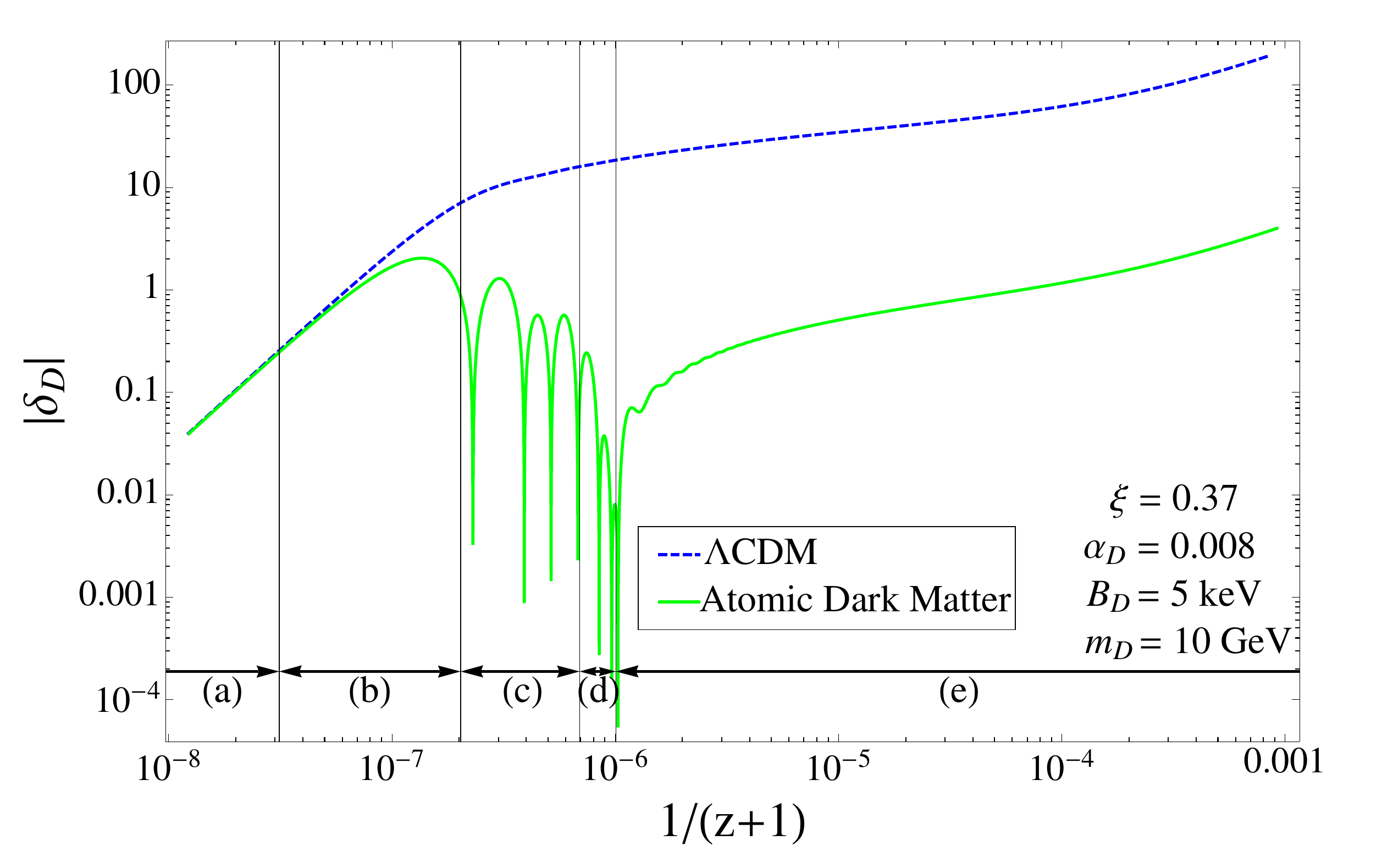}
\caption[Redshift evolution of an atomic DM fluctuation with $k=70$ Mpc$^{-1}$ in synchronous gauge.]{Redshift evolution of an atomic DM fluctuation with $k=70$ Mpc$^{-1}$ in synchronous gauge (green solid line). We identify on the figure the different regimes that the DM fluctuation encounters during its evolution. These are: (a) superhorizon; (b) DAO; (c) Acoustic damping; (d) gravitationally-dominated; and (e) kinetically-decoupled. For the case displayed here, the fluctuation enters the horizon while the dark plasma is weakly-coupled and therefore transitions to the acoustic damping regime very rapidly. This damping ceases to be effective when the rate of momentum transfer between the DR and DM falls below the oscillation frequency of dark-photon fluctuations. When this happens, the fluctuation enters the gravitationally-dominated regime where the perturbation is further damped before it settles into the logarithmic growing mode.  For comparison, we also show the behavior of a CDM fluctuation with the same wavenumber.}
\label{evolution_adm_w}
\end{figure}

We solve numerically Eqs.~(\ref{delta_D}-\ref{FgDl}) together with the standard Boltzmann equations describing the evolution of baryons, photon and neutrinos \cite{Ma:1995ey}. We use a modified version of the publicly-available code \texttt{CAMB} \cite{Lewis:1999bs} assuming that all of the DM is made of dark atoms. We modify the perturbed Einstein equation to include the new contributions from DM and DR. We use a flat background cosmology compatible with WMAP 7-year release \cite{Komatsu:2010fb}: $\Omega_bh^2=0.0226$, $\Omega_Dh^2= 0.1123$, $H_0=70.4\,\text{km/s/Mpc}$, $\Delta_{\mathcal{R}}^2=2.3\times10^{-9}$, $n_s=0.963$, and $\tau_{\rm re}=0.088$. We consider pure adiabatic initial conditions
\be
\de_D(z_{\rm i}) = \de_b(z_{\rm i})\qquad \de_{\tilde{\gamma}}(z_{\rm i})=\de_{\gamma}(z_{\rm i}),
\ee
\be
\theta_D(z_{\rm i})=\theta_{\tilde{\gamma}}(z_{\rm i})=\theta_{\gamma}(z_{\rm i}),
\ee
\be
F_{\tilde{\gamma}l} = 0,\quad l\geq2.
\ee
where $z_{\rm i}$ is the initial redshift which is determined such that all modes are superhorizon at early times, $k\tau(z_{\rm i})\ll1$. We first pre-compute the ionization and thermal history of the dark sector as described in section \ref{thermal} and use the result to compute the opacity of the dark plasma as given in Eq.~(\ref{opacity}) above. The ionization history of the baryon-photon sector is pre-computed in the usual way using \texttt{RecFast} \cite{Seager:1999bc}. The linear perturbation equations are then evolved forward in time from $z_{\rm i}$ to $z=0$. At early times when $k\tau_D\ll1$ and $\tau_D/\tau\ll1$, Eqs.~(\ref{theta_D}) and (\ref{theta_gD}) are very stiff and we use a second-order tight-coupling scheme similar to that used for the baryon-photon plasma at early times \cite{1970ApJ...162..815P,2011PhRvD..83j3521C,Pitrou:2010ai}. 

In Fig.~\ref{evolution_adm_s}, we show the time evolution of a single Fourier mode for relatively strongly-coupled dark atoms. We clearly identify the different regimes  that the fluctuation encounters from its horizon crossing to its late-time growth. For the particular choice of parameters displayed here, this Fourier mode enters the horizon in the DAO regime and oscillates until dark-photon diffusion exponentially suppresses its amplitude. Once DR kinematically decouples from DM, the fluctuation can start growing like regular CDM but from a much-reduced amplitude. 

To contrast, we show in Fig.~\ref{evolution_adm_w} the redshift evolution of the same Fourier mode for a weakly-coupled model of dark atoms. In this case, $k\tau_D\sim1$ shortly after the mode enters the horizon and therefore it only briefly experiences the DAO regime. It then rapidly transitions to the acoustic damping regime where its amplitude decays, though not as quickly or strongly as in the diffusion damping regime. After kinetic decoupling, the mode finally settles into the logarithmic growing mode.
\begin{figure*}
\subfigure[]{\includegraphics[width=0.495\textwidth]{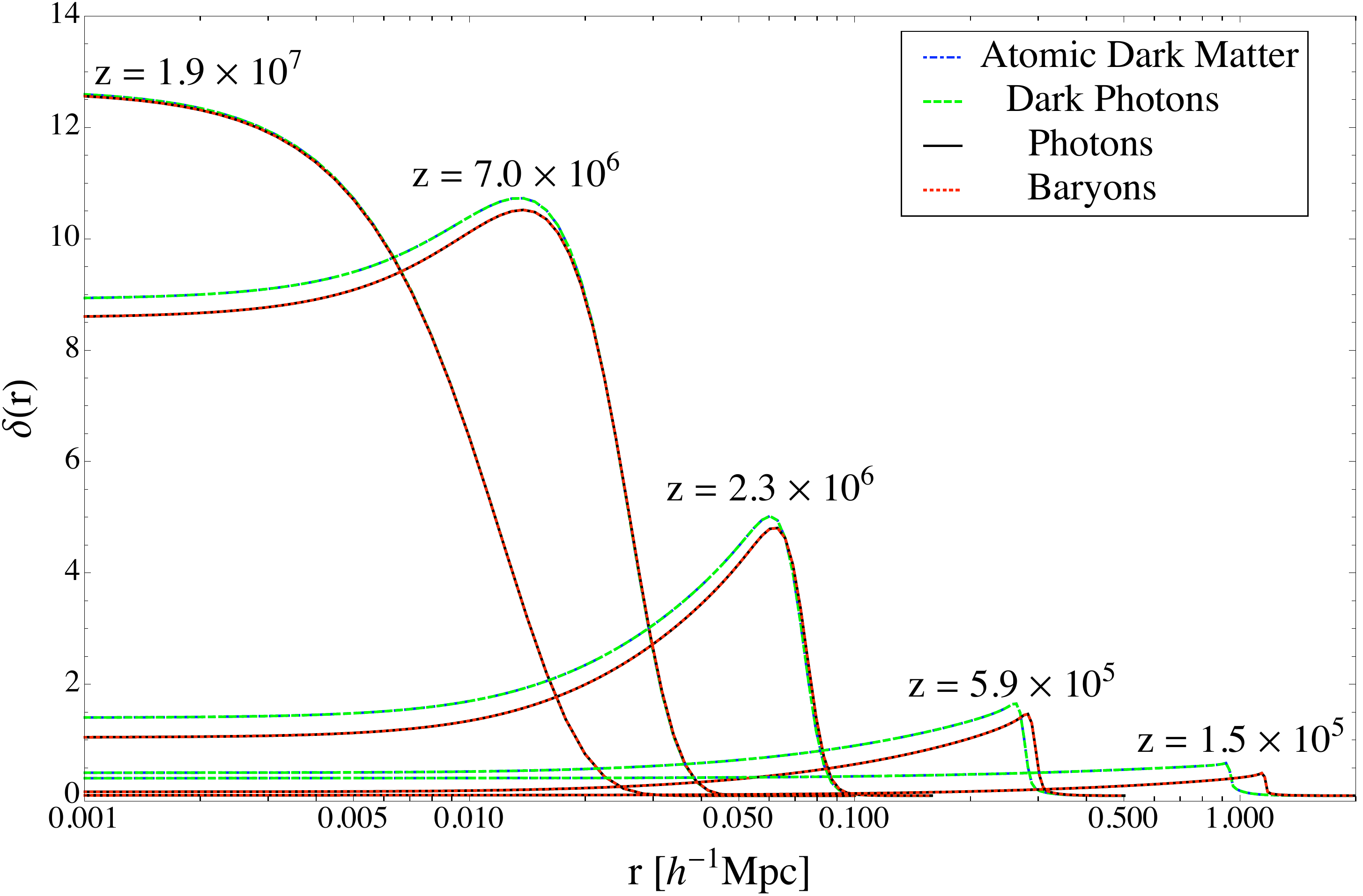}}
\subfigure[]{\includegraphics[width=0.495\textwidth]{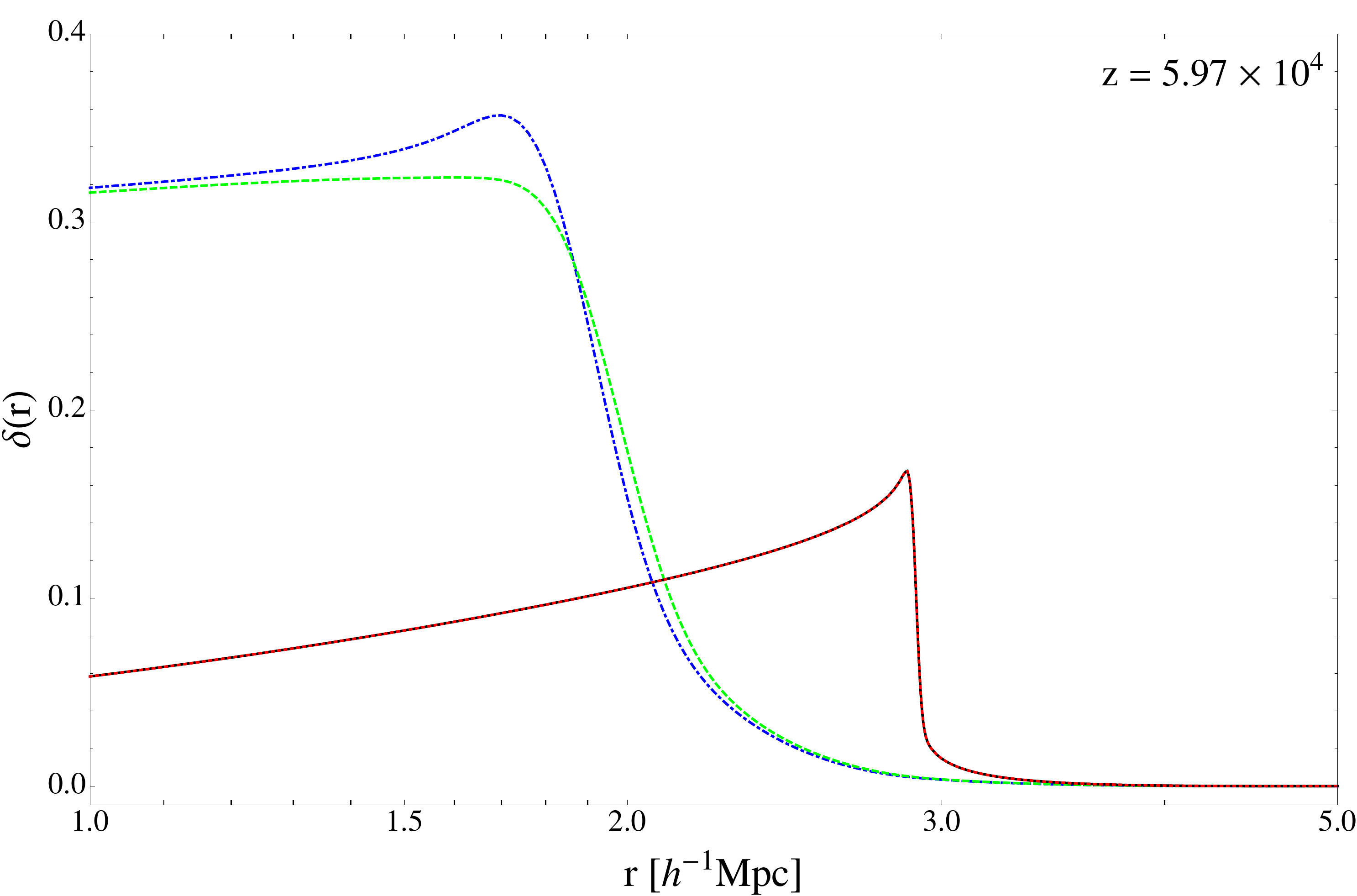}}\\
\subfigure[]{\includegraphics[width=0.495\textwidth]{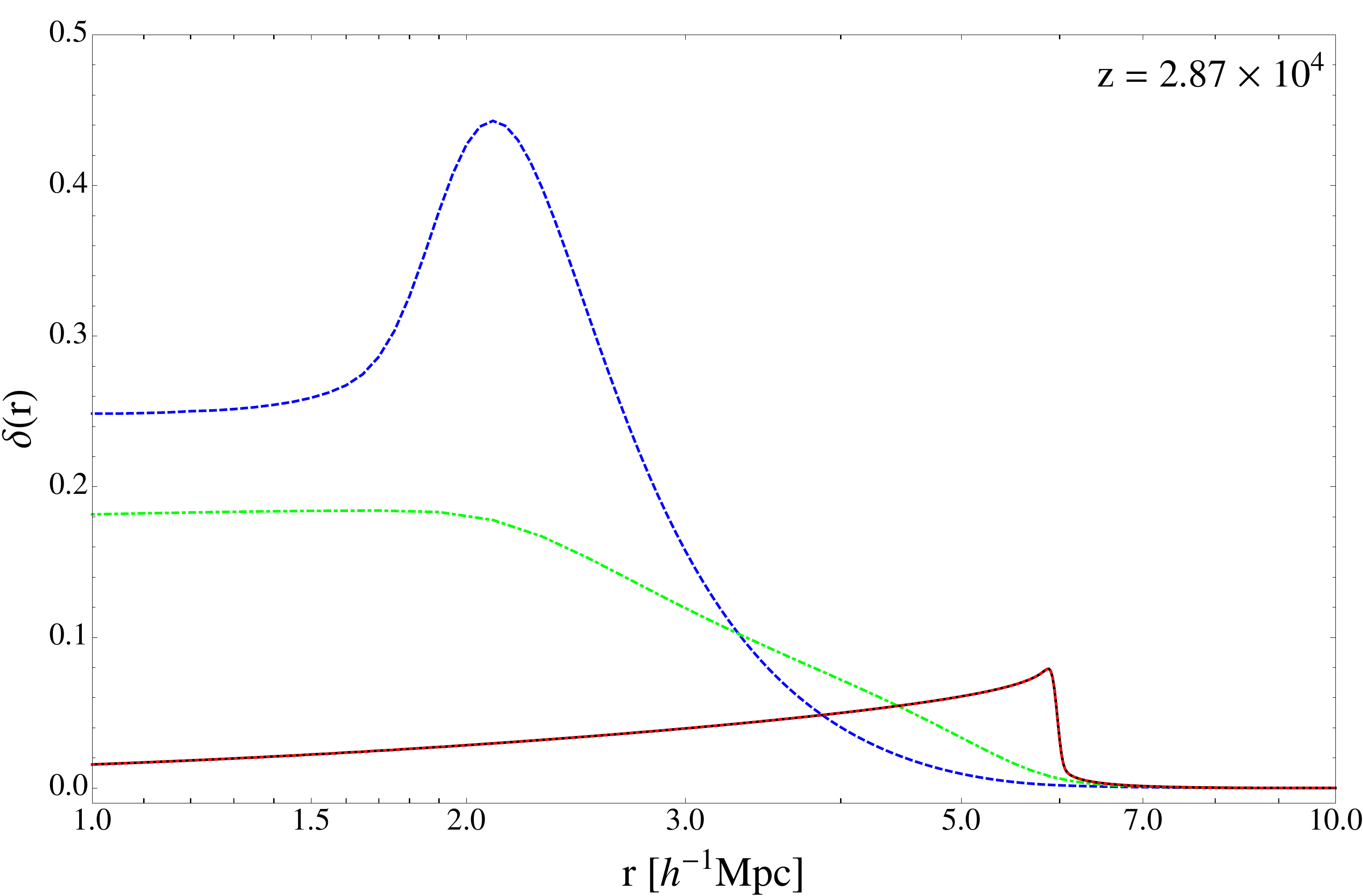}}
\subfigure[]{\includegraphics[width=0.49\textwidth]{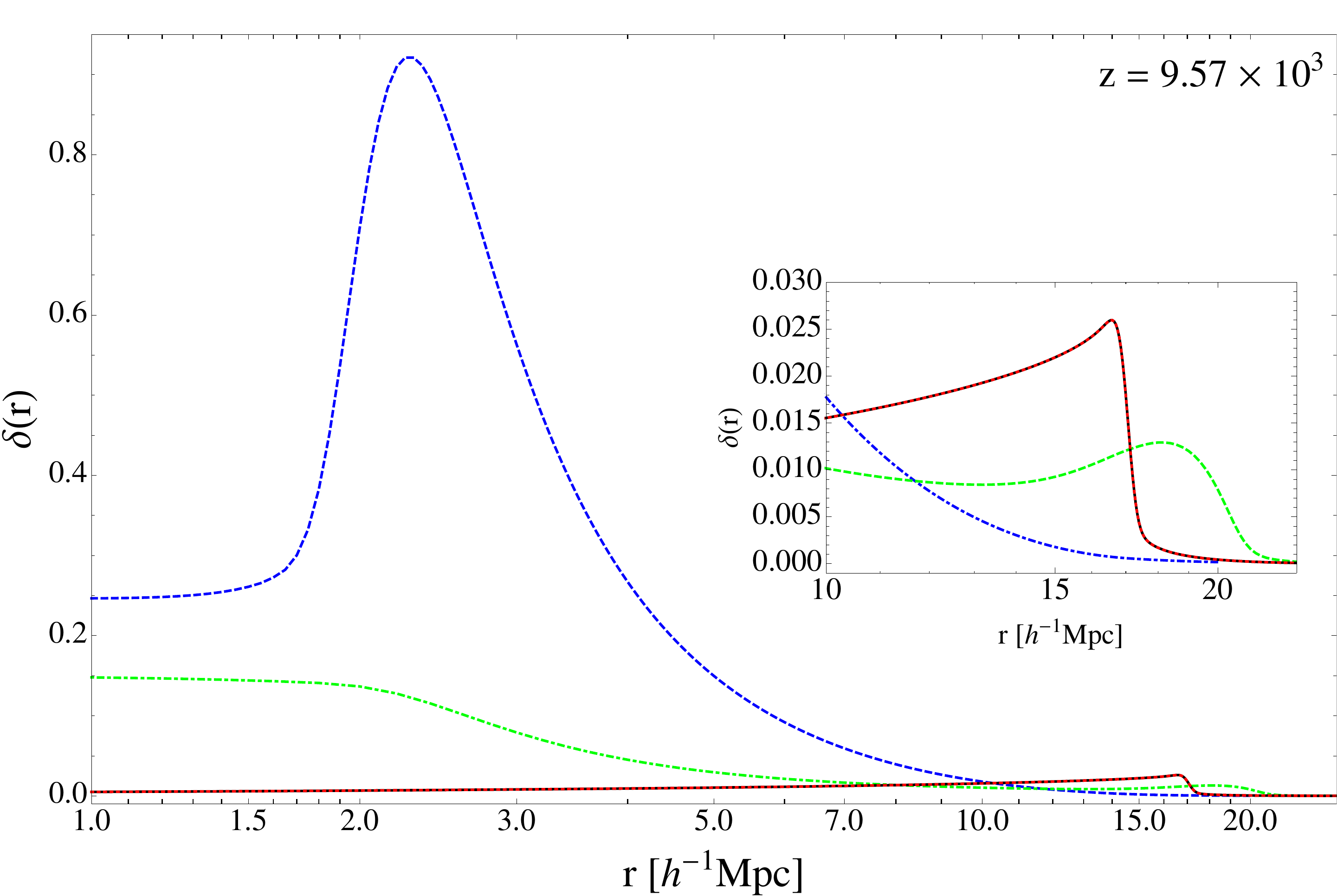}}\\
\subfigure[]{\includegraphics[width=0.495\textwidth]{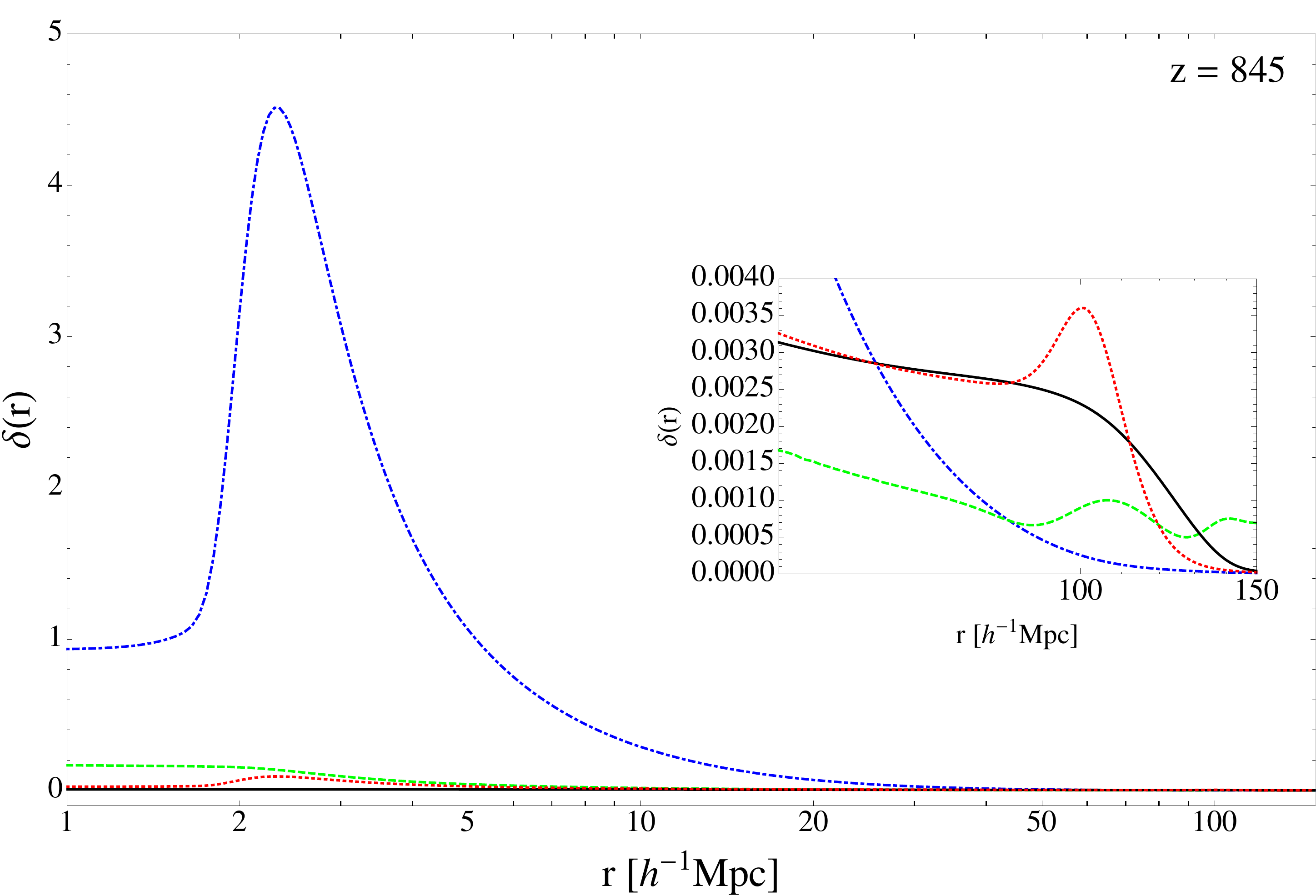}}
\subfigure[]{\includegraphics[width=0.495\textwidth]{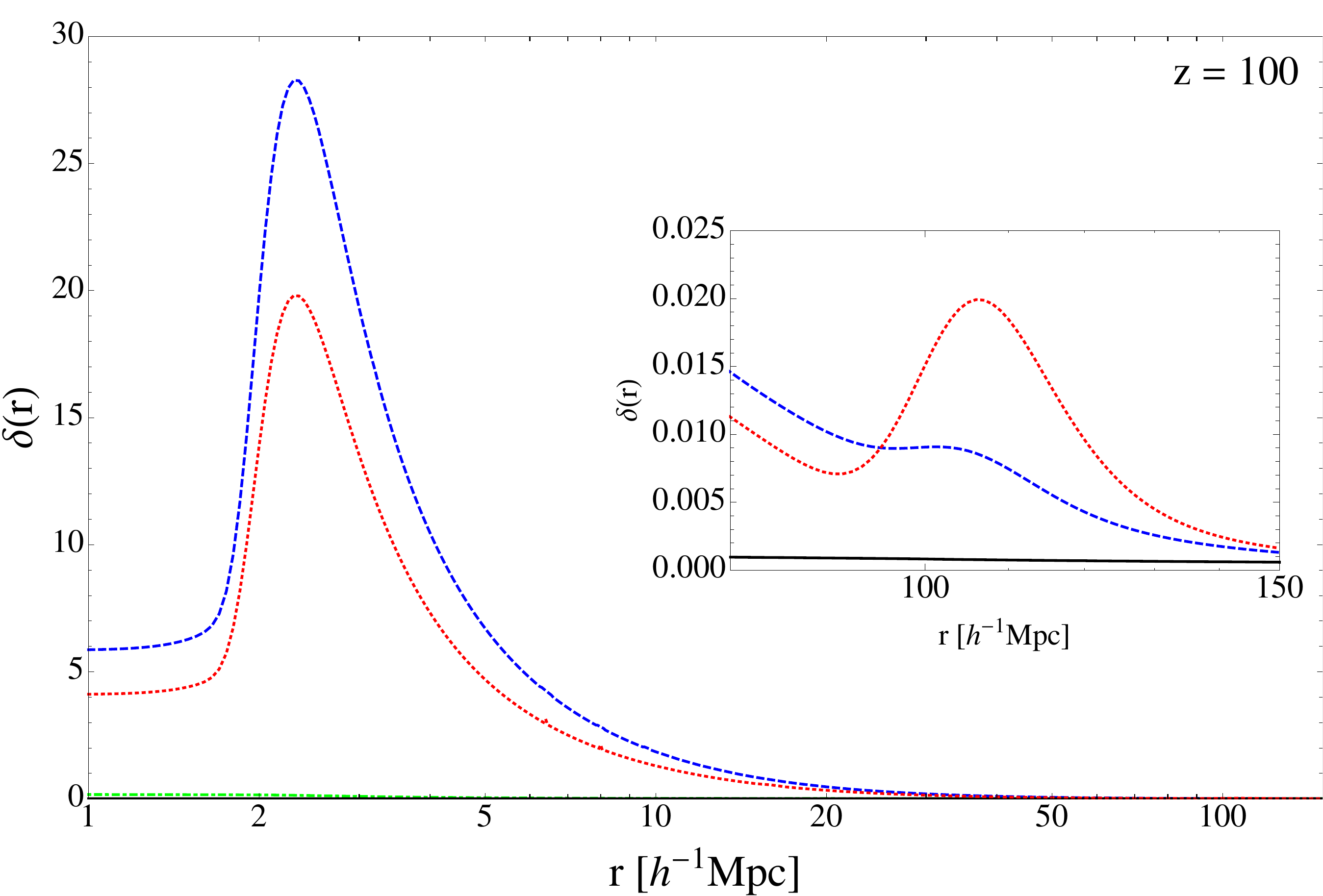}}
\caption{Redshift evolution of an initially Gaussian-shaped density fluctuation in configuration space. We take the fluctuation to be adiabatic. Here, $\alpha_D = 0.08$, $B_D= 1$ keV, $m_D=1$ GeV, $\xi =0.37$, and $r$ stands for the comoving spatial separation. We display the evolution of the DM (dot-dashed, blue), DR (dashed, green), baryons (dotted, red), and photons (solid, black). Panel (a) shows the initial Gaussian density fluctuation together with four snapshots of the outgoing density waves at lower redshifts. The next five panels display different stages of the evolution of the dark and baryon-photon plasmas. Note the changing axes from panel to panel. The insets in panels (d), (e) and (f) focus on the progression of the baryon-photon sound horizon. At late times, two key lengthscales emerge: the standard BAO scale at $r_{\text{BAO}}\simeq 147$ Mpc and the new DAO scale at $r_{\text{DAO}}\simeq 3.3$ Mpc (for the model plotted here).  }
\label{real_space_evolution}
\end{figure*}
%

%
\subsection{Real-Space Evolution: DAO Scale}

The Fourier-space description of cosmological fluctuations allows one to qualitatively understand the different stages of their evolution and to obtain accurate numerical solutions to their equations of motion. Ultimately however, physical density fluctuations evolve in configuration space and it is thus important to translate their Fourier-space behavior into this latter space. 

In Fig.~\ref{real_space_evolution}, we display the configuration-space redshift evolution of a Gaussian adiabatic density fluctuation. In panel (a), we see both a dark-plasma and a baryon-photon plasma density wave traveling outward from the initial over-density. Since the dark plasma generically has a lower sound speed than the regular baryon-photon plasma, density waves propagating in the former do not travel as far in a given time interval as waves propagating in the latter. This results in the dark-plasma density wave lagging behind its baryon-photon counterpart. Panel (a) of Fig.~\ref{real_space_evolution} also clearly shows the damping of the initial density fluctuation at short length-scales resulting from the outward propagation of acoustic waves.

Panel (b) presents a snapshot of the outward-moving waves shortly after dark photons cease to be tightly-coupled to the DM. Panel (c) further shows the DR diffusing out of the DM fluctuation after they kinematically decouple from each other. After the dark decoupling epoch, DM fluctuations begin to grow while DR continues to propagate out of the initial overdensity, eventually overtaking the baryon-photon sound horizon.  Note that the sound horizon of the dark plasma remains imprinted on the DM fluctuations, resulting in a preferred scales in the late-time density field similar to the standard BAO scale. This so-called DAO scale constitutes a tell-tale signature of the presence of a dark plasma distinct from the standard baryon-photon plasma in the early Universe. We will discuss its cosmological implications in the next section.
 
The inset inside panel (d) shows that the dark-photon density wave can actually overtake the baryon-photon plasma wave after the former decouples from the DM. As in the case of free-streaming neutrinos, this effectively establishes gravitational-potential perturbations beyond the sound horizon of the baryon-photon plasma. It has been shown \cite{2004PhRvD..69h3002B} that such supersonic gravitational-potential perturbations can have a measurable impact on the CMB power spectrum. This indicates that DR fluctuations could in principle have an impact on the CMB, which we consider section \ref{cosmology}.
 
Panels (e) and (f) of Fig.~\ref{real_space_evolution} show the late-time decoupling of photons and baryons. As in the standard CDM scenario, baryons fall in the gravitational potential wells established by DM once they cease to be dragged along by the photons. In the atomic DM scenario however, the DAO scale becomes imprinted on the baryons as they fall toward DM in a process similar to how the BAO scale is imprinted on the DM as it falls toward the baryons at late times.

In summary, compared to a vanilla $\Lambda$CDM cosmology, the atomic DM scenario has an additional cosmological length scale corresponding to the size of the dark-plasma sound horizon when the DR kinematically decouples from the DM. Since the dark plasma decouples earlier and has a lower sound speed than the baryon-photon plasma, this DAO scale is generically much smaller than the BAO scale. Furthermore, the amplitude of density fluctuations on length scales smaller than the DAO scale is severely suppressed, leading to a minimal mass for the first objects that collapse at late times. We discuss the cosmological implications of atomic DM in the next section.

\section{Cosmological Implications}\label{cosmology}
The atomic DM scenario alters cosmological observables in two possible ways. On the one hand, the presence of dark relativistic degrees of freedom modifies the cosmological \emph{expansion history} of the Universe. On the other hand, the gravity and pressure of the dark-photon \emph{perturbations} impacts the evolution of DM and baryon-photon fluctuations, hence affecting late-time observables such as the CMB and the matter power spectrum. Unfortunately, the former case does not lead to unique effects on observable cosmological probes since the background cosmology of atomic DM is indistinguishable from that of a standard $\Lambda$CDM model that contains extra relativistic species. As the impact of relativistic neutrinos on the CMB and the matter power spectrum has been extensively studied in the literature (see e.g. \cite{2004PhRvD..69h3002B,2011arXiv1104.2333H}), we shall only briefly review their key effects in the two following subsections. 

Conversely, dark-photon \emph{perturbations} can affect cosmological observables in a way that distinguish them from $\Lambda$CDM models containing extra neutrinos. This stems from the fact that relativistic neutrinos can free-stream from a very early epoch while dark photons can only do so after they decouple from the DM.   Cosmological modes entering the horizon while the dark photons are free-streaming are expected to behave similarly to a $\Lambda$CDM model with an equivalent number of relativistic neutrinos. On the other hand, modes crossing the horizon while the dark photons are tightly-coupled to the DM do not experience the damping and phase shift of acoustic oscillations \cite{2004PhRvD..69h3002B} usually associated with the presence of extra radiation. Therefore, we expect the atomic DM scenario to leave a \emph{distinct} imprint on the CMB if dark photons begin free-streaming when the length scales relevant for this cosmological probe are crossing into the Hubble horizon. A corollary of this statement is that a dark-atom model for which dark photons decouple very early is, as far as the CMB is concerned, indistinguishable from a $\Lambda$CDM universe containing extra neutrinos. 

Atomic DM itself alters cosmological observations through the modified growth of its density fluctuations. As explained in section \ref{DA-regime}, DM fluctuations entering the horizon before kinetic decoupling in the radiation-dominated era miss out on the amplitude boost due to the rapidly decaying gravitational potential. Further, atomic DM perturbations on scales smaller than the dark-photon diffusion length are strongly damped, leading to an absence of cosmological structure at these scales. The acoustic oscillations in the DM plasma which are one of the key feature of the atomic DM scenario remain imprinted on the late-time matter power spectrum. In configuration space, these oscillations point to an important length scale, the DAO scale, at which the clustering of DM is enhanced and below which it is suppressed when compared with an equivalent $\Lambda$CDM model. For most atomic DM models that are in agreement with current observations, the DAO scale must lie today in the highly non-linear regime of cosmological fluctuations, hence making its impact on cosmology difficult to observe.

\subsection{Matter Power Spectrum}\label{mps}
%
\begin{figure}[t]
\includegraphics[width=0.5\textwidth]{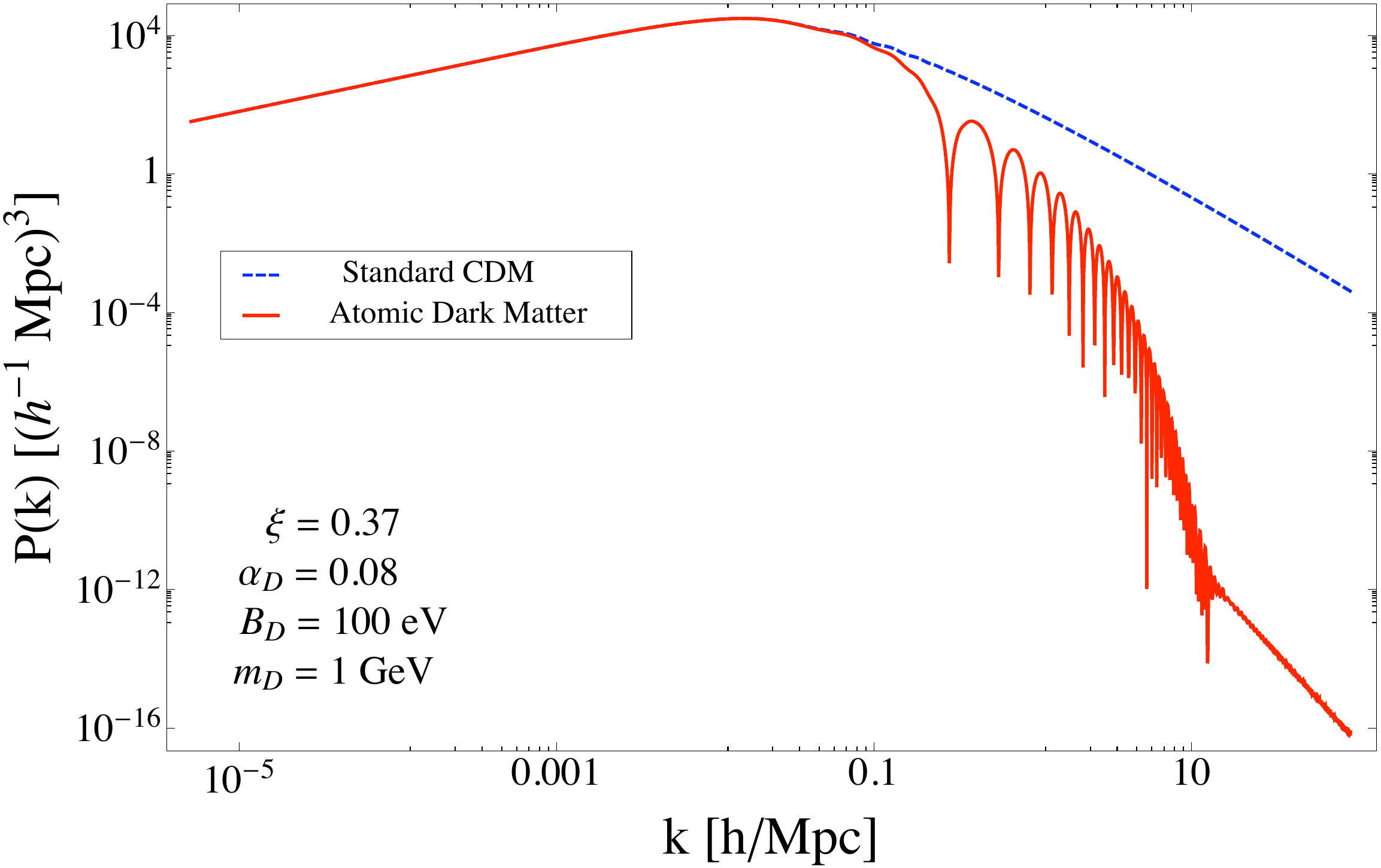}
\includegraphics[width=0.5\textwidth]{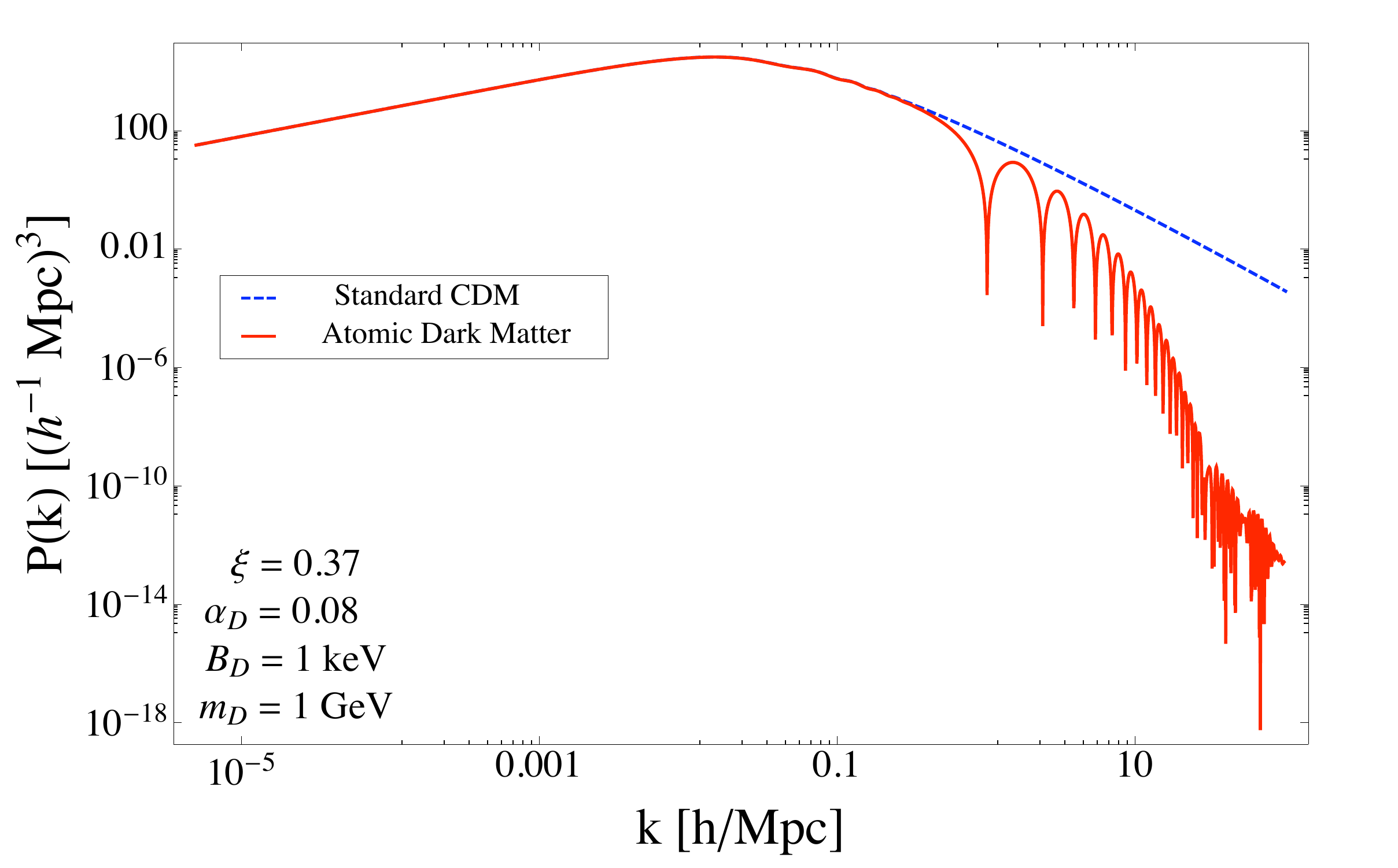}
\includegraphics[width=0.5\textwidth]{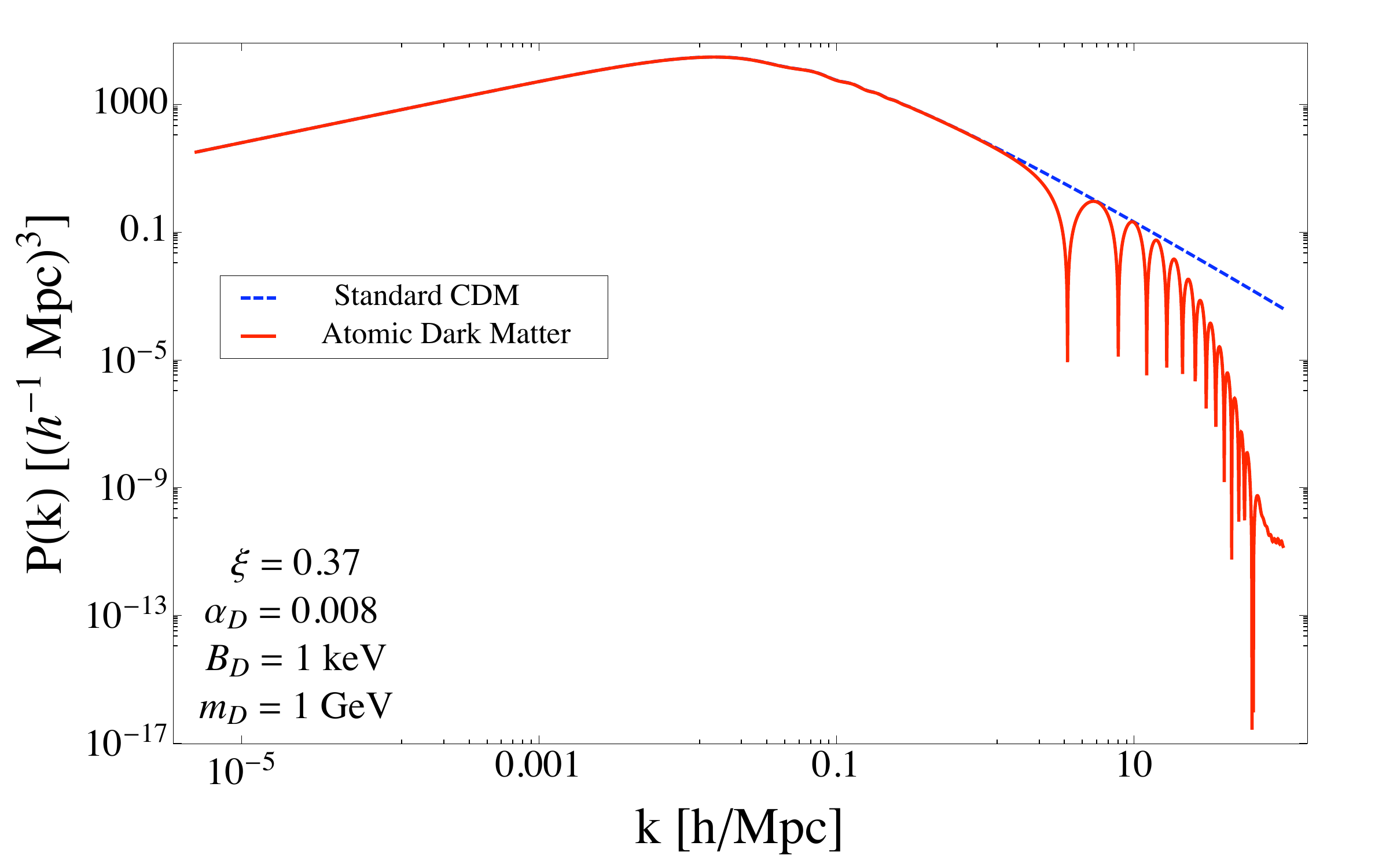}
\caption{Total linear matter power spectrum at $z=0$ for three atomic DM models. For reference, we also display the linear matter power spectrum for a vanilla $\Lambda$CDM model.}
\label{matterpower1}
\end{figure}
\begin{figure}[b]
\begin{centering}
\includegraphics[width=0.5\textwidth]{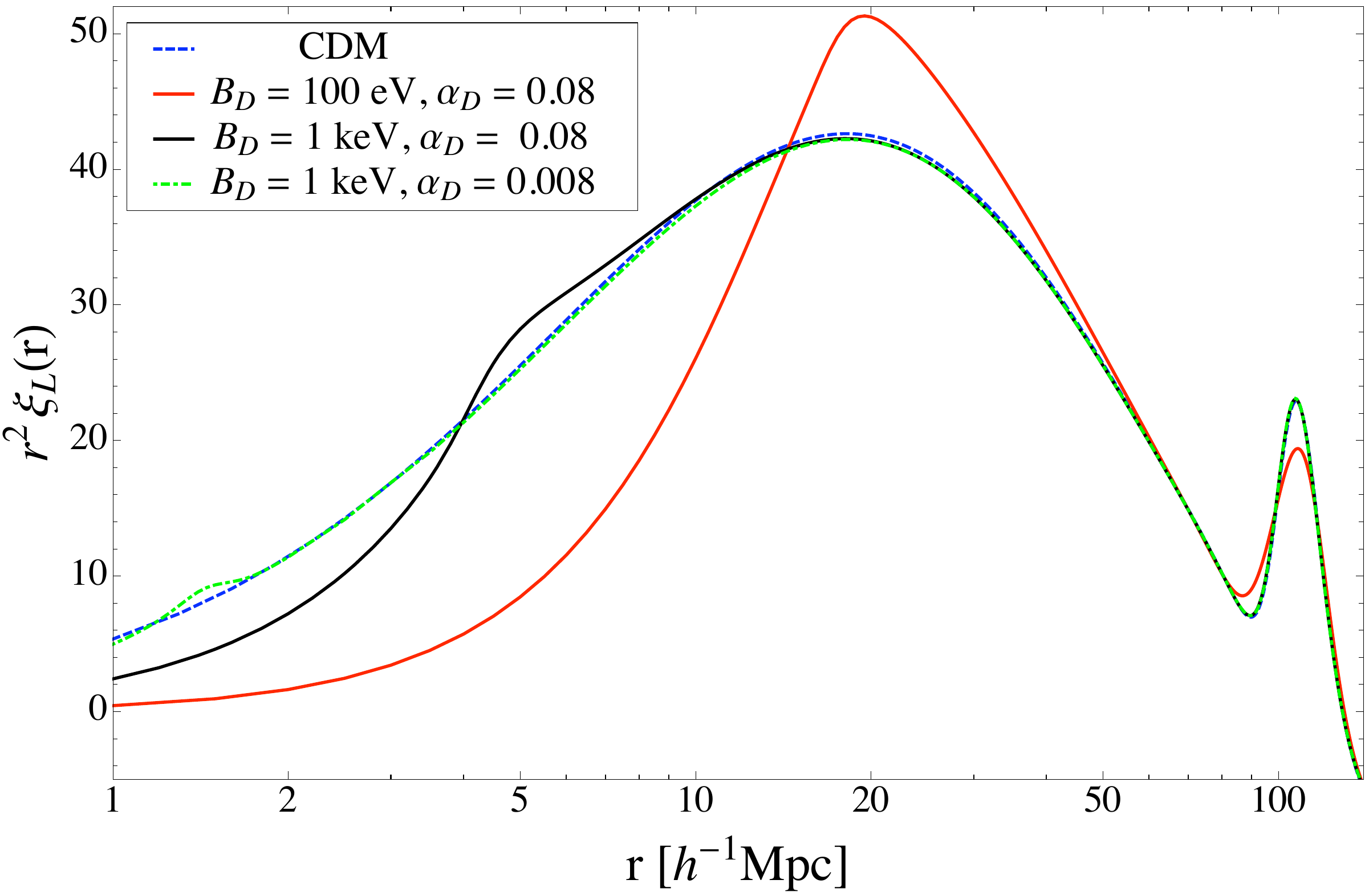}
\caption[Linear matter correlation function.]{Linear correlation function for the three atomic DM models plotted in Fig.~\ref{matterpower1}.}
\label{correlation_function}
\end{centering}
\end{figure}
The most dramatic and distinct cosmological implication of the atomic DM scenario is the modification of the small-scale matter power spectrum. Ultimately, this is a consequence of the relatively late kinetic decoupling of atomic DM compared to a standard WIMP CDM model. Indeed, the delayed kinetic decoupling of the dark plasma considerably impedes the growth of DM fluctuations for all subhorizon modes. Furthermore, fluctuations on length scales shorter than the diffusion distance of dark photons are \emph{exponentially} damped compared to a CDM model, effectively prohibiting the formation of any late-time structure at these scales. Subhorizon scales that exceed the diffusion damping scale display acoustic oscillations that remain clearly imprinted on the late-time matter power spectrum, since DM forms the bulk of the non-relativistic matter. Fluctuations on length scales that cross into the horizon after DM kinematically decouples have a growth history similar to that of a vanilla $\Lambda$CDM model and we therefore expect no particular signature at these scales (see however a caveat at the end of this subsection). 

In Fig.~\ref{matterpower1}, we show examples of late-time linear matter power spectra for different models of atomic DM. The features discussed above are clearly visible in the spectra. We also observe that as the binding energy of the dark atoms is increased, the comoving wavenumber at which the power spectrum significantly departs from the $\Lambda$CDM case is increased. This makes sense since a higher binding energy implies an earlier dark recombination and kinetic decoupling, therefore pushing the impact of the DM-DR coupling toward higher comoving wavenumbers. 

In the two lowest panels of Fig.~\ref{matterpower1}, we keep the dark-atom binding energy and mass constant but vary the dark fine-structure constant $\alpha_D$. We observe that as  $\alpha_D$ is decreased, the smallest comoving wavenumber affected by the DAOs moves toward higher values. At first, this seems counterintuitive as a higher value of the dark coupling constant generally leads to a lower residual ionization fraction which in turns allows the dark photons to rapidly decouple from the DM. We must however remember that dark photons also interact with neutral dark atoms through Rayleigh scattering. For a fixed binding energy and DM mass, the Compton-scattering contribution to the dark-plasma opacity is roughly independent of $\alpha_D$ since $x_D(T_D\ll B_D)\propto \alpha_D^{-6}$ and $\sigma_{{\rm T},D}\propto\alpha_D^6$. On the other hand, the Rayleigh scattering contribution to $\tau_D$ is a steep function of $\alpha_D$ with $\tau_{\rm R}^{-1}\propto\alpha_D^6$ after the onset of dark recombination. Thus, an increase of the dark fine-structure constant can considerably boost the Rayleigh-scattering contribution to the opacity of the dark plasma, hence significantly postponing its epoch of kinetic decoupling.

It is also instructive to consider the correlation function of matter fluctuations in configuration space. The linear correlation function is related to the matter power spectrum via a 3D Fourier transform which, after simplification, can be reduced to
\be
\xi_L(r) =\frac{1}{2\pi^2} \int dk\,k^2\,P(k)\, j_0(k r),
\ee
where $j_0(kr)$ is the Spherical Bessel function of order 0. In Fig.~\ref{correlation_function}, we display the linear correlation function computed from the three matter power spectra shown in Fig.~\ref{matterpower1} as well as the correlation expected from a standard $\Lambda$CDM model. In all cases, the usual BAO scale is clearly visible around $r\sim104h^{-1}$Mpc. In a similar manner, the novel DAO length scale appears as a local enhancement of the correlation function at the scale corresponding to the sound horizon of the dark plasma at dark decoupling. On scales smaller than this sound horizon, the correlation function is significantly damped compared to the $\Lambda$CDM case, a consequence of the damping of small-scale fluctuations discussed above.

At late times, the key signature of these new features in the matter power spectrum and correlation function is a minimal DM halo mass. Indeed, since most of primordial fluctuations on scales smaller than the dark-plasma sound horizon are effectively wiped out by the diffusion of dark photons, no self-bound object can form at late times at these scales. The first regions that can collapse into self-bound DM halos must then have a minimal initial comoving size $\sim r_{\rm DAO}$. Therefore, the first DM halos have a minimal mass given approximately by
\be
M_{\rm min}\approx\frac{4\pi}{3}r_{\rm DAO}^3\Omega_{D}\rho_{\text{crit}}.
\ee
In the hierarchical model of structure formation, these first halos are then accreted into larger, more massive halos. While some of these minimal-mass halos are destroyed through tidal stripping in larger halos, it is expected that a certain fraction of them remains as discrete sub-halos in larger bound objects such as galactic halos. These could potentially be detected through strong-lensing studies of substructures inside galactic halos (see e.g. \cite{2002ApJ...572...25D}). Indeed, there is currently a growing scientific effort aimed at developing analytical methods and experimental techniques geared toward the detection of small-scale substructures inside galactic halos \cite{Moustakas:2008ib,Keeton:2008gq,Moustakas:2009na,Koopmans:2009mq,Marshall:2009eu,2009MNRAS.400.1583V,2009MNRAS.392..945V}. While current results mostly point out the existence of individual massive subhalos in galactic strong lenses \cite{2012Natur.481..341V,2012MNRAS.419..936F,2010MNRAS.408.1969V}, statistical analyses of multiple-image lenses have the potential to lead to strong constraints on the possible minimal subhalo mass. Thus, it is not unreasonable to think that this tell-tale signature of atomic DM might be detected in the near future. 

Currently, the most stringent constraints on the small-scale matter power spectrum come from the Ly-$\alpha$ forest data \cite{Tegmark:2002cy,McDonald:2004eu,Seljak:2004xh}. There is however considerable systematic uncertainties in converting the Ly-$\alpha$ flux power spectrum to the actual linear matter power spectrum \cite{Bird:2010mp}. Most studies have assumed a power-law spectrum with a running spectral index. Since atomic DM predicts a much more complex shape of the matter power spectrum, it is not straightforward to apply these constraints to this scenario. In reality, hydrodynamical simulations of the Ly-$\alpha$ flux power spectrum in the atomic DM scenario will be required to derive the appropriate Ly-$\alpha$ constraints. We can nevertheless use the current measurements to obtain rough guidelines. Ref.~\cite{Seljak:2006qw} has found no deviation from the standard $\Lambda$CDM scenario on scales $k < 2\,h$ Mpc$^{-1}$, while Refs.~\cite{Croft:2000hs,Gnedin:2001wg,Tegmark:2002cy} have determined that the linear power spectrum is consistent with CDM on scales $k \lesssim 5\,h$ Mpc$^{-1}$. Pending unforeseen physical effects, it is unlikely that atomic DM can have a large impact on these scales while leaving the  Ly-$\alpha$ flux power spectrum unchanged. We therefore demand that the atomic DM linear power spectrum does not significantly deviate from that of a vanilla $\Lambda$CDM cosmology on those scales. More precisely, we compute two constraints: $k_{\rm DAO} > 1\,h$ Mpc$^{-1}$ and $k_{\rm DAO} > 5\,h$ Mpc$^{-1}$, where we have defined $k_{\rm DAO}\equiv \pi/r_{\rm DAO}$. These bounds correspond to $M_{\rm min}\approx 10^{13}\text{M}_{\odot}$ and $M_{\rm min}\approx9.3\times 10^{10}\text{M}_{\odot}$, respectively.

We show this constraint in Fig.~\ref{halo_constraint} where we display countours of constant minimal halo mass (dotted white lines) in the $\alpha_D$-$m_D$ plane for various values of the atomic binding energy. The dark and light gray regions correspond to $M_{\rm min}> 10^{13}\text{M}_{\odot}$ and $M_{\rm min}>9.3\times 10^{10}\text{M}_{\odot}$, respectively.  We see that the Ly-$\alpha$ forest data only constrain models with $B_D\lesssim10$ keV. For higher values of the atomic binding energy, the kinetic decoupling of the DM happens very early (even for $\alpha_D\sim\mathcal{O}(1)$) and leads to a minimal halo mass $M_{\rm min}\lesssim10^7$ M$_{\odot}$ which is unconstrained by data.

To modify the faint-end of the galaxy luminosity function and bring it in agreement with the data, atomic DM needs to modify the properties of halos in the range $10^{8}\text{M}_{\odot}\lesssim M_{\rm min}\lesssim10^{10}\text{M}_{\odot}$ \cite{Aarssen:2012fx}. According to the constraints in Fig.~\ref{halo_constraint}, this is possible for model with $B_D\sim10$ keV, $m_D\sim100$ TeV and $\alpha_D\sim0.2$. Detailed $N$-body simulations will be required to assess how the atomic DM scenario exactly affects the halo mass function and density profile, but it is clear that there are allowed values of the dark parameters that can directly address some of the dwarf-galaxy problems.
\begin{figure}[t]
\includegraphics[width=0.5\textwidth]{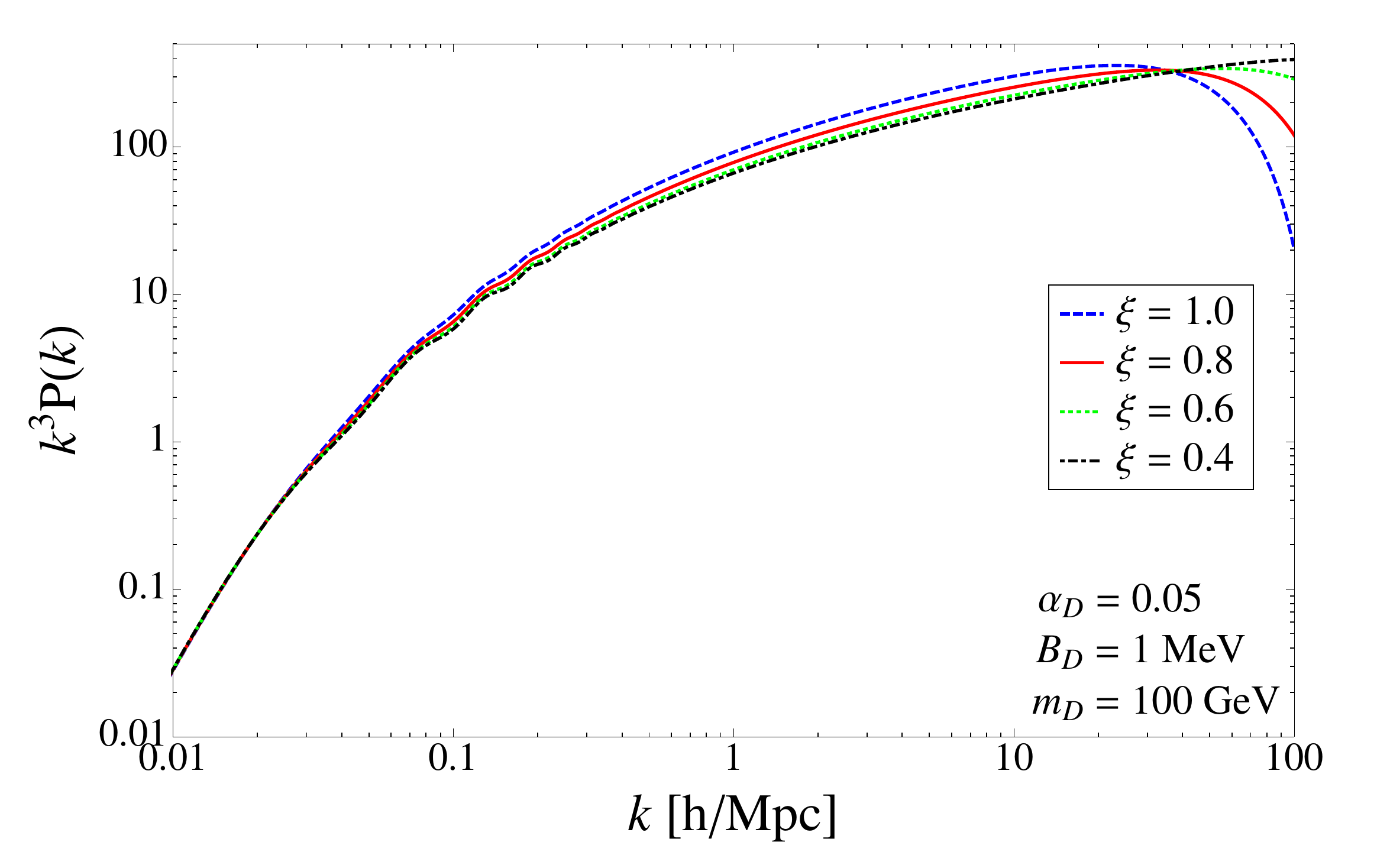}
\caption[Dimensionless linear matter power spectrum at $z=0$ for a single dark-atom model for different values of $\xi$.]{Dimensionless linear matter power spectrum at $z=0$ for a single dark-atom model for different values of $\xi$. We plot $k^3P(k)$ to magnify the small-scale region of the spectrum. Throughout the plot, we fix $z_{\rm eq}$, $\Omega_bh^2$, and the angular scale of the CMB sound horizon at decoupling ($\theta_s$). }
\label{matterpower2}
\end{figure}

For completeness, we also mention in passing that the dark radiation has a small impact on the evolution of DM perturbations even for modes that enter the horizon after dark kinetic decoupling in the radiation era. Indeed, the free-streaming dark radiation amplifies the DM fluctuations as they enter the horizon due to their impact on the rapidly decaying gravitational potential. To illustrate this effect, we must however be careful since the presence of extra radiation shifts the epoch of matter-radiation equality $z_{\rm eq}$ and the angular scale of the CMB sound horizon at decoupling $\theta_s$. Since these quantities have been precisely measured by experiments, we should keep them fixed as we vary the amount of dark radiation. This entails to adjusting the physical DM density ($\Omega_Dh^2$) and the Hubble constant to keep the redshift of equality and the angular scale at decoupling unchanged. We keep $\Omega_bh^2$ fixed throughout since the CMB tightly constrains its value.

We illustrate in Fig.~\ref{matterpower2} the linear matter power spectrum for different values of the ratio of the dark sector temperature to the CMB temperature. We clearly observe that as $\xi$ is increased, the amplitude of Fourier modes crossing into the Hubble horizon in the radiation era are enhanced. In addition to the previously mentioned effect caused by the decaying gravitational potential, this enhancement is also caused by the larger DM density of model with high values of $\xi$. Indeed, the matter fluctuation spectrum is sensitive to the ratio $\Omega_b/\Omega_m$, since baryons are withheld from gravitational collapse prior to hydrogen recombination. As the total matter density is increased for a fixed amount of baryons, more matter can form gravitationally-bound structure and the fluctuation spectrum is therefore enhanced.  Ref.~\cite{2004PhRvD..69h3002B} determined that about a third of the amplification comes from the increase in the radiation density while the rest can be attributed to the larger DM fraction.

\subsection{Cosmic Microwave Background}\label{cmb}
We begin our study of the impact of the atomic DM scenario on the CMB by considering how the modified evolution of \emph{cosmological perturbations} alter the spectra of temperature and polarization anisotropies. The CMB features caused by changes in the fluctuation evolution are more interesting than those resulting from the modified background cosmology since they are potentially \emph{unique} to the atomic DM scenario (or, more generally, to theories incorporating dark plasmas). To isolate the impact of the fluctuations, it is important to keep fixed quantities that only depend on the background cosmology. These are $z_{\rm eq}$, $\theta_s$, the dark energy equation of state parameter $w$, the CMB acoustic damping scale $r_d$, and physical baryon density $\Omega_bh^2$. With these quantities fixed, we wish to determine if it is possible to distinguish an atomic DM model from a $\Lambda$CDM model containing an equivalent number of relativistic degrees of freedom. As we discuss below, the answer is positive. 

To address the issue at hand, it is instructive to first review the impact of relativistic neutrinos on the CMB in the radiation-dominated era. The key point that distinguishes relativistic neutrinos from regular photons is their \emph{free-streaming} nature. Indeed, while photons can only begin free-streaming after they decouple from the baryons at redshift $z\sim1100$, neutrinos are generally assumed to have free-streamed since a very early epoch. This has two important consequences for the physics of the CMB \cite{2004PhRvD..69h3002B}. First, the free-streaming of neutrinos causes a phase shift of the acoustic oscillations upon their entry into the horizon. Second, the impact of free-streaming radiation on the gravitational potential generates a uniform suppression of the CMB oscillation amplitude across all multipoles. Both outcomes can be traced back to the facts that: (1) neutrinos propagate supersonically with respect to the baryon-photon plasma and can thus establish metric fluctuations beyond the sound horizon of the CMB; and (2) the free propagation of neutrinos sources the growth of anisotropic stress on all scales (including superhorizon modes), hence affecting the gravitational source terms in the photon equations of motion.

We would like to determine how the above repercussions on the CMB change when we substitute a dark photon for a relativistic neutrino. As with the regular photons, the fundamental difference between the dark photons and the neutrinos is that the former can only start free-streaming after they decouple from the DM (see \cite{Bell:2005dr} for a similar phenomena involving only neutrinos). This immediately suggests a possible way to distinguish the atomic DM scenario from a $\Lambda$CDM model containing extra relativistic neutrinos. Indeed, if the dark photons begin free-streaming while the length scales relevant to the CMB are entering the horizon, then the phase shift and amplitude suppression associated with the free-streaming of radiation will not be uniform across all the CMB multipoles. Small-scale modes entering the horizon while the dark photons are still coupled to the DM will not be affected by the phase shift and amplitude suppression while larger scales entering after dark-photon decoupling will be affected, as long as they become subhorizon during radiation domination. These non-uniform phase shifts and suppressions of power across multipoles constitute the tell-tale signature of a relativistic degrees of freedom decoupling from the plasma while the Fourier modes important to the the CMB are subhorizon. 
\begin{figure}[t!]
\includegraphics[width=0.5\textwidth]{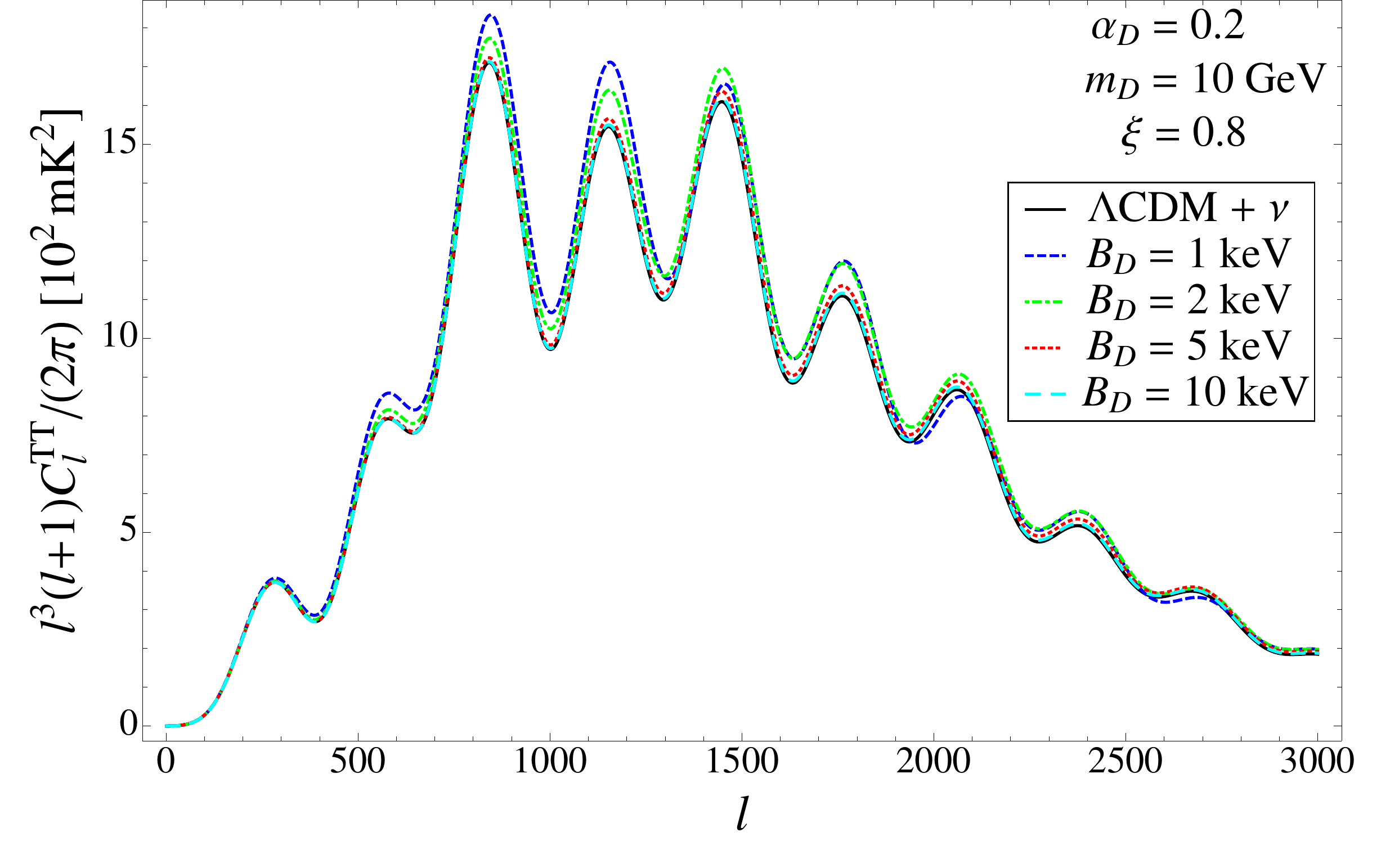}
\includegraphics[width=0.5\textwidth]{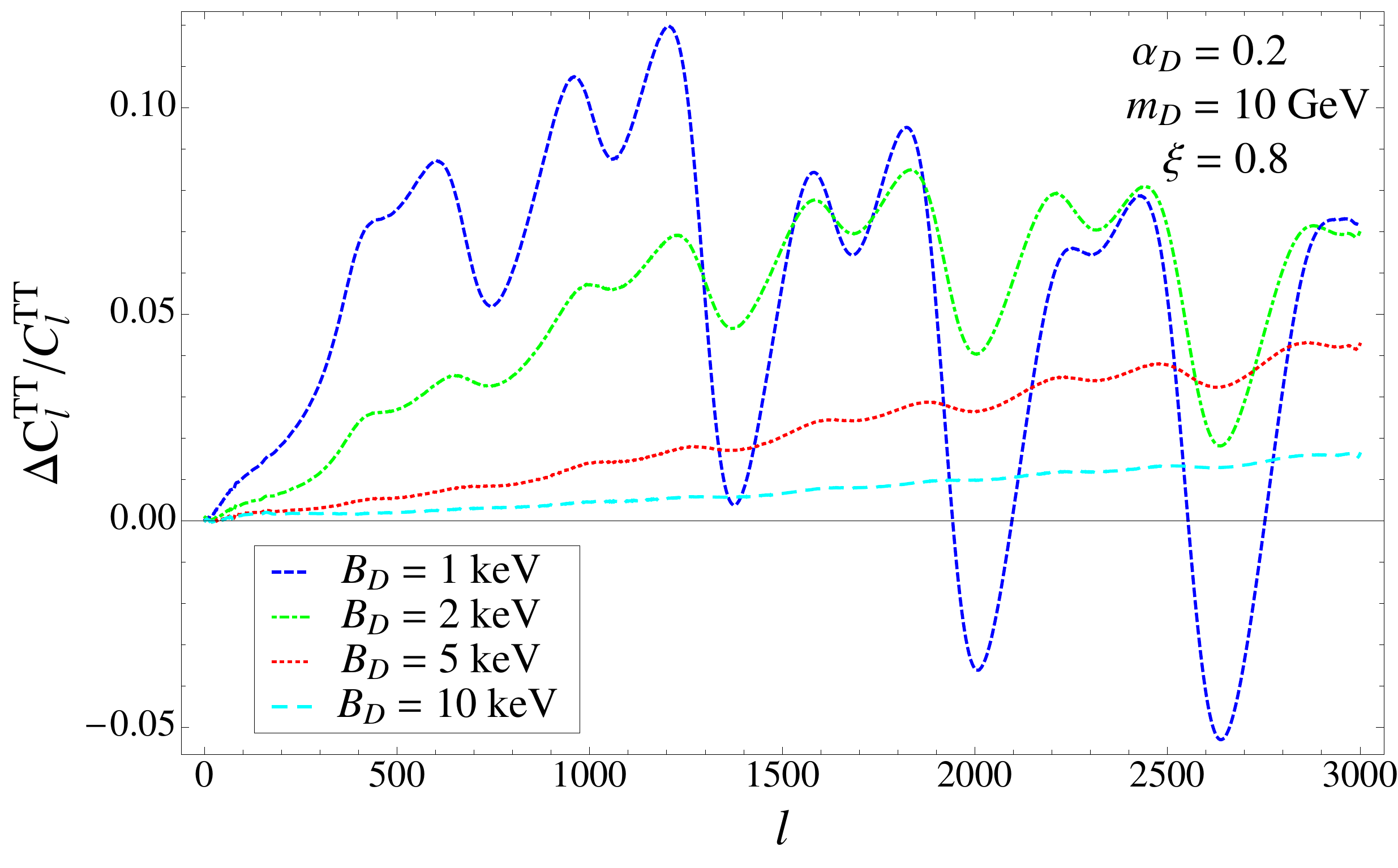}
\caption[Comparison between the CMB temperature angular power spectrum of atomic DM and that of a $\Lambda$CDM model with an equivalent number of extra neutrinos.]{Comparison between the CMB angular power spectrum of atomic DM and that of a $\Lambda$CDM model with an equivalent number of extra neutrinos ($N_{\nu}=4.849$ here). The various lines illustrate different values of the atomic binding energy, $B_D$. We fix all other dark parameters to the values indicated on the plot. The upper panel displays the TT spectra while the lower panel shows the fractional difference between the TT spectra of atomic DM and that of the $\Lambda$CDM model containing extra neutrinos. All other cosmological parameters are held fixed. Here, the helium fraction is fixed to $Y_p=0.24$.}
\label{changing_B}
\end{figure}
\begin{figure}[t!]
\begin{centering}
\includegraphics[width=0.5\textwidth]{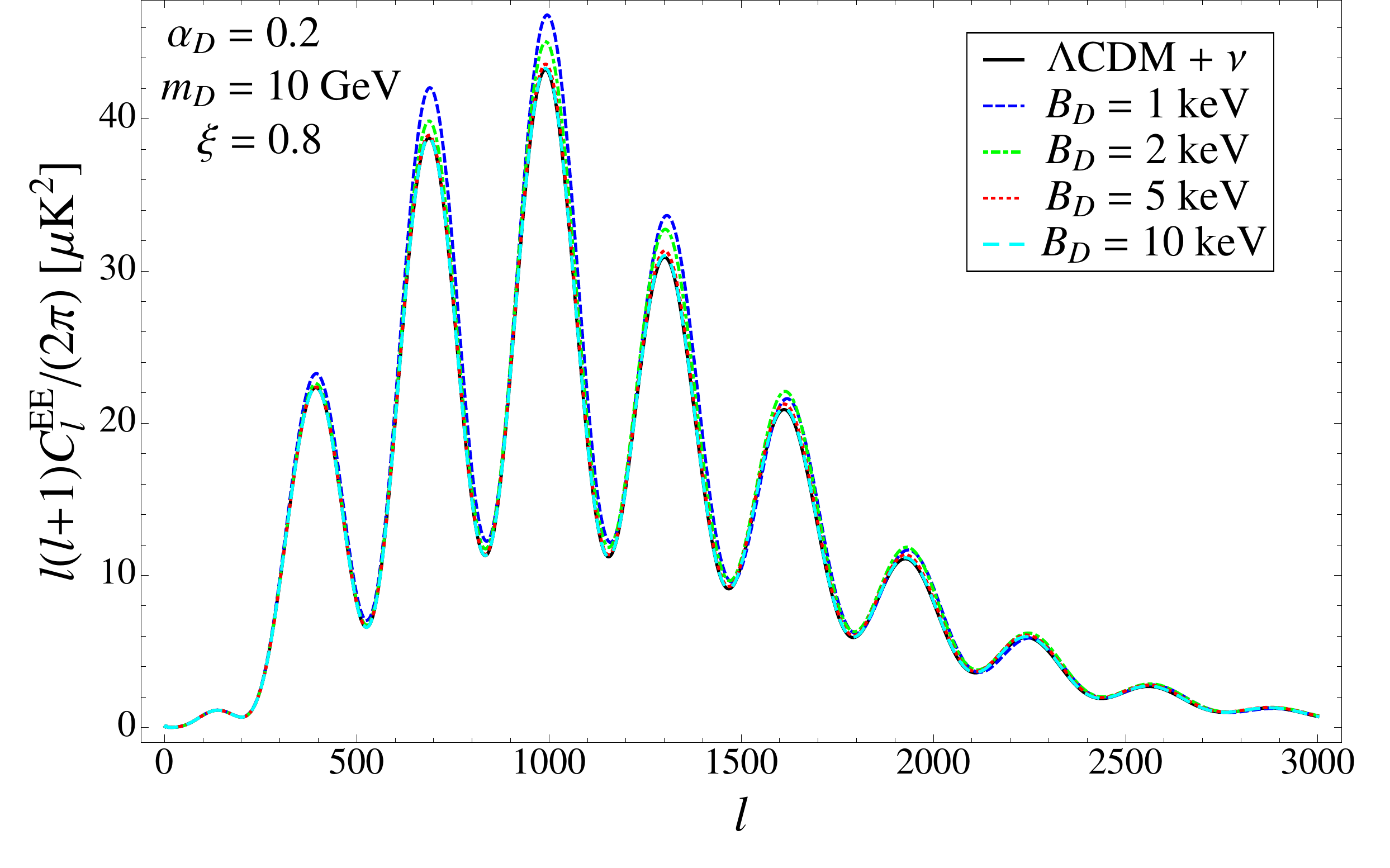}
\includegraphics[width=0.5\textwidth]{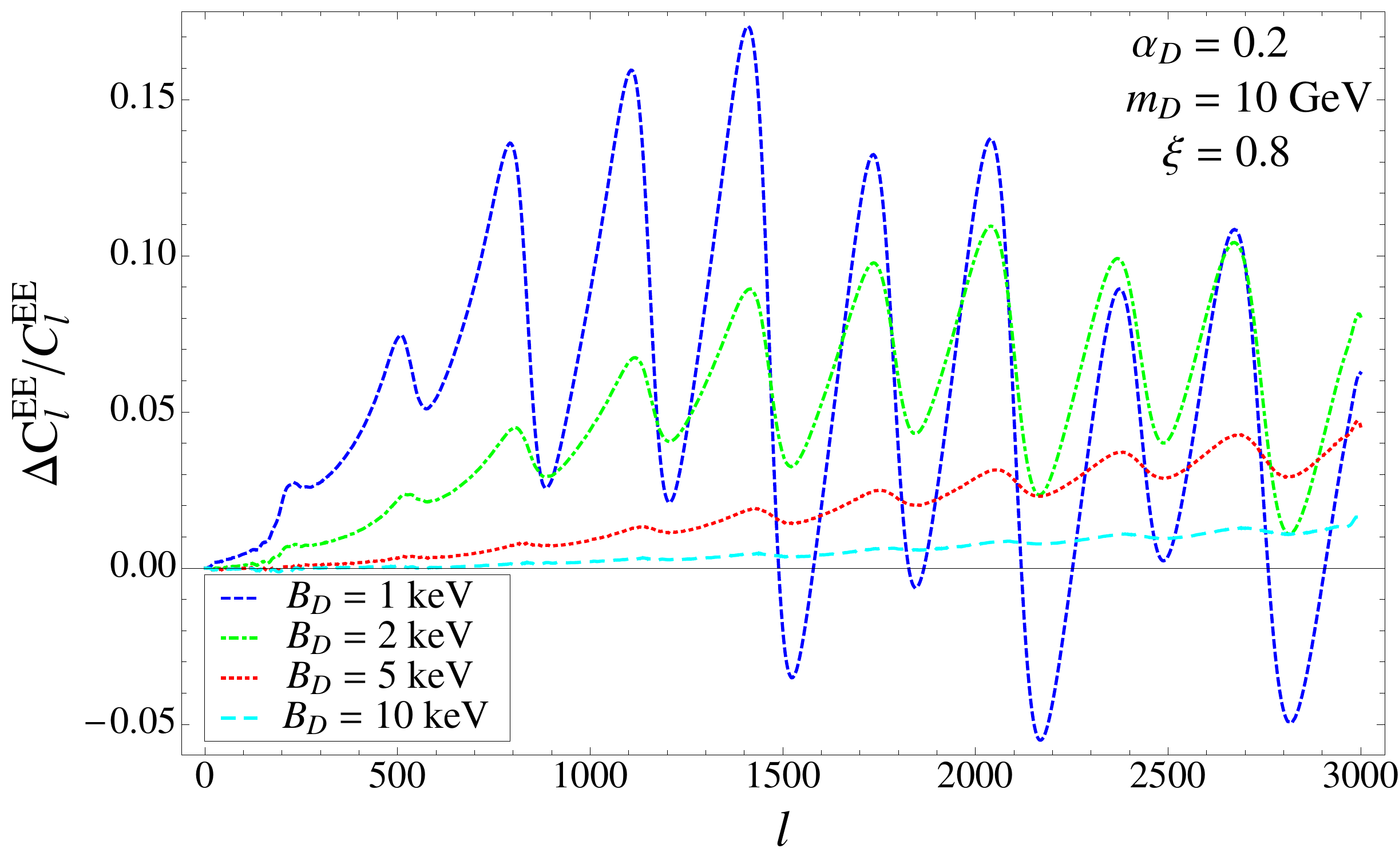}
\caption[Comparison between the CMB polarization angular power spectrum of atomic DM and that of a $\Lambda$CDM model with an equivalent number of extra neutrinos.]{Similar to Fig.~\ref{changing_B} but for the CMB EE polarization spectra.}
\label{EEchanging_B}
\end{centering}
\end{figure}

In Figs.~\ref{changing_B} and \ref{EEchanging_B}, we compare the temperature and polarization CMB power spectra for atomic DM models having different binding energy. We also show the CMB spectra of a $\Lambda$CDM model incorporating the same amount of additional relativistic degrees of freedom as the atomic DM models such that all the spectra shown have the same cosmological background evolution. The upper panels show the spectra themselves while the lower panels display the relative differences between the atomic DM spectra and the corresponding $\Lambda$CDM model. We immediately notice that as the binding energy is increased, both temperature and polarization power spectra converge toward the $\Lambda$CDM model containing extra neutrinos. Indeed, as the binding energy of dark atoms is made larger, dark photons decouple earlier from the DM and can therefore begin free-streaming at an earlier epoch, making them hard to distinguish from relativistic neutrinos. 

For both temperature and polarization spectra, the key physical signatures that distinguish dark photons from neutrinos are clearly visible. First, let us discuss the amplitude suppression associated with the free-streaming radiation. We see that, compared with the $\Lambda$CDM model containing extra radiation, the amplitude of the CMB spectra for the atomic DM scenario are less suppressed, with the high-$l$ multipoles being the least affected by the suppression.  This is in line with our expectations since high multipoles enter the horizon before the dark photons have the chance to significantly free-stream and are therefore more immune to the suppression. The atomic DM models with $B_D=5$ keV and $10$ keV clearly display this behavior. On the other hand, the scenarios with the lowest binding energies ($B_D=1$ and $2$ keV) exhibit a more complex $l$-dependence when compared with the $\Lambda$CDM model. To understand this difference, we need to invoke the important phase shifts between the atomic DM and $\Lambda$CDM models as well as the different growth history of the DM fluctuations in the two scenarios. 

\begin{figure}[t!]
\includegraphics[width=0.5\textwidth]{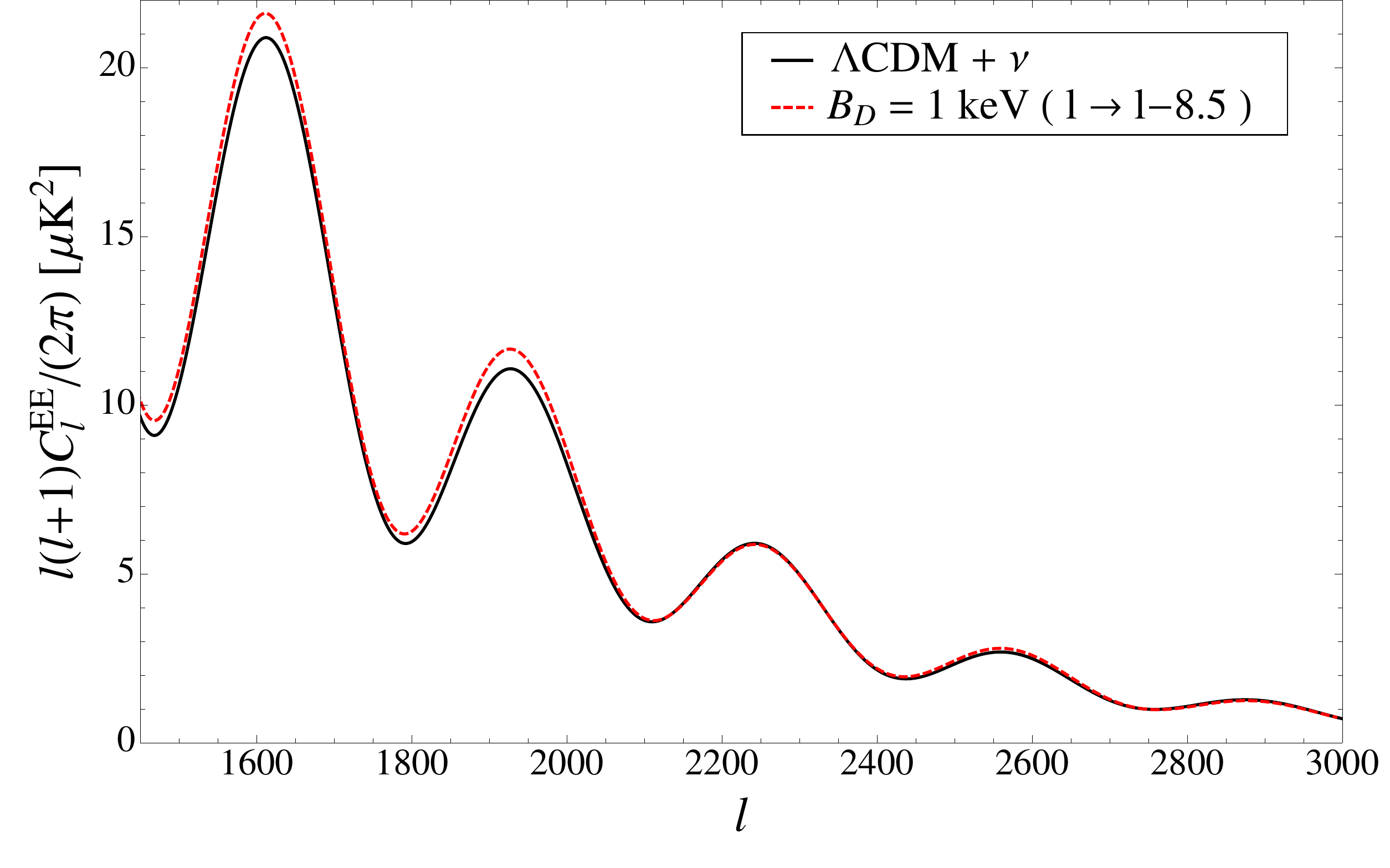}
\caption[Illustration of the constant phase shift between an atomic DM model and a $\Lambda$CDM model containing an equivalent number of relativistic species.]{Illustration of the constant phase shift between an atomic DM model and a $\Lambda$CDM model containing an equivalent number of relativistic species. We show here that, in the atomic DM scenario, multipoles that enter the Hubble horizon before dark photons begin free-streaming do not experience the constant phase shift toward smaller $l$ that usually characterizes models with extra relativistic degrees of freedom. To illustrate this, we have shifted the atomic DM spectrum by $\Delta l=-8.5$ and observed that the phases of both spectra coincide.}
\label{phase_shift_CMB}
\end{figure}

The effect of the phase shift is most visible for the $B_D=1$ keV model. Indeed, since dark photons in this model are just beginning to free-stream when most Fourier modes contributing to the CMB enter the horizon, these do not experience the same phase shifts as the $\Lambda$CDM model. This can be most clearly discerned in the polarization spectrum (Fig.~\ref{EEchanging_B}). There, we see that the phase difference between the atomic DM models and the $\Lambda$CDM containing extra neutrinos becomes progressively larger toward higher multipoles. This is exactly what we expect since the high multipoles enter the horizon before they can be affected by dark-photon free-streaming, while smaller multipoles experience a phase shift that progressively converges toward the pure neutrino case as $l$ is lowered. One of the key feature of this drifting phase shift is that it converges toward constant values for both $l\gg l_{\rm dec}$ and $l\ll l_{\rm dec}$, where $l_{\rm dec}$ corresponds to the multipole that crosses into the horizon as dark photons begin free-streaming. This contrasts with phase shifts caused by a change in the angular scale of the sound horizon which act multiplicatively $l\rightarrow\alpha l$. We show in Fig.~\ref{phase_shift_CMB} the damping tail of the $EE$-polarization spectrum for the $B_D=1$ keV model. To illustrate the constant phase shift at $l\gg l_{\rm dec} ( \sim 500$ here), we have shifted its spectrum by $\Delta l\sim-8.5$. We see that the phase of the shifted spectrum matches very well that of the $\Lambda$CDM model containing extra relativistic neutrinos, hence showing the phase shift does asymptote to a constant at high multipoles. 

There is an important ramification to the above observations. Since the CMB temperature anisotropies are sensitive to the DM-dominated gravitational potential at the epoch of last scattering, any significant modification to the growth history of DM density perturbations will be reflected in the $C_l^{TT}$ power spectrum. As we discussed in section \ref{DA-regime}, DM fluctuations cannot grow as long as they are coupled to the DR. In the matter-dominated era, growing DM fluctuations are usually responsible for establishing the gravitational potentials that act as a restoring force to counterbalance the pressure of the photons. If DM perturbations are prohibited to grow by their coupling to the DR, then the gravitational potentials cannot be established and the gravitational source term acting on photon fluctuations is much weaker. Observationally, this has the consequence of altering the ratios of the odd and even peaks of the CMB spectrum, with the odd peaks being suppressed on scales where DM fluctuations cannot grow. Indeed, the odd peaks correspond to gravity-driven compression waves which are very sensitive to the size of the DM fluctuations. The lower panel of Fig.~\ref{changing_B} clearly shows the deep troughs associated with the damping of the odd $C_l^{TT}$ peaks caused by the late kinetic decoupling of DM for the models with $B_D=1$ and 2 keV.  Unfortunately, models displaying such significant suppression of odd $C_l^{TT}$ peaks at high multipoles have very low values of $\sigma_8$ and violate the Ly-$\alpha$ bound on $k_{\rm DAO}$ are therefore ruled out by observations.

We now turn briefly our attention on how the atomic DM scenario affects the CMB through its impact on the \emph{background} cosmology. Since atomic DM modifies the background evolution by the presence of the DR, the effects on the CMB are not unique but common to any theory incorporating extra relativistic degrees of freedom. For completeness, we nevertheless discuss their effects here. These include a modification of the primordial helium abundance, a change to the angular scale of the sound horizon at decoupling, a shift of the matter-radiation equality epoch, an enhanced early ISW effect and a modified CMB damping tail.

\noindent\emph{Primordial Helium Abundance} The impact of the relativistic dark components on the expansion rate during BBN tends to increase the primordial helium fraction through the approximate relation \cite{Simha:2008zj}
\ba\label{Yp}
Y_p &\approx& 0.2485 + 0.0016\Bigg[(273.9\,\Omega_bh^2-6)\en
&&\left.\qquad\quad+ \,\,100\left( \sqrt{1+\frac{4}{43}g_{*,D}^{\text{BBN}}\xi_{\text{BBN}}^4}-1\right)\right],
\ea
 where we have used the relation 
 \be
 \Delta N_{\nu}^{\text{BBN}} =  \frac{8}{7}\frac{g_{*,D}^{\text{BBN}}}{2}\xi_{\text{BBN}}^4
 \ee
between the number of additional neutrinos at BBN ($ \Delta N_{\nu} \equiv N_{\nu}-3$) and the relativistic degrees of freedom of the dark sector at that epoch. Since helium recombines before hydrogen at late times, a larger helium fraction leads to a net decrease in the free-electron fraction around $z\simeq1100$. Consequently, the photons can diffuse more easily out of inhomogeneities and damp temperature and polarization anisotropies on scales smaller than the diffusion length. Therefore, we generically expect the CMB to display less power on small angular scales for atomic DM scenarios predicting a large primordial helium fraction \cite{2004PhRvD..69h3002B,2011arXiv1104.2333H}. 
\begin{figure}[t!]
\begin{centering}
\includegraphics[width=0.5\textwidth]{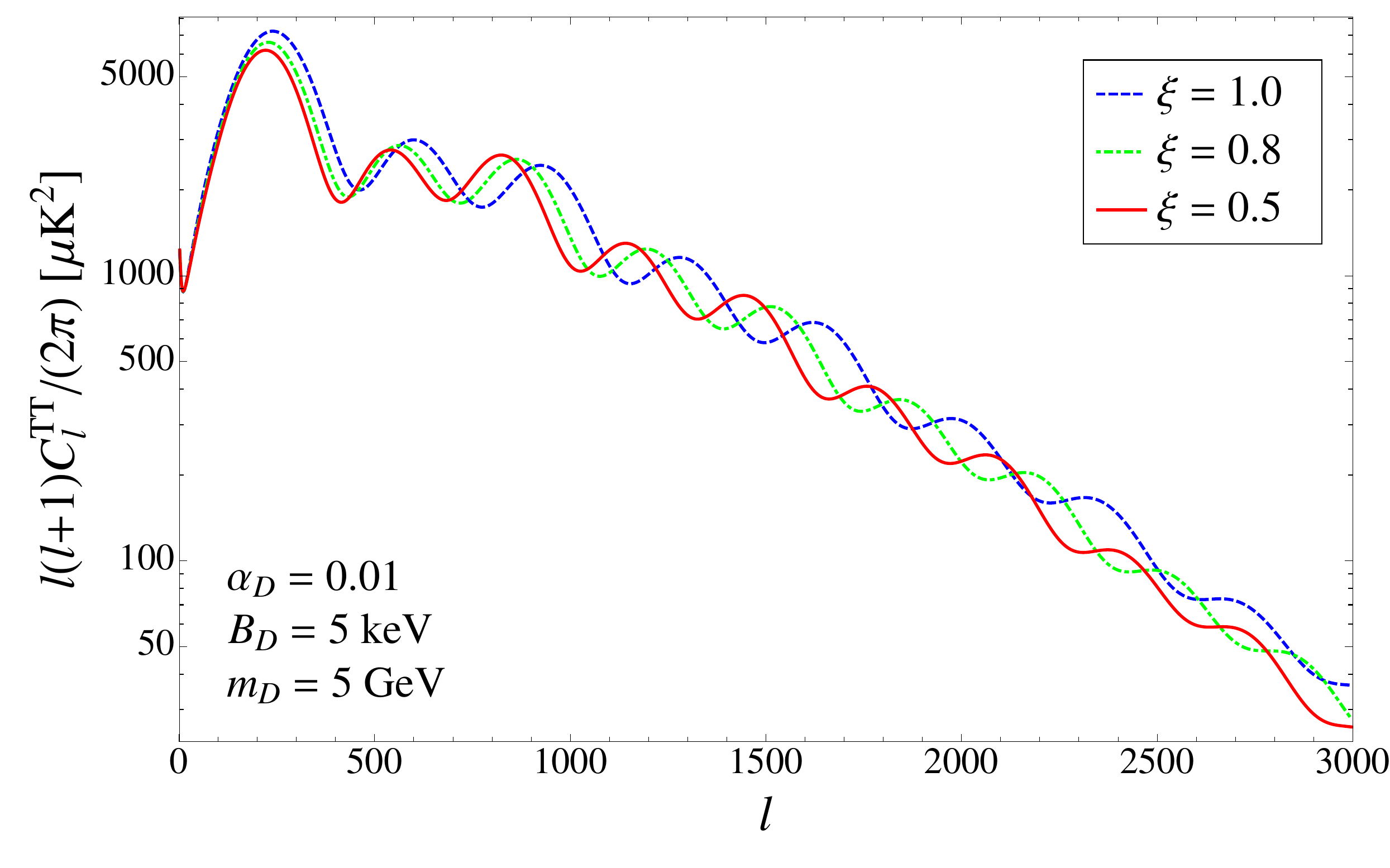}
\includegraphics[width=0.5\textwidth]{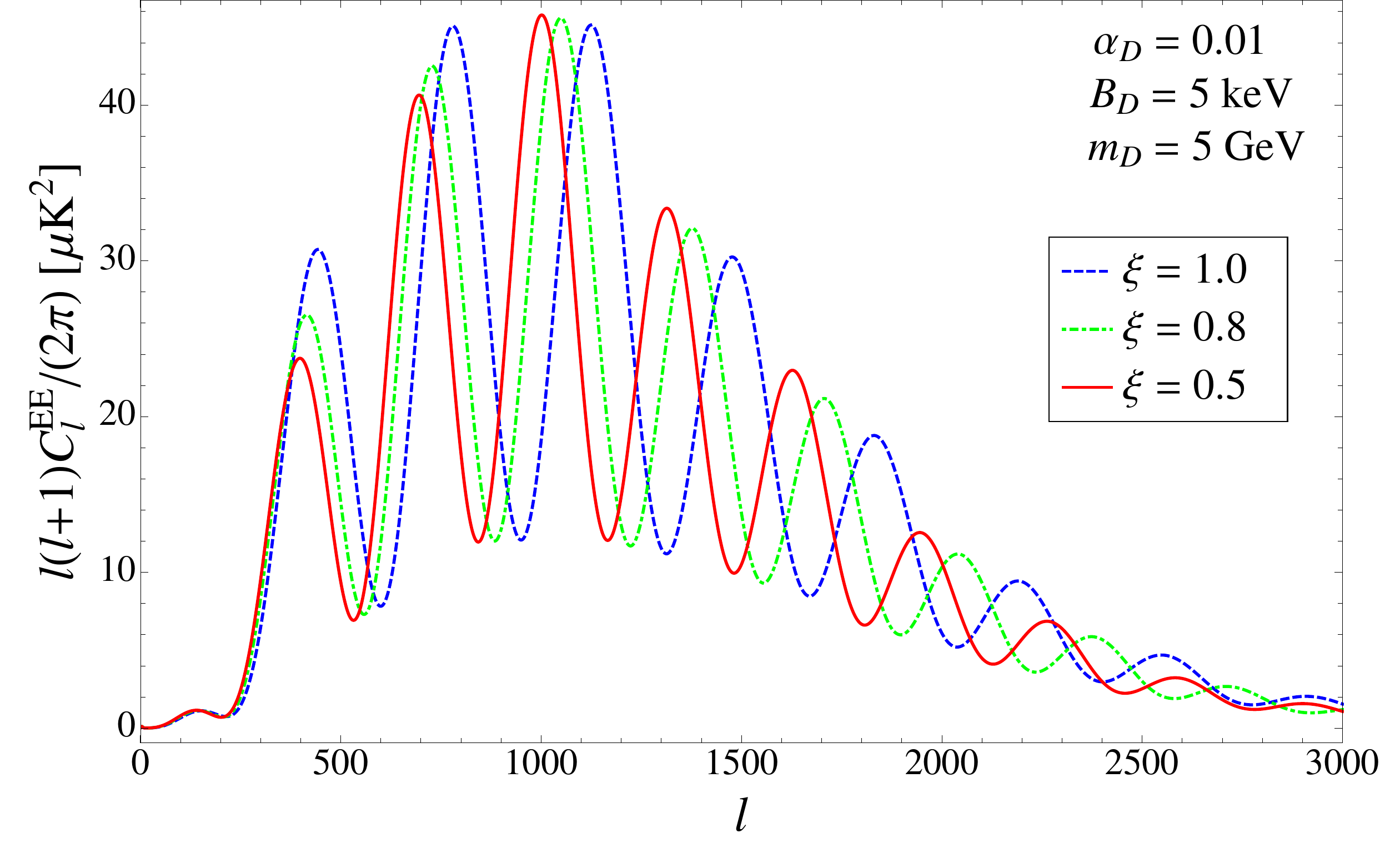}
\caption[CMB angular power spectra in the atomic DM scenario for different values of $\xi$. ]{CMB angular power spectra in the atomic DM scenario for different values of $\xi$. We fix all other dark parameters to the values indicated on the plots. The upper panel displays the TT spectra while the lower panel shows the EE polarization spectra. All other cosmological parameters are held fixed. Here, the helium fraction is fixed to $Y_p=0.24$ to isolate the effect from the changing sound horizon.}
\label{changing_xi}
\end{centering}
\end{figure}

\noindent\emph{Angular size of Sound Horizon, Matter-Radiation Equality and Early ISW} The presence of the dark photons affects the size of the baryon-photon sound horizon through its impact on the Hubble expansion rate prior to hydrogen recombination. The sound horizon of the baryon-photon plasma is given by
\be
r_{s}^{(b-\gamma)}=\int_0^{a_*}\frac{c_{s}^{(b-\gamma)} \,da}{a^2H},
\ee
where $a_*$ is the scale factor at recombination and $c_{s}^{(b-\gamma)}$ is the sound speed of the baryon-photon plasma. Since the extra DR works to increase the Hubble rate, the net effect is a smaller value of $r_{s}^{(b-\gamma)}$. As the angular size of sounds horizon is given by $\theta_s=r_{s}^{(b-\gamma)}/D_{\rm A}$, where $D_{\rm A}$ is the angular-diameter distance to the last-scattering surface, we therefore expect the CMB acoustic peaks to be shifted toward smaller angular scales (higher multipoles $l$). This effect is illustrated in Fig.~\ref{changing_xi} where we display the temperature and E-polarization spectra for different values of $\xi$. The shift of the acoustic peaks to smaller angular scales is clearly visible for both types of spectra. Further, we see that the temperature anisotropies are amplified around the first acoustic peak as $\xi$ increases. This is the result of the integrated Sachs-Wolfe (ISW) effect caused by the extra DR through its impact on the changing gravitational potential after recombination. Indeed, increasing $\xi$ brings the epoch of matter-radiation equality closer to that of recombination, hence increasing the impact of radiation on the gravitational potential at late times.      

\begin{figure}[t]
\includegraphics[width=0.5\textwidth]{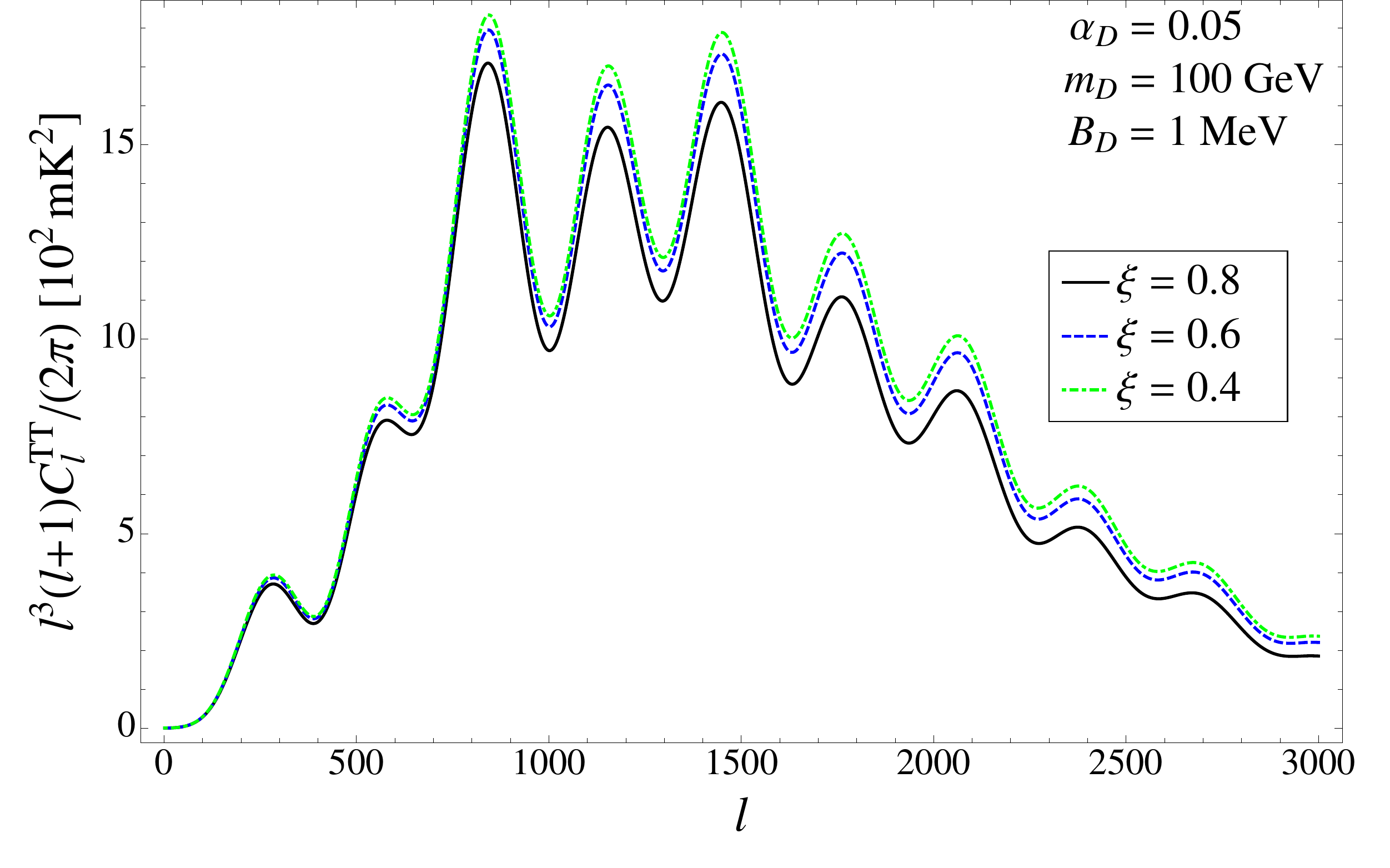}
\caption[CMB temperature power spectra in the atomic DM scenario for different values of $\xi$.]{CMB temperature power spectra in the atomic DM scenario for different values of $\xi$. We fix all other dark parameters to the values indicated on the plots. We keep fixed throughout the redshift of matter-radiation equality and the angular size of the baryon-photon sound horizon at decoupling. Here, the helium fraction is fixed to $Y_p=0.24$ to isolate the effect from the changing damping scale.}
\label{damping_tail}
\end{figure}

\noindent\emph{Silk Damping Tail} If we fix the epoch of matter-radiation equality, the primordial helium fraction and the angular scale of the sound horizon at decoupling, increasing the energy density of the dark photons leads to an enhanced damping of the CMB anisotropies \cite{2011arXiv1104.2333H}. To understand the origin of this effect, we need to remember that the photon diffusion distance scales as $r_d\propto H^{-0.5}$ (see Eq.~(\ref{k_D})) while the angular diameter distance scales as $H^{-1}$. Thus, the damping angular scale $\theta_d\equiv r_d/D_{\rm A}$ effectively increases if the Hubble rate is sped up due to the presence of extra radiation. We therefore expect that as the value of $\xi$ is raised, the CMB spectrum will be increasingly affected by Silk damping. This effect is shown in Fig.~\ref{damping_tail} where we clearly observe the decline in amplitude associated with the increasing DR density. In addition, if the primordial helium fraction was allowed to vary according to Eq.~(\ref{Yp}), this would further increase the amount of Silk damping. Therefore, it is clear that measurements of the CMB damping tail provide strong constraint on $\xi$.    

In summary, beyond the impact of the atomic DM scenario on the background cosmology caused by the DR, we have identified four key cosmological signatures that distinguish the atomic DM scenario from a $\Lambda$CDM model containing extra relativistic neutrinos. First, the emergence of the new DAO length scale in the late-time density field results in a minimal mass for the first DM protohalos that is generically larger than in the standard WIMP paradigm. Also, as the dark photons transition from being tightly-coupled to the dark plasma to a free-streaming state, they impart varying phase shifts and amplitude suppressions to the CMB multipoles entering the horizon. Importantly, these suppressions and phase shifts asymptote to constant values for $l\gg l_{\rm dec}$ and $l\ll l_{\rm dec}$, a distinct feature of atomic DM that is not easily reproduced in the $\Lambda$CDM scenario. Furthermore, we have shown that the odd $C_l^{TT}$ peaks are suppressed on scales that enter the causal horizon before DM kinematically decouples. It is therefore clear that precise measurements on the CMB damping tail could provide meaningful constraints on the parameter space of atomic DM. We should however keep in mind that the modified evolution of DM and DR \emph{fluctuations} can only affect the CMB if the dark sector kinetic decoupling happens close enough to the epoch of last scattering. As such, a non-detection of these signatures effectively puts a lower bound on the redshift of kinetic decoupling which itself depends on a combination of $\alpha_D$, $B_D$, $m_D$, and $\xi$.

\section{Astrophysical Constraints on Atomic Dark Matter}\label{astro}
As the Universe expands and cools down, non-linear structures begin to emerge and eventually form present-day astrophysical objects such as galaxies and clusters of galaxies. The internal dynamics of these objects is deeply influenced by the microphysics governing DM because the latter contributes the vast majority of the mass inside these objects. Since the atomic DM scenario naturally incorporates new interactions in the dark sector, it is important to discuss its implications for the dynamics of DM halos on a wide range of scales. 

On one side, observations indicate that the DM halos of elliptical galaxies and clusters display a triaxial ellipsoidal shape \cite{MiraldaEscude:2000qt,Buote:2002wd,Inada:2003vc,Hoekstra:2003pn,Mandelbaum:2005nf,Koopmans:2006iu,Morokuma:2006qq,Fang:2008ad,Evans:2008mp,Host:2008gi}. This indicates that the relaxation (thermalization) time of their DM halos is much longer than the typical dynamical time of these celestial objects. From this observation, one can obtain a bound on the elastic collisional rate of DM particles. Similarly, detailed studies of the Bullet Cluster provide direct constraint on the self-interaction cross section of DM \cite{2004ApJ...606..819M}. However, these constraints have been shown to be weaker than those derived from the halo ellipticity \cite{Feng:2009mn} and we therefore do not consider them here.     

On the other side, due to the rich internal structure of dark atoms, the atomic DM scenario inherently includes dissipation mechanisms that could potentially have dramatic effects on the dynamics of DM halos. Indeed, collisional excitations of dark atoms followed by dark-photon emissions provide a cooling mechanisms for DM that could drastically alter the internal structure of halos. Further, the \emph{chemistry} of the dark sector naturally provides other heat-dissipation mechanisms such as molecular cooling. Fortunately, demanding that DM particles be effectively collision-less to preserve the structure of halos strongly hinder the efficiency of the atomic dissipative processes, rendering them mostly irrelevant. Similarly, the requirement that very few actual particle collisions take place in the dark sector most likely shut off any chemical reactions inside halos. We therefore do not further consider the possibility of dark chemistry.  
 
\subsection{Ellipticity of DM Halos}
Collisions among the different constituents of the dark sector lead to a redistribution of the linear and angular momentum inside a DM halo. This tends to erase velocity correlations on halo scales and brings the DM halo closer to isothermality, effectively altering the halo structure and making it roughly isotropic. Since observations indicate that halo iso-density contours are elliptical and thus significantly deviate from isotropy, we conclude that the DM thermalization time is much longer than the dynamical timescale of halos. This in turns tightly constrains the rate of particle collisions in which a significant amount of momentum is exchanged.  

In the atomic DM scenario, we need to consider collisions between the dark atoms, the dark electrons, and the dark protons. In total, this amounts to six different types of collisions: {\bf H-H}, {\bf H-e}, {\bf H-p}, {\bf e-e}, {\bf e-p}, and {\bf p-p}. Here, {\bf H} stands for the dark atoms. Since we are most interested in dark sector that are mostly neutral when non-linear structures form, the contribution from the first three types of collision is expected to dominate. To compute the rate of collisions with large momentum exchange, we need the momentum-transfer cross section which is defined by
\be\label{mt-definition}
\sigma_{\rm mt}\equiv\int d\Omega \frac{d\sigma}{d\Omega}(1-\cos{\theta}),
\ee
where $d\sigma/d\Omega$ is the elastic differential cross section. A detailed computation of the {\bf H-H}, {\bf H-e} and {\bf H-p} momentum-transfer cross section is beyond the scope of this paper, but we can use the extensive literature (see e.g.~\cite{0953-4075-32-14-317}) on the corresponding visible-sector cross sections to derive some general properties. We refer the reader to appendix \ref{atomic_xs} for further details. To a good approximation, the {\bf H-H}, {\bf H-p}, and {\bf H-e} momentum-transfer cross sections are
\ba
\sigma_{\rm mt}^{\bf  H-H}(v)&\approx&\frac{30\pi\alpha_D^2}{B_D^2v^{1/4}}\left[\frac{\mu_{\rm H}}{\mu_D}\frac{m_D}{B_D}\right]^{-\frac{1}{8}}e^{-\frac{\mu_{\rm H}}{\mu_D}\frac{m_D}{B_D}\frac{v^2}{300}},
\ea
\ba
\sigma_{\rm mt}^{\bf  H-p}(v)&\approx&\frac{60\pi\alpha_D^2}{B_D^2}\sqrt{\frac{m_{\bf p}}{m_D}}e^{-\frac{\mu_{\rm H}}{\mu_D}\frac{\mu_{D{\bf p}}}{B_D}\frac{v^2}{200}},\\
\sigma_{\rm mt}^{\bf  H-e}(v)&\approx&\frac{60\pi\alpha_D^2}{B_D^2}\sqrt{\frac{m_{\bf e}}{m_{D}}} e^{-\frac{\mu_{\rm H}}{\mu_D}\frac{\mu_{D{\bf e}}}{B_D}\frac{v^2}{200}},
\ea
where $\mu_{D{\bf e}}=m_Dm_{\bf e}/(m_D+m_{\bf e})\simeq m_{\bf e}$ and $v$ is the relative velocity of the colliding dark ions or atoms. The ion-ion differential cross sections are simply given by the Rutherford scattering cross section. Performing the integral in Eq.~\ref{mt-definition} leads to 
\ba
\sigma_{\rm mt}^{\bf p-p}(v)&=&\frac{8\pi\alpha_D^2}{m_{\bf p}^2v^4}\ln{[\csc^{2}{(\theta_{\rm min}^{\bf p-p}/2)}]},\\
\sigma_{\rm mt}^{\bf e-e}(v)&=&\frac{8\pi\alpha_D^2}{m_{\bf e}^2v^4}\ln{[\csc^{2}{(\theta_{\rm min}^{\bf e-e}/2)}]},\\
\sigma_{\rm mt}^{\bf e-p}(v)&=&\frac{2\pi\alpha_D^2}{\mu_D^2v^4}\ln{[\csc^{2}{(\theta_{\rm min}^{\bf e-p}/2)}]},
\ea
where $\theta_{\rm min}^{\bf i-j}$ is the minimum scattering angle. We take this angle to be given by the Debye screening length $\lambda_{\rm De}$ of the dark plasma through the relation
\be
\csc^2{(\theta_{\rm min}^{\bf i-j}/2)}=\left(\frac{\lambda_{\rm De}\mu_{\bf ij}v^2}{\alpha_D}\right)^2+1,
\ee
where
 \be
 \lambda_{\rm De}\simeq \sqrt{\frac{\mu_Dv^2}{8\pi\alpha_Dn_{\bf e}}},
 \ee
and where $n_{\bf e}$ is the number density of free dark electrons inside a DM halo. The momentum-loss rate of dark species {\bf i} upon collisions with species {\bf j} in a DM halo is given by
\be
\dd{p}_{\bf ij}=n_{\bf j}\int dv \,v f(v) \,\sigma_{\rm mt}^{\bf i-j}(v) \Delta p_{\bf i},
\ee
where $n_{\bf j}=n_{\bf j}(r)$ is the number density of specie {\bf j} inside a halo, $v$ is the relative velocity of the collision, $\Delta p_{\bf i}$ is the momentum loss in a single collision and $f(v)$ is the velocity distribution which we take to be locally Maxwellian with velocity dispersion $v_0$
\be\label{maxwell_dist}
f(v)dv=\frac{4}{\sqrt{\pi}v_0^3}v^2e^{-v^2/v_0^2}dv.
\ee
The momentum lost by particle {\bf i} upon colliding with particle {\bf j} is 
\be
\Delta p_{\bf i}=p_{\bf i}\frac{m_{\bf j}}{m_{\bf i}+m_{\bf j}}.
\ee
where $p_{\bf i}$ is the momentum of particle {\bf i}. Normally, the momentum loss is weighted by a factor $(1-\cos{\theta})$, but we have absorbed this factor into the momentum-transfer cross section. The rate of momentum-changing collisions between dark species {\bf i} and {\bf j} is defined as
\be
\Gamma_{\bf ij}\equiv  \frac{\dd{p}_{\bf ij}}{\bar{p}_{\bf i}}.
\ee
where $\bar{p}_{\bf i}$ is the momentum of particle {\bf i} averaged over the velocity distribution given in Eq.~(\ref{maxwell_dist}). With the help of the above momentum-transfer cross sections, these collisions rates are
\ba
\Gamma_{\bf HH}&\simeq&\frac{15\pi\,\Gamma(\frac{19}{8})\alpha_D^2v_0^{3/4}n_{\bf H}}{B_D^2}\left[\frac{\mu_{\rm H}}{\mu_D}\frac{m_D}{B_D}\right]^{-\frac{1}{8}}\en
&&\qquad\qquad\times\left[1+\frac{\mu_{\rm H}}{\mu_D}\frac{m_D}{B_D}\frac{v_0^2}{300}\right]^{-\frac{19}{8}},\\
\Gamma_{\bf ep}&\simeq&\frac{2\pi^{\frac{3}{2}} \alpha_D^2n_{\bf p}}{\mu_D^2v_0^3}\frac{m_{\bf p}}{m_{\bf p}+m_{\bf e}}\ln{\left[1+\frac{\mu_D^3v_0^6}{4\pi^{\frac{3}{2}} \alpha_D^3n_{\bf e}}\right]},\\
\Gamma_{\bf pp}&\simeq&\frac{4\pi^{\frac{3}{2}} \alpha_D^2n_{\bf p}}{m_{\bf p}^2v_0^3}\ln{\left[1+\frac{\mu_Dm_{\bf p}^2v_0^6}{32\pi\alpha_D^3n_{\bf e}}\right]},\\
\Gamma_{\bf ee}&\simeq&\frac{4\pi^{\frac{3}{2}} \alpha_D^2n_{\bf e}}{m_{\bf e}^2v_0^3}\ln{\left[1+\frac{\mu_Dm_{\bf e}^2v_0^6}{32\pi\alpha_D^3n_{\bf e}}\right]},\\
\Gamma_{\bf eH}&\simeq&\frac{45\pi^{\frac{3}{2}} \alpha_D^2v_0n_{\bf H}}{B_D^2}\sqrt{\frac{m_{\bf e}}{m_D}}\frac{m_D}{m_D+m_{\bf e}}\en
&&\qquad\qquad\times\left[1+\frac{\mu_{\rm H}}{\mu_D}\frac{\mu_{D{\bf e}}}{B_D}\frac{v_0^2}{200}\right]^{-\frac{5}{2}},\\
\Gamma_{\bf pH}&\simeq&\frac{45\pi^{\frac{3}{2}} \alpha_D^2v_0n_{\bf H}}{B_D^2}\sqrt{\frac{m_{\bf p}}{m_D}}\frac{m_D}{m_D+m_{\bf p}}\en
&&\qquad\qquad\times\left[1+\frac{\mu_{\rm H}}{\mu_D}\frac{\mu_{D{\bf p}}}{B_D}\frac{v_0^2}{200}\right]^{-\frac{5}{2}}.
\ea
In the above, $\Gamma(x)$ is the Gamma function. The rates for the opposite processes are obtained by rescaling the above rates with the appropriate mass densities 
\ba
\Gamma_{\bf pe}&=&\frac{n_{\bf e}m_{\bf e}}{n_{\bf p}m_{\bf p}}\Gamma_{\bf ep},\\
\Gamma_{\bf He}&=&\frac{n_{\bf e}m_{\bf e}}{n_{\bf H}m_D}\Gamma_{\bf eH},\\
\Gamma_{\bf Hp}&=&\frac{n_{\bf p}m_{\bf p}}{n_{\bf H}m_D}\Gamma_{\bf pH}.
\ea
In the case for which $m_{\bf e} \ll m_{\bf p},\,m_D$, we see that $\Gamma_{\bf pe} \ll \Gamma_{\bf ep}$ and $\Gamma_{\bf He} \ll \Gamma_{\bf He} $ if the abundance of dark atoms and dark ions is roughly similar. This is reasonable: a heavy dark proton needs to scatter off many light dark electrons before its momentum is significantly affected. Conversely, a dark electron's momentum can be dramatically changed by a single collision with a dark proton or a dark atom. However, since the dark electrons are usually much lighter than the dark protons and therefore carry a small fraction of the overall halo mass, these high-momentum-transfer collisions do not affect the ellipticity of DM halos. Therefore, we do not consider any further the  scattering of dark electrons (that is, we neglect {\bf ee}, {\bf ep} and {\bf eH} collisions). In the case where the dark proton and the dark electron are nearly degenerate in mass, constraints on dark proton collisions naturally engulf the dark electron constraints so we can still neglect them. 

Detailed simulations of self-interacting DM halos have shown that DM particles forming the bulk of the matter density can undergo up to 10 hard scatters in a Hubble time before the ellipticity of halos is adversely affected \cite{2012arXiv1208.3026P,2012arXiv1208.3025R}. Therefore, the ellipticity of DM halos is preserved if we have
\be
\Gamma_{\text{coll}} < 10 H_0,
\ee
where $\Gamma_{\text{coll}}$ is the overall hard-scatter rate of the dark sector. The total collision rate inside a halo is the sum of the individual rates weighted by the relative abundance of each dark species
\be
\Gamma_{\text{coll}}\simeq \bar{x}_D\Gamma_{\bf p}+(1-\bar{x}_D)\Gamma_{\bf H},
\ee
where
\be
\Gamma_{\bf p}= \Gamma_{\bf pp} + \Gamma_{\bf pH} +\Gamma_{\bf pe},
\ee
and
\be
\Gamma_{\bf H}= \Gamma_{\bf HH} + \Gamma_{\bf Hp} +\Gamma_{\bf He}.
\ee
To compute the above constraint, we need to specify the density profile of both dark ions and dark atoms as well as the velocity dispersion $v_0(r)$. The main difficulty here is that the actual density profile of each dark constituents is itself determined by their collisional rates. For example, if the dark ions undergo many hard scatters among themselves within the typical dynamical time of a galaxy, they settle into their own isothermal density profile \cite{Kaplan:2011yj}, while the neutral dark atoms maintain their CDM-like profile. This implies that the local ionized fraction inside a DM halo can vastly differs from the background values computed in section \ref{dark_recombination} (by several order of magnitudes in some cases). To complicate matters further, collisions between dark atoms in the central region of halos likely lead to the formation of a cored density profile. Therefore, it is clear that detailed $N$-body simulations would be required to determine the actual density profile of each dark constituent.  

We can nevertheless derive conservative constraints on atomic DM by making some simple assumptions. We first assume halos to be locally neutral such that  $n_{\bf e}(r)=n_{\bf p}(r)$. We further assume the ionized component of the DM follows the total density profile, that is, $n_{\bf e}(r)=n_{\bf p}(r)=\bar{x}_Dn_D(r)$. Similarly, $n_{\bf H}(r)=(1-\bar{x}_D)n_{D}(r)$. We take the halo ionized fraction ($\bar{x}_D$) to be equal to the late-time ionized fraction of the background, that  is $\bar{x}_D=x_D(z=0)$. It is important to emphasize that the above prescription is not accurate since the ionized and neutral components generally have different density profiles \cite{Kaplan:2011yj} and the ionized fraction inside halos may significantly differ from the overall background cosmological values. As such, the bounds derived from these assumptions should therefore be considered as upper limits on how limiting the ellipticity constraint can be. The actual bounds on the parameter space of atomic DM are potentially \emph{much less} constraining. Nevertheless, the above treatment is expected to be fairly accurate for $\bar{x}_D\ll1$ and $\bar{x}_D\sim1$.

While it is true that the most stringent constraints on halo ellipticity (see e.g.~\cite{Feng:2009mn}) come from the inner part of galactic halos, it is unclear how these bounds apply to atomic DM. Indeed, ellipticity constraints strongly depend on the local DM density in the inner central region of halos. The latter is usually obtained by fitting a predefined halo profile (e.g. NFW) to the data \cite{Humphrey:2006rv}. Since atomic DM generally admits a different halo shape, we cannot blindly apply these results to the atomic DM scenario. We therefore resort to typical values of the DM density and velocity dispersion inside galactic halos. Explicitly, we evaluate the above constraints using $\rho_D\simeq3, 1$, and $0.3$ GeV/cm$^{-3}$. We take the velocity dispersion to be $v_0\simeq250$ km/s. The number density of DM particles is then given by $n_D=\rho_D/m_D$. 

\begin{figure*}[h!]
\subfigure{\includegraphics[width=0.4\textwidth]{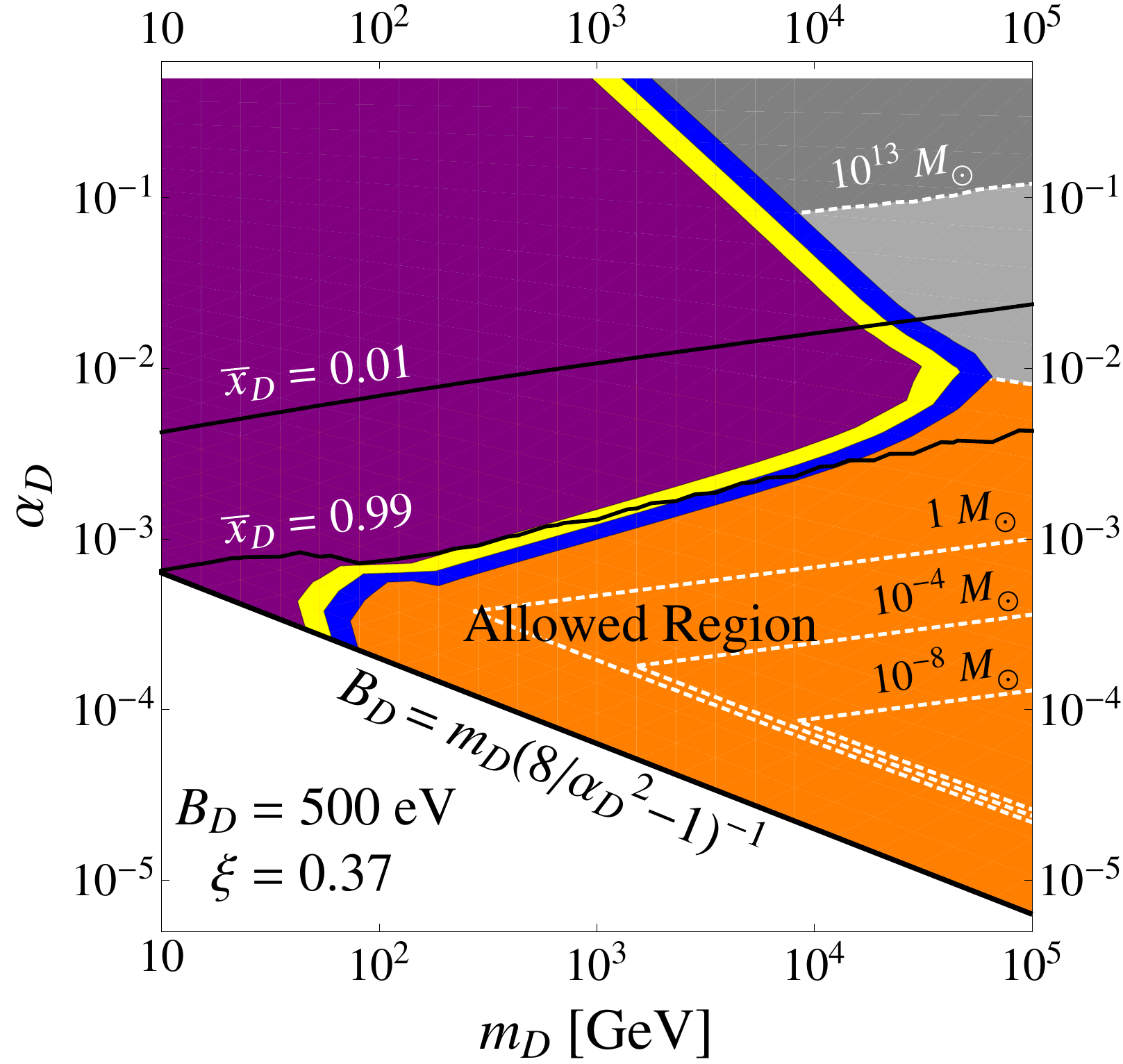}}
\subfigure{\includegraphics[width=0.4\textwidth]{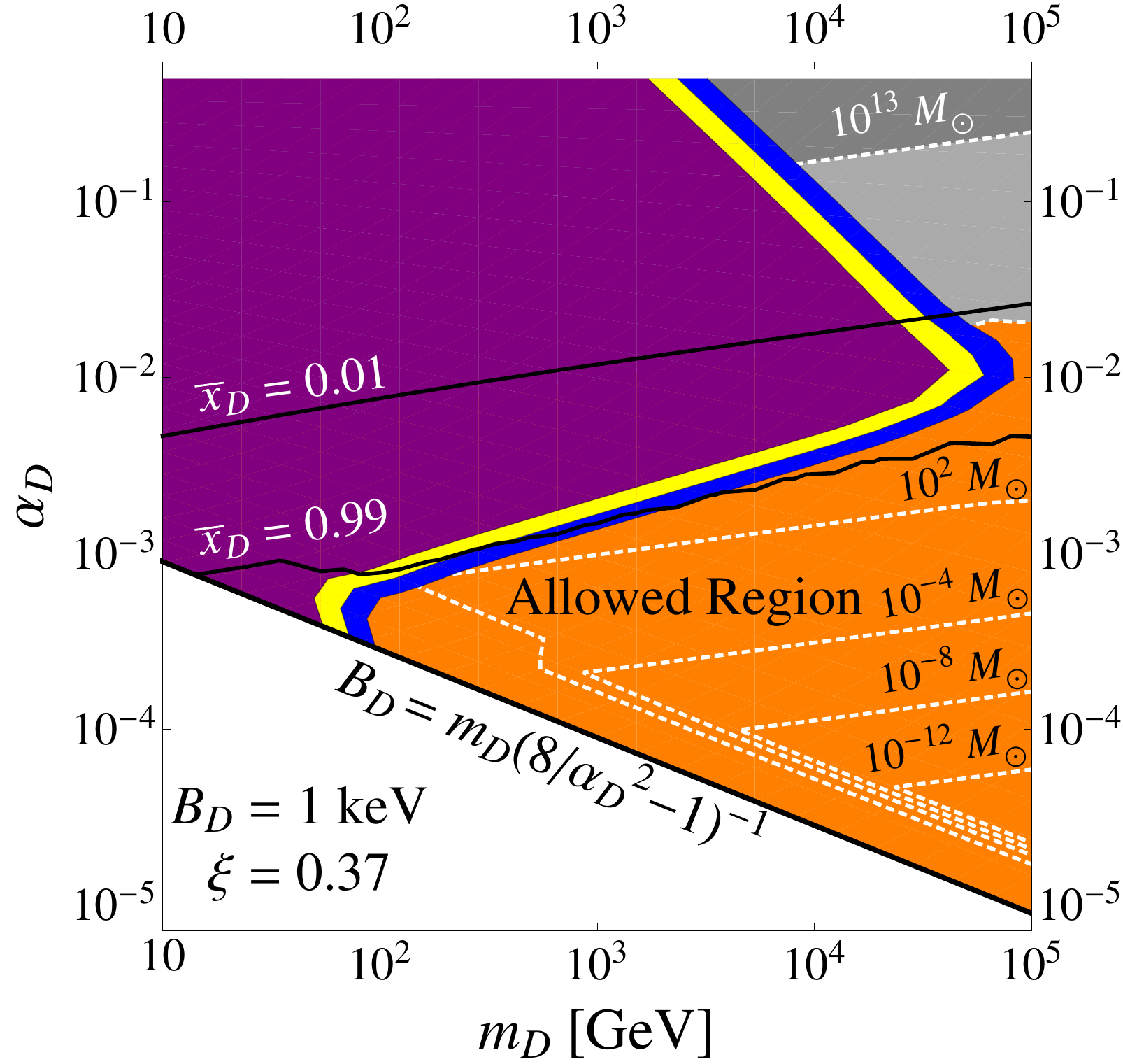}}\\
\subfigure{\includegraphics[width=0.4\textwidth]{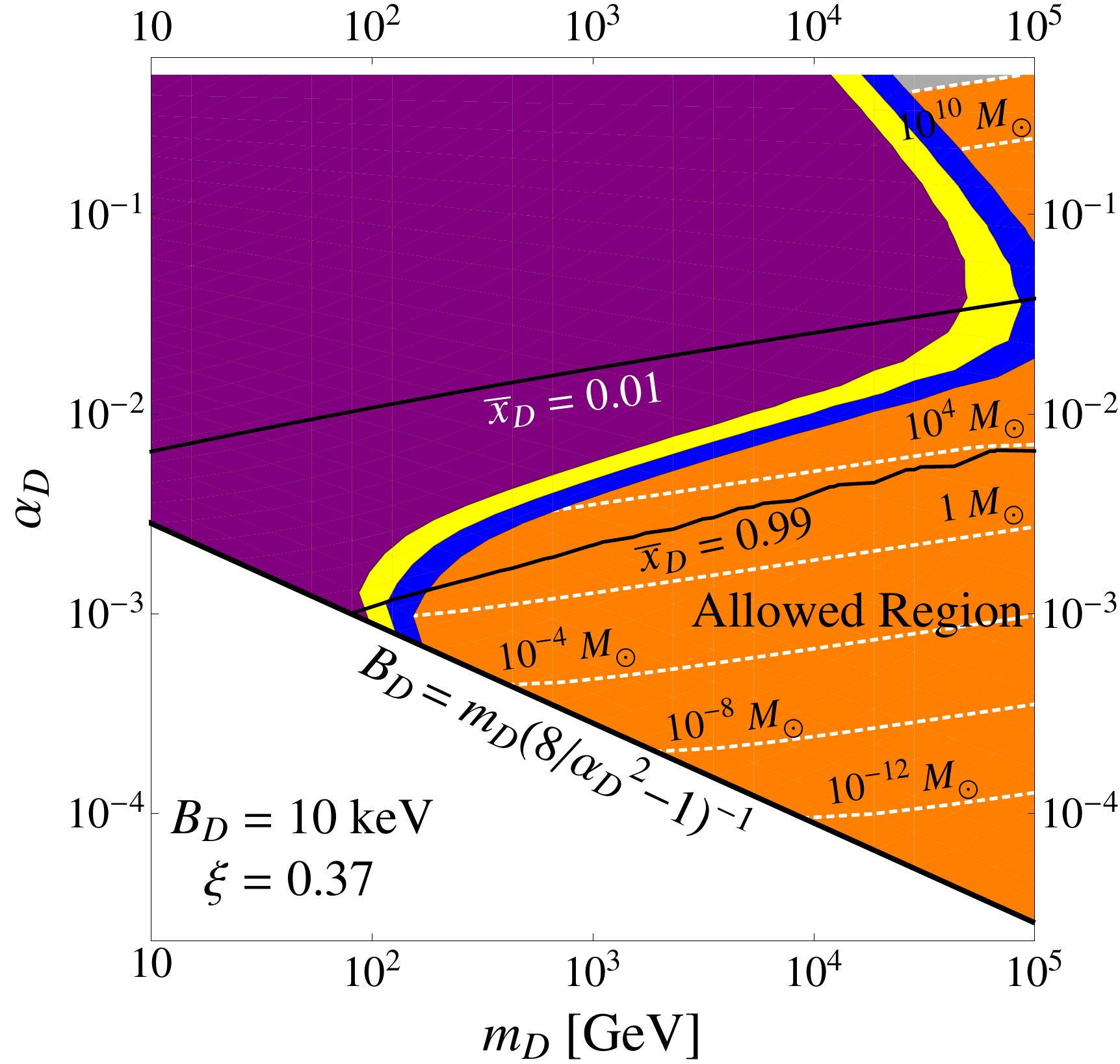}}
\subfigure{\includegraphics[width=0.4\textwidth]{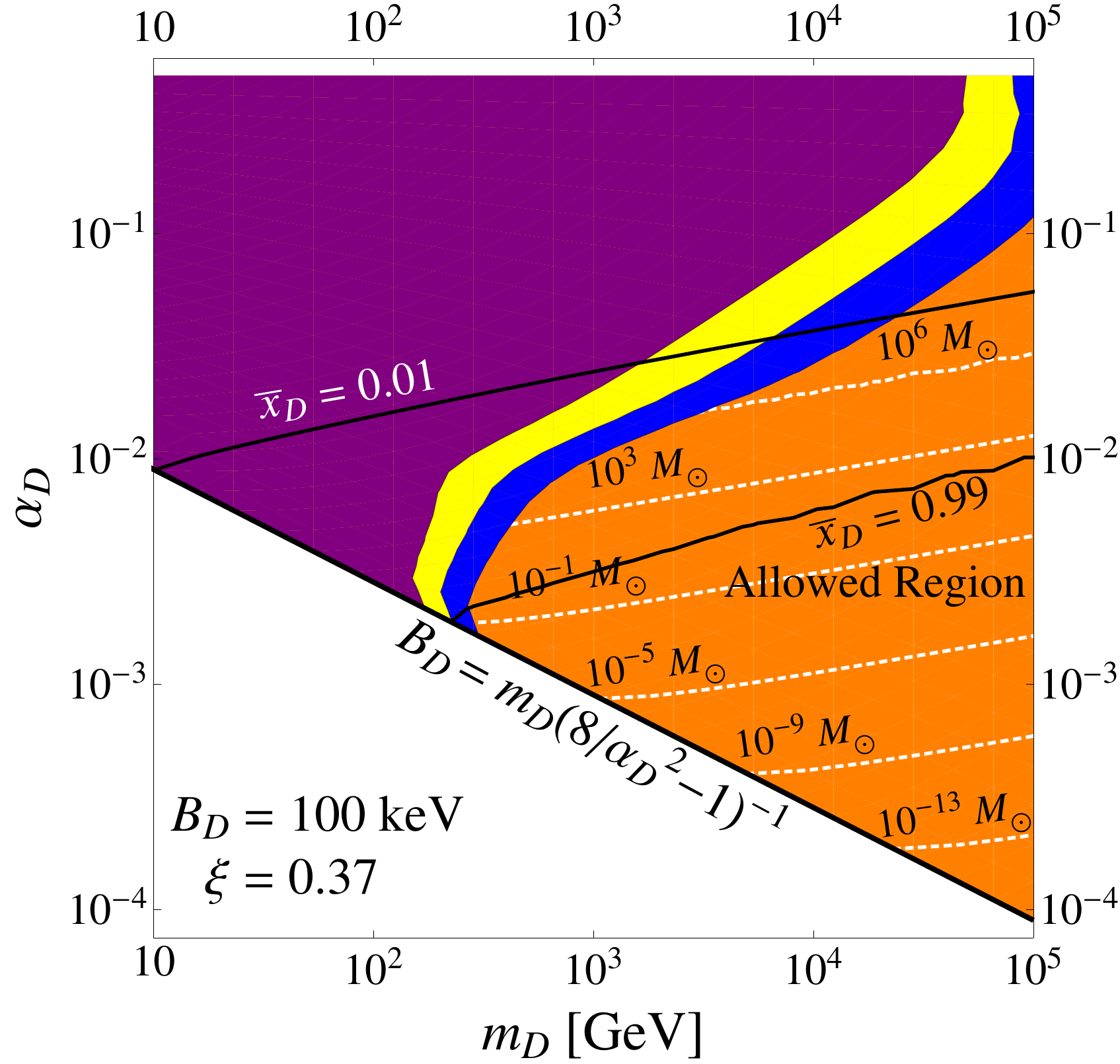}}\\
\subfigure{\includegraphics[width=0.4\textwidth]{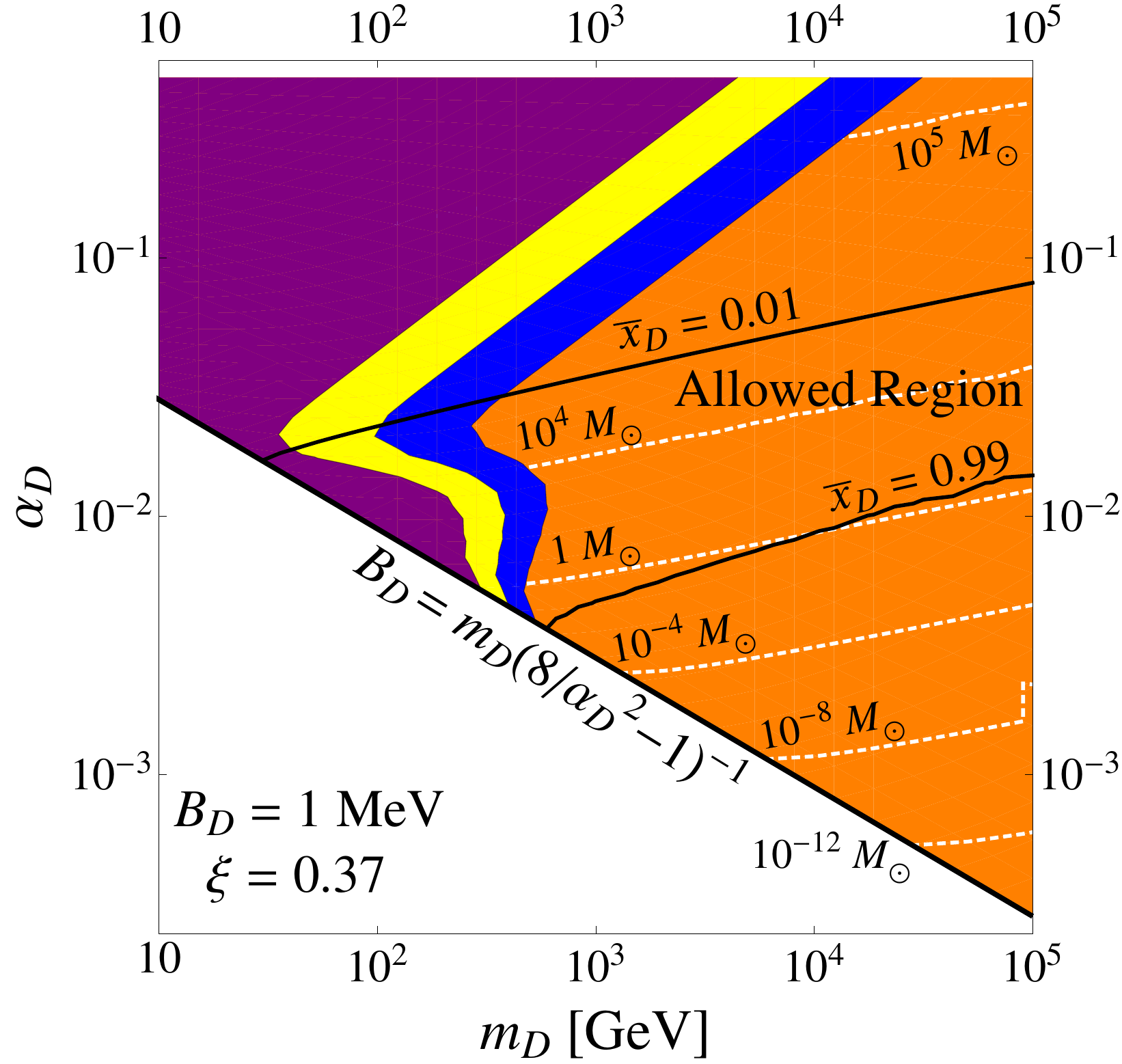}}
\subfigure{\includegraphics[width=0.4\textwidth]{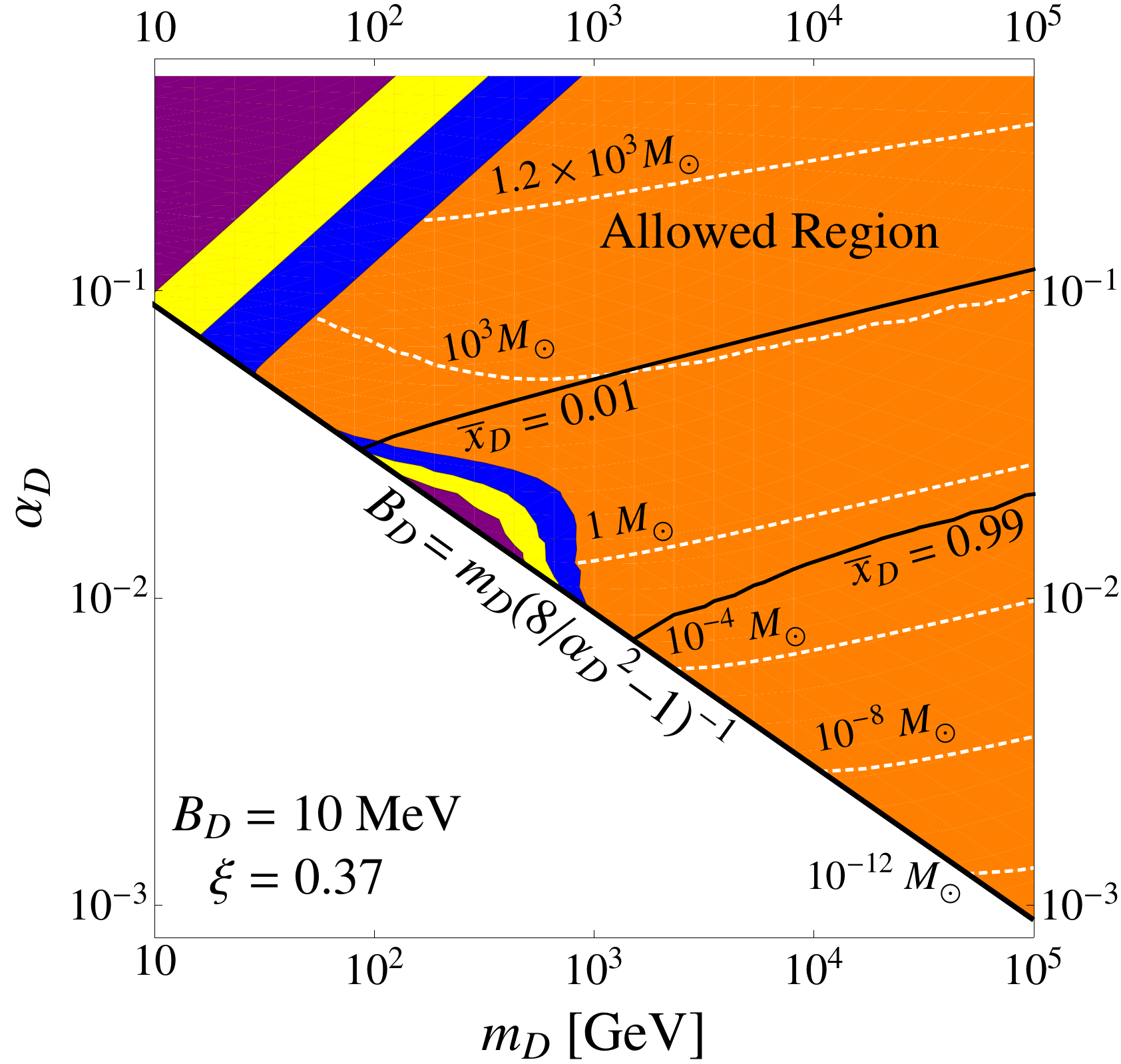}}
\caption{Halo ellipticity and Ly-$\alpha$ forest constraints on the parameter space of atomic DM. We display the constraint for two values of the atomic binding energy. The allowed region (orange) is clearly indicated on the plots. The blue, yellow and purple contours (outermost to innermost) display the disfavored regions when $\rho_D\simeq3, 1$, and $0.3$ GeV/cm$^{-3}$, respectively. The light and dark gray regions correspond to minimal halo masses $M_{\rm min}> 9.3\times10^{10}$ M$_{\odot}$ and $M_{\rm min}> 10\times10^{13}$ M$_{\odot}$, respectively. The white dashed lines indicate contours of constant minimal halo mass. We also show the contours (solid black) of constant background ionization fraction for $\bar{x}_D=1\%$ and  $\bar{x}_D=99\%$.}
\label{halo_constraint}
\end{figure*}

We display in Fig.~\ref{halo_constraint} the constraints on $\alpha_D$ and $m_D$ for six different values of the atomic binding energy. We display the three disfavored contours corresponding $\rho_D\simeq3, 1$, and $0.3$ GeV/cm$^{-3}$. We observe that for $10\,\text{keV}\lesssim B_D\lesssim 100$ keV, there is little parameter space for which the dark sector is mostly neutral \emph{and} is not in tension with the ellipticity constraint. As the binding energy is increased above this threshold, a large allowed parameter space opens up since the atomic geometric cross section rapidly decreases as the binding energy climbs up in value. Below $10$ keV, another unconstrained region opens up at large coupling constant values and masses. This is caused by the collision energy approaching the excitation threshold of the dark atoms. As this limit is approached, the momentum-transfer cross section for atom-atom scattering become more and more suppressed, while the inelastic channels start to growth in importance. The inelastic cross sections are however much smaller than the elastic ones ($\sim \pi a_{0,D}^2$ instead of $\sim 10^2  \pi a_{0,D}^2$) and we thus do not expect these collisions to severely affect the ellipticity of haloes. 

Overall, the ellipticity constraint unambiguously disfavors dark atoms with mass $m_D\lesssim 1$ TeV, unless $B_D\gtrsim 1$ MeV. We however reiterate that the constraints shown in Fig.~\ref{halo_constraint} are \emph{conservative}, especially in the regions where $ 1\%\lesssim\bar{x}_D\lesssim 99\%$ for which detailed simulations of dark-atomic halos will likely be required to assess the validity of the bounds. Indeed, if the typical hard-scatter timescale of dark ions is shorter than the dynamical time of a galaxy, we can safely assume that the dark ions settle into a separate isothermal density profile \cite{Kaplan:2011yj}. In the presence of a sizable population of neutral dark atoms, the overall DM halo can however still display significant ellipticity if the dark atoms are themselves mostly collisionless. Since the isothermal ionic halo is typically much more diffuse \cite{Kaplan:2011yj} than the neutral one, ion-ion and ion-atom interactions are probably much more suppressed than our naive estimate suggests. This is likely to open up much of the parameter space for $ 1\%\lesssim\bar{x}_D\lesssim 99\%$.

In the case where the dark ions form the majority of the matter density inside halos, the ellipticity constraints derived in the context of hidden charged DM apply \cite{Ackerman:2008gi,Feng:2009mn}. In this scenario, dark ions cannot settle into an isothermal profile without utterly violating the ellipticity of halos. Viable models can nevertheless be constructed by suppressing the dark fine-structure constant and by considering very large ions masses (such that the number density is low). Fortunately, as $\alpha_D$ is increased and $m_D$ is decreased, more and more ions are bound into dark atoms hence softening the ion-scattering bound. In fact, since the early recombination rate scales as $\sim\alpha_D^3$ while the ion-ion hard-scatter rates scale as $\sim\alpha_D^2$, the ionized fraction drops faster than the ion-ion momentum-transfer rate is increasing as the the dark fine-structure constant is dialed up. Thus, the atomic DM scenario naturally provides a way to evade the halo ellipticity constraints on hidden-charged DM.

In summary, while detailed simulations of DM halo formation in the atomic DM scenario are likely required to determine the exact constraints, it is clear that dark atoms lighter than $1$ TeV are likely to lead to collisional DM that is in tension with the observed internal structure of halos, unless $B_D\gtrsim1$ MeV. 
\subsection{Cooling of DM Halos}
Collisional excitation of dark atoms followed by the emission of a dark photon provides a natural cooling mechanism for DM. Since DM is observed to have very different properties than baryons (which are allowed to cool), collisional cooling of dark atoms must be suppressed. The easiest channel to excite a dark atom is through the hyperfine transition. The hyperfine splitting is given by
\be
E_{\rm hf}\approx\frac{4}{3}g_{\bf e}g_{\bf p}\alpha_D^2\frac{m_{\bf e}}{m_{\bf p}}B_D,
\ee
where $g_{\bf e}\simeq2$ and $g_{\bf p}\simeq2$ are the gyromagnetic ratio of the dark electron and dark proton, respectively. This transition can be excited when a dark atom in a spin singlet state collides with a dark ion or atom and undergoes a spin-flip to the triplet state. The cross section for this process is somewhat smaller than the elastic scattering cross section \cite{0953-4075-32-14-317}. Therefore, it is most likely that all regions of parameters space where hyperfine emission leads to significant cooling are already ruled by the ellipticity constraint. To verify this, we compute the typical timescale for a dark atom to lose an $\mathcal{O}(1)$ fraction of its kinetic energy due to hyperfine emission. Assuming that this process is approximately governed by the geometric cross section, this timescale is
\be
\tau_{\rm hf}\simeq\frac{9m_DB_Dm_{\bf p}v_0}{256\sqrt{\pi}\alpha_D^4m_{\bf e}n_{D}}.
\ee
Demanding that this timescale be longer than the age of the Universe leads to no new constraints beyond those already plotted in Fig.~\ref{halo_constraint}.

Beyond the hyperfine transition, the other dissipative process that is relevant for dark atoms is the $1s\rightarrow2s,2p$ collisional excitation. Heuristically, such inelastic collisions can only happen when the timescale of the collision is shorter than the typical timescale of the dark atoms (i.e. adiabaticity is violated). The typical velocity of a dark electron inside an atom is $v_{\bf e}\sim\alpha_D$, leading to an atomic timescale $\tilde{\tau}_D\sim  a_{0,D}/v_{\bf e}$. On the other hand, the collision timescale is of the order $\tau_{\rm coll}\sim a_{0,D}/v$, where $v$ is the relative velocity of the collision. Taking $\tau_{\rm coll} < \tilde{\tau}_{D}$, inelastic collisions are only possible when
\be
v\gtrsim\alpha_D. 
\ee
According to Fig.~\ref{halo_constraint}, all allowed regions of parameter space displaying a large fraction of neutral dark atoms have $\alpha_D\gtrsim10^{-2}$. Taking $v$ to be approximately equal to the typical velocity dispersion inside a DM halo, we have $v\sim10^{-3}$ for galactic halos while $v\sim10^{-2}$ for galaxy clusters. It is thus clear that inelastic collisions can play no major role inside galactic halos while they could play a marginal role inside clusters. In the latter case however, the small number density of dark atoms coupled with the typically small inelastic cross sections (see e.g. \cite{PhysRev.183.241}) likely render inelastic collisions completely negligible inside clusters. We therefore conclude that inelastic processes do not constrain the atomic DM scenario beyond the regions of parameter space already ruled out by the ellipticity and the matter power spectrum bounds.
\section{Direct Detection}\label{direct-detection}
Most previous works on atomic DM \cite{Kaplan:2009de,Kaplan:2011yj,Cline:2012is,Cline:2012ei} have focused their attention on the potential direct-detection signatures. Since atomic DM can naturally scatter inelastically, it offers a mechanism to reconcile the annual modulation seen by DAMA \cite{Bernabei:2008yi} and CoGeNT \cite{Aalseth:2010vx} with the null signal of the CDMS \cite{Ahmed:2008eu,Ahmed:2010wy} and Xenon10 \cite{Angle:2007uj,Angle:2009xb,Angle:2011th} experiments. The actual direct-detection results strongly depend on how atomic DM couples to the SM. In this section, we compare the atomic DM models which provide a good fit to the direct-detection data to our cosmological and astrophysical constraints derived above.

 In Ref.~\cite{Kaplan:2009de}, the authors considered a model in which dark fermions are axially coupled to a broken $U(1)$ which mixes with the SM hypercharge. They found that dark atoms having an hyperfine splitting $E_{\rm hf}\simeq100$ keV, $70 \lesssim m_{\bf p}\lesssim 200$ GeV  and $2.2 \gtrsim m_{\bf e}\gtrsim 1.6$ GeV can provide a good fit to the modulated spectrum of DAMA while evading the constraints from other direct-detection experiments. This corresponds to an atomic DM scenario with $27\lesssim B_D\lesssim45$ MeV, $0.16\lesssim \alpha_D\lesssim0.24$ and $72\lesssim m_D\lesssim 201$ GeV. Extrapolating the constraints shown in Fig.~\ref{halo_constraint} to the appropriate binding energies, we find that these values lie on the edge of the ellipticity bound and are therefore marginally allowed by our constraints.

In Ref.~\cite{Kaplan:2011yj}, the authors considered a similar broken axial $U(1)$ model which mixes with the SM hypercharge. In this work, they show that an atomic DM scenario with $E_{\rm hf}\sim5-15$ keV with dark fermion masses $m_{\bf e}\sim m_{\bf p} \sim5 $ GeV could reconcile the CoGeNT data with the count rate seen by the CRESST experiment \cite{Angloher:2011uu}. This corresponds to an atomic DM scenario with $\alpha_D\sim0.03$, $B_D\sim 1$ MeV and $m_D\sim10$ GeV. This model is in serious tension with the halo constraint shown in Fig.~\ref{halo_constraint} and is therefore likely to lead to collisional DM which would dramatically affect the structure of halos. 

In Refs.~\cite{Cline:2012is,Cline:2012ei}, the authors considered atomic DM models where the dark photon mixes with the regular photon. This effectively gives small standard electromagnetic charges to the dark fermions. The authors concluded that scenarios with $\alpha_D\sim0.04-0.06$, $B_D\sim 2-5$ MeV and $m_D\sim6-10$ GeV predict the right cross section to explain the DM events recorded by the CoGeNT collaboration. These values are however clearly in tension with the halo ellipticity bound. Nevertheless, since they lie close to the boundary with the allowed region and since our constraints are conservative, detailed $N$-body simulations will be required to assess whether this model is ruled out or not.
 
We observe that dark-atom models capable of explaining the current direct-detection experiments are at best marginally allowed by our conservative astrophysical constraints. This stems from the fact that direct-detection experiments generally favor $m_D\sim1-10$ GeV while the halo constraint typically imposes $m_D>1$ TeV, unless $B_D\gtrsim1$ MeV. From the point of view of direct detection, the main challenge for building a successful atomic DM theory is to avoid strong elastic scattering off nuclei while allowing for enough inelastic collisions to explain both DAMA and CoGeNT. The study of atomic DM models that could be in agreement with both the direct-detection data and the astrophysical constraints are left for future work.

\section{Discussion}\label{discussion}
We have presented in this work a thorough study of a dark sector made up of atom-like bound states. This model naturally incorporates extra relativistic degrees of freedom whose presence are currently favored by CMB experiments. The observed primordial abundances of light elements constrain the temperature of DR to be somewhat cooler than the CMB, with an upper bound given approximately by $T_D/T_{\rm CMB}<0.8$ at $z=0$. We have revisited the atomic physics necessary to describe the processes of dark recombination, thermal decoupling, and kinetic decoupling. We find that in some cases, the inclusion of physics beyond that of standard atomic hydrogen is required to properly describe these transitions. In particular, we find that bound-free processes such as photo-ionization heating and photo-recombination cooling are key to determine the thermal decoupling epoch of weakly-coupled atomic models. For strongly-coupled dark atoms, we have shown that the addition of Rayleigh scattering can significantly delay the kinetic and thermal decoupling of DM. 

We have solved the linear cosmological perturbation equations, taking into account the interaction between DM and DR and showed that atomic DM can go through various regimes as time evolves and as the dark parameters are varied. In particular, the DR pressure leads to strong acoustic oscillations for Fourier modes entering the Hubble horizon prior to the dark drag epoch. Further, diffusion and acoustic damping severely suppress the amplitude of DM fluctuations on scales shorter than the sound horizon at kinetic decoupling. At late times, these features remain imprinted on the matter power spectrum. Importantly, the atomic DM scenario introduces the new DAO length scale in the density field which basically determines the minimal DM halo mass. We have shown that observations of Ly-$\alpha$ forest flux power spectrum put an upper bound on the size of the DAO scale and rule out a large fraction of atomic DM models with $B_D\lesssim 1$ keV. 

We have performed a detailed study of the impact of the atomic DM scenario on the CMB. We have determined that the largest impact on the CMB in this model is due to the presence of dark photons. If dark photons begin free-streaming at a very early epoch, their impact on the CMB is likely indistinguishable from that of  extra relativistic neutrinos. On the other hand, if dark photons decouple from DM when Fourier modes relevant for the CMB are inside the horizon, then the atomic DM scenario predicts CMB signatures that are difficult to reproduce with only relativistic neutrinos. These signatures include non-uniform phase shifts and amplitude suppressions of the temperature anisotropy power spectrum. The clearest CMB signature of the atomic DM scenario is a non-uniform phase shift of the polarization power spectrum that asymptotes to a constant at both large and small scales. 

These signatures can only be present if the dark photons form a sizable fraction of the radiation energy density ($\xi\gtrsim0.6$) and if they decouple at a late enough redshift, which usually requires $B_D\lesssim10$ keV. These relatively low atomic binding energies are however strongly constrained by both the Ly-$\alpha$ data and the halo ellipticity requirement. For the remaining allowed parameter space, dark photons generically decouple too early to have an impact on the CMB that is noticeably different from that of standard relativistic neutrinos. It is therefore unlikely that the CMB can be used to learn about the simplest atomic DM model discussed here. Nevertheless, in a more general model where dark radiation only couples to a certain fraction of the DM, it is possible that the former could significantly affect the CMB without modifying the small-scale matter power spectrum. This is an intriguing possibility that we leave for future work.

The strongest constraint on atomic DM comes from requiring that DM is effectively collision-less inside galactic halos. This stems from the fact that atoms naturally have much larger geometric cross sections than point particles. Since this geometric cross section scales as $B_D^{-2}$, the halo constraint favors a large binding energy, with models having $B_D\gtrsim 10$ MeV largely unconstrained. Unsurprisingly, this corresponds to the regime where atomic DM closely resembles a standard WIMP particle. For $B_D\lesssim1$ keV, the ellipticity and Ly-$\alpha$ constraints favor a dark sector that is mostly ionized at all times. Our constraints in these regions of parameter space are similar to those found in Refs.~\cite{Ackerman:2008gi,Feng:2009mn}. Importantly,  we have shown that inelastic processes are inefficient at dissipating the energy of dark atoms in all the regions that are not already ruled out by the halo and the small-scale power spectrum bounds. Overall, it is clear that preserving the observed ellipticity of halos severely limits the possible interactions between DM particles. 

Since the atom-atom, ion-atom and ion-ion cross sections are all velocity dependent, atomic DM can provide enough interactions to smooth out the central region of small satellites galaxies while retaining the ellipticity of large galactic and cluster halos. Furthermore, atomic DM scenarios predicting a minimal halo mass in the range $M_{\rm min}\sim10^8-10^{10}$ M$_{\odot}$ have the potential to affect the faint-end of the galaxy luminosity function and bring it in line with observations. In addition, the presence of an ionized component naturally makes the halo more diffuse and could potentially alleviate the so-called ``too big to fail'' problem \cite{BoylanKolchin:2011de,BoylanKolchin:2011dk}. Interestingly, an atomic DM model with $B_D\sim5$ keV, $m_D\sim80$ TeV and $\alpha_D\sim0.02$ could possibly address all three problems affecting dwarf galaxies. Indeed, this model has a late-time global ionized fraction of $\bar{x}_D\simeq0.2$, a minimal halo mass of $M_{\rm min}\simeq 8\times10^8$ M$_{\odot}$ and lies at the boundary of the collisional constraint, meaning it could contain enough interaction to form cores in galactic halos while retaining their overall ellipticity.  A detailed numerical study will be necessary to assess the success of this model. 

One issue that we have not touched upon in this work is the possibility of dark magnetic fields. These have the potential to significantly alter structure formation unless there is a mechanism that naturally suppresses their amplitude and range. The simplest way to achieve such suppression is to break the $U(1)_D$ gauge force by introducing a small dark-photon mass. This is an intriguing possibility since, although a minute mass is needed to quench large-scale magnetic fields, a mass whose Yukawa length is comparable to the size of a dark atom could considerably alleviate the halo constraint by shrinking the size of the atomic cross section and by limiting the range of ion-ion interactions. One would however have to revisit the atomic physics described in section \ref{thermal} to make quantitative predictions about the thermal history of such a dark sector. We leave the study of such a model to future work.

Direct-detection data favor light atomic DM candidates which are in tension with the halo constraint. Nevertheless, the direct-detection signatures strongly depend on how dark atoms couple to the SM. Whether a successful model which agrees with both the halo and direct-detection constraint could be constructed remains an open question. While an explicit model would have to be specified in order to make quantitative predictions, we emphasize that dark atoms are still a viable DM candidate from the point of view of both astrophysics and direct detection.

\acknowledgments
We sincerely thank Antony Lewis for collaboration at early stages of this project, and Yacine Ali-Ha{\"i}moud and Jens Chluba for letting us adapt their codes. We thank Douglas Scott and Yacine Ali-Ha{\"i}moud for comments on an early version of this draft. We thank Daniel Grin, Jim Cline, Jesus Zavala, Pat Scott, Latham Boyle and Guy Moore for useful discussions. The work of FYCR is supported by the National Science and Engineering Research Council of Canada and by the UBC Four-Year Fellowship program. KS would like to thank Perimeter Institute for Theoretical Physics for their hospitality. The research of KS is supported by NSERC and CFI/BCKDF.  Infrastructure funded by the CFI/BCKDF Leaders Opportunity Fund was essential to completing this work.

\appendix
\section{Relations between Dark Parameters}\label{m_ep}
In this appendix, we relate the dark electron mass $m_{\bf e}$, the dark proton mass $m_{\bf p}$, and the atomic reduced mass $\mu_D$ to the dark sector parameters.
\be
\mu_D=\frac{2B_D}{\alpha_D^2}\qquad m_{\bf e} = \frac{\mu_D m_{\bf p}}{m_{\bf p}-\mu_D}
\ee
\be
m_{\bf p} =  \frac{m_D+B_D+\sqrt{(m_D+B_D)^2-4(m_D+B_D)\mu_D}}{2}
\ee
%
\section{Effective Number of Relativistic Degrees of Freedom}\label{gstar}
The effective number of degrees of freedom contributing to the entropy density of the DS at a temperature $T_D$ is \cite{1990eaun.book.....K}
\ba\label{exact_g_*D}
g_{*S,D}(T_D) &=& 2 +\frac{45}{\pi^4}\Bigg[\int_{y_{\bf e}}^\infty\frac{(u^2-y_{\bf e}^2)^\frac{1}{2}}{e^u+1}du\\
&&\qquad\qquad+\frac{1}{3}\int_{y_{\bf e}}^\infty\frac{(u^2-y_{\bf e}^2)^\frac{3}{2}}{e^u+1}du\Bigg],\nonumber
\ea
where $y_{\bf e}\equiv\frac{m_{\bf e}}{T_D}$ and the first term corresponds to the dark-photon contribution while the term in square brackets denote the dark-electron contribution. Performing the integrals, we obtain, to a very good approximation,
\be
 g_{*S,D}(T_D)\simeq2+\frac{7}{2}(1+y_{\bf e}^{1.394})^{0.247}e^{-0.277y_{\bf e}^{1.384}}.
 \ee
 %
\section{Thermal Rates}\label{thermal_rates_app}
In this appendix, we compute the rates governing the energy exchange between the DM and the DR.  
\subsection{Photo-Ionization Heating Rate}
\begin{figure}[h]
\begin{centering}
\includegraphics[width=0.45\textwidth]{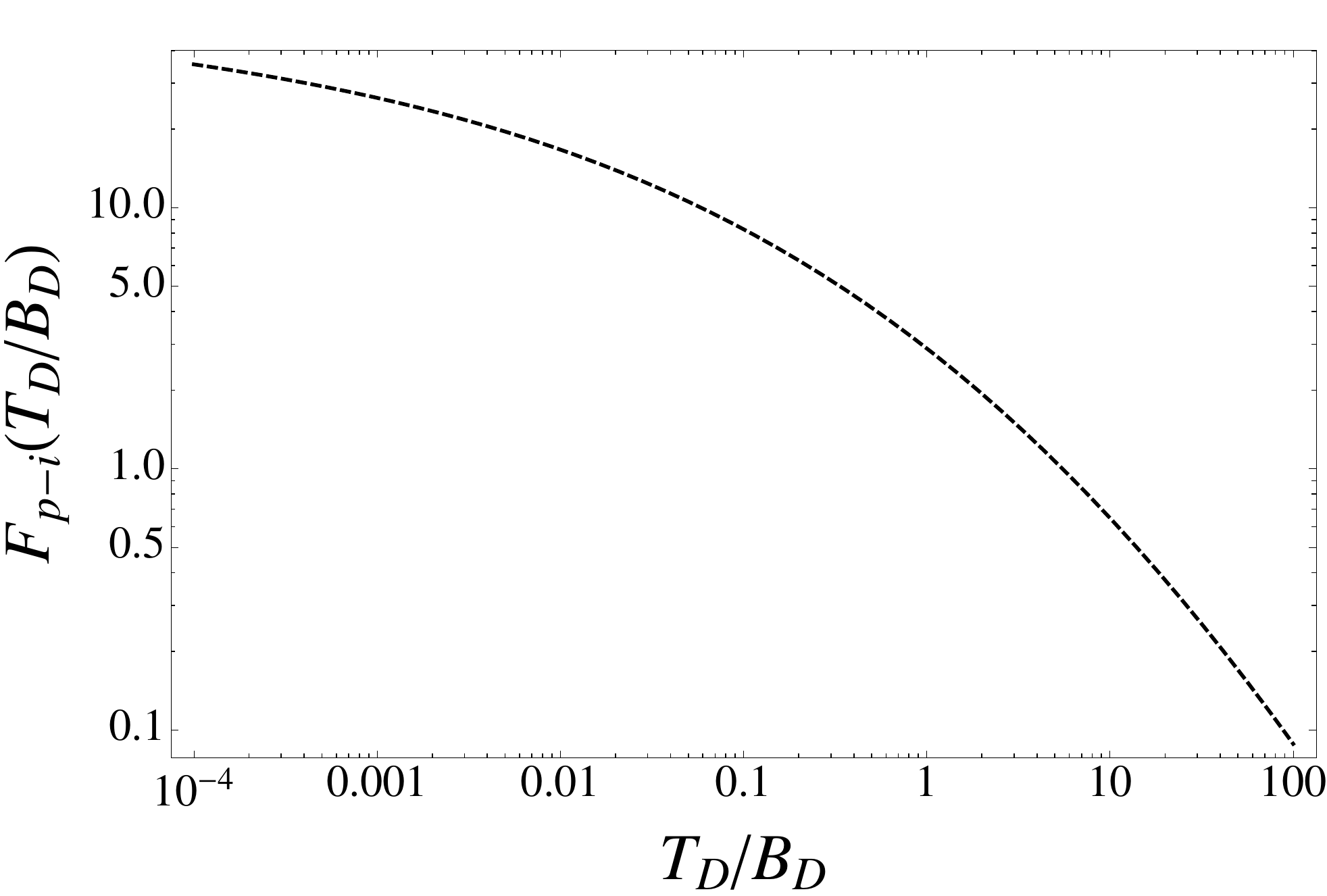}
\caption[Universal dimensionless fitting functions for the photo-ionization heating rate.]{Universal dimensionless fitting functions $F_{\rm p-i}$ for the photo-ionization heating rate, plotted as a function of $T_D/B_D$.}
\label{F_p-i}
\end{centering}
\end{figure}

The photo-ionization heating rate is given by \cite{Seager:1999km}:
\be\label{raw_heating_rate}
\Pi_{\rm p-i}=\sum_{n,l}n_{nl}B_D\int_0^{\infty}\frac{\kappa^2\gamma_{nl}(\kappa)}{e^{B_D(\kappa^2+n^{-2})/T_D}-1}d(\kappa^2),
\ee
where the sum runs over all atomic states specified by the quantum numbers $n\geq2$ and $l\leq n-1$, $n_{nl}$ represents the number density of dark atoms in state $nl$ and $\gamma_{nl}(\kappa)$ is defined in Eq.~(\ref{gamma_nl}). We do not include bound-free transitions to and from the ground state since recombination directly to the $1s$ state results in the emission of an energetic photon that is immediately reabsorbed by a nearby neutral atom, hence resulting in no net cooling or heating. We assume that the occupations numbers of excited states are in Boltzmann equilibrium with the $2s$ state
\be\label{n_nl}
n_{nl}=n_{2s}\frac{g_{nl}}{g_{2s}}e^{-B_D(1/4-n^{-2})/T_D},
\ee
where $g_{nl}$ is the degeneracy of the energy level with quantum numbers $nl$. This approximation is valid as long as $\mathcal{B}_D(T_D) \gtrsim H$. Performing the momentum integration in Eq.~(\ref{raw_heating_rate}) and computing the sum over atomic states up to $n_{\rm max} = 250$ yield a photo-ionization heating rate of the form
\be
\Pi_{\rm p-i}(T_D)=\frac{\alpha_D^3T_D^2}{3\pi}x_{2s}n_{D}e^{-\frac{B_D}{4T_D}}F_{\rm p-i}(T_D/B_D),
\ee
$F_{\rm p-i}(y)$ is a dimensionless universal function encoding the remaining of the temperature dependence of the photo-ionization heating rate. It is shown in Fig.~\ref{F_p-i}. This function is well-fitted by
\be
F_{\rm p-i}(y)=\frac{3973.6\,y^{-0.0222}}{(2.012+y^{0.2412})^{6.55}},
\ee
for $4\times10^{-4}\lesssim y \lesssim10^{2}$. 
%

\subsection{Photo-Recombination Cooling Rate}
\begin{figure}[h]
\begin{centering}
\includegraphics[width=0.5\textwidth]{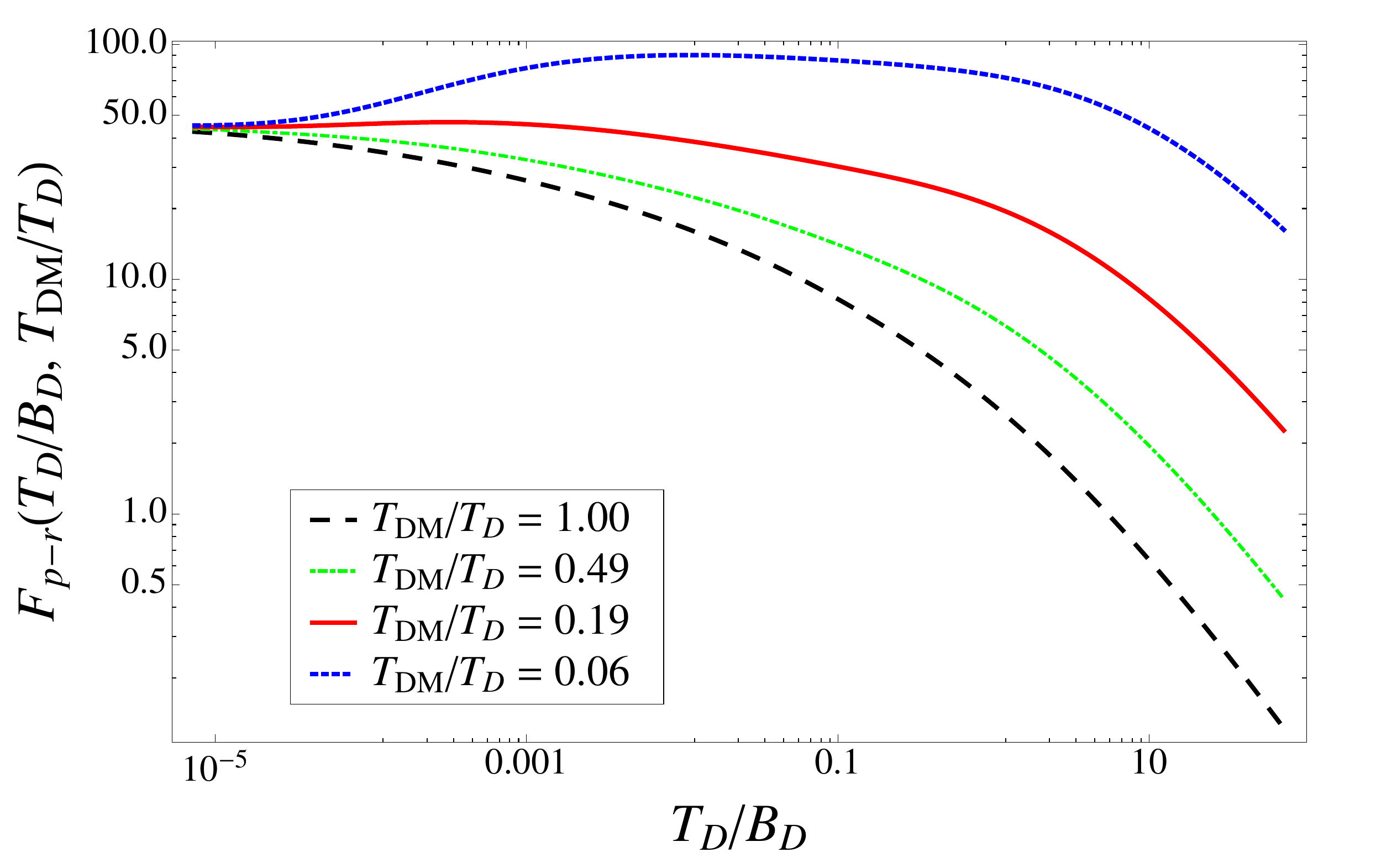}
\caption[Universal dimensionless fitting function for the photo-recombination cooling rate.]{Universal dimensionless fitting function $F_{\rm p-r}$ for the photo-recombination cooling rate, plotted as a function of $T_D/B_D$ for 4 different constant values of $T_{DM}/T_D$.}
\label{H_pr}
\end{centering}
\end{figure}

The rate of photo-recombination cooling is given by \cite{Seager:1999km}
\ba
\Pi_{\rm p-r}&=&\sum_{n,l}x_D^2n_{D}^2\frac{(2\pi)^{3/2}B_D}{(\mu_DT_{DM})^{3/2}}\\
&&\times\int_0^{\infty}e^{-B_D\kappa^2/T_{DM}}\kappa^2\gamma_{nl}(\kappa)\en
&&\times\left[1+f_{\rm BB}(B_D(\kappa^2+n^{-2}),T_D)\right]d(\kappa^2).\nonumber
\ea
As before, we do not consider recombination directly to the ground state. Computing the momentum integral and summing over all atomic states up to $n_{\rm max}=250$ yield a photo-recombination cooling rate of the form
\be
\Pi_{\rm p-r}=\frac{2\alpha_D^3\sqrt{2\pi T_{DM}}}{3\mu_D^{3/2}}x_D^2n_{D}^2F_{\rm p-r}(\frac{T_D}{B_D},\frac{T_{DM}}{T_D}),
\ee
where $F_{\rm p-r}$ is a dimensionless universal function. We illustrate its behavior in Fig.~\ref{H_pr}. 
%
\subsection{Free-Free Cooling and Heating}
The free-free emission (absorption) process refers to the emission (absorption) of a dark photon by a dark electron due to its acceleration in the Coulomb field created by a dark proton. For atomic DM, we are mostly interested in the emission and absorption of dark radiation by thermal dark electron obeying a Maxwell-Boltzmann velocity distribution. The rate for thermal Bremsstrahlung cooling is \cite{1975ctf..book.....L,1979rpa..book.....R}
\be
\Pi_{\rm ff-c}=\frac{16\alpha_D^3\bar{g}_{\rm ff}\sqrt{2\pi T_{DM}}x_D^2n_{D}^2}{(3\mu_D)^{3/2}},
\ee
where $\bar{g}_{\rm ff}$ is the free-free velocity-averaged and frequency-averaged Gaunt factor, which is of order unity. The inverse process of thermal free-free absorption is characterized by the following absorption coefficient \cite{1979rpa..book.....R}
\be
a_{\nu}^{\rm ff}(T_{DM})=\frac{2\alpha_D^3n_{D}^2x_D^2(1-e^{-2\pi\nu/T_{DM}})g_{\nu}^{\rm ff}}{(3\mu_D)^{3/2}\sqrt{2\pi T_{DM}}\nu^3},
\ee
where $g_{\nu}^{\rm ff}$ is the velocity-averaged free-free Gaunt factor and $\nu$ is the frequency of the incoming dark photons. The total amount of energy absorbed through free-free interactions per unit volume per unit time is then
\ba
\Pi_{\rm ff-h}&=&4\pi\int_0^{\infty}  a_{\nu}^{\rm ff}(T_{DM})B_{\nu}(T_D) d\nu\en
&\simeq&\frac{2^5\pi\sqrt{2\pi}\alpha_D^3\bar{g}_{\rm ff}x_D^2n_{D}^2}{(3\mu_D)^{3/2}\sqrt{T_{DM}}}\\
&&\times\int_0^{\infty} e^{-2\pi\nu/T_{DM}}\frac{e^{2\pi\nu/T_{DM}}-1}{e^{2\pi\nu/T_{D}}-1}d\nu,\nonumber
\ea
where $B_{\nu}(T_D)$ is the Planck function. To evaluate this integral, we first note that the integrand only depends on $T_{DM}$ itself and on the fractional temperature difference between the DM and DR,  $\epsilon\equiv(T_D-T_{DM})/T_D$ 
\be
e^{-2\pi\nu/T_{DM}}\frac{e^{2\pi\nu/T_{DM}}-1}{e^{2\pi\nu/T_{D}}-1}=\frac{e^{-2\pi\nu/T_{DM}}-e^{-4\pi\nu/T_{DM}}}{e^{-2\pi\nu\epsilon/T_{DM}}-e^{-2\pi\nu/T_{DM}}}.
\ee
In the limit of quasi thermal equilibrium, $\epsilon\ll1$, we can expand the integrand as a power series in $\epsilon$ and compute the frequency integral order-by-order in the fractional temperature difference. This leads to
\begin{align}
\int_0^{\infty}e^{-2\pi\nu/T_{DM}}\frac{e^{2\pi\nu/T_{DM}}-1}{e^{2\pi\nu/T_{D}}-1}d\nu&\approx\\
&\hspace{-2.5cm}\frac{T_{DM}}{12\pi}(6+\pi^2\epsilon(1+2\epsilon)-6\epsilon^2\zeta(3)+\mathcal{O}(\epsilon^3)),\nonumber
\end{align}
where $\zeta(x)$ is the Riemann zeta function. The net rate at which DM gains energy due to free-free interactions, $\Pi_{\rm ff}\equiv\Pi_{\rm ff-h}-\Pi_{\rm ff-c}$, is the given by
\ba
\Pi_{\rm ff}&\simeq&\frac{16\alpha_D^3\bar{g}_{\rm ff}\sqrt{2\pi T_{DM}}x_D^2n_{D}^2}{(3\mu_D)^{3/2}}\en
&&\qquad\times\left(\frac{\pi^2\epsilon(1+2\epsilon)-6\zeta(3)\epsilon^2}{6}\right).
\ea
For numerical computation, we take $\bar{g}_{\rm ff}\simeq1.3$. 
%
\subsection{Rayleigh Heating}
After the onset of recombination, dark photons can transfer energy to the DM sector via Rayleigh scattering off neutral dark atoms. The scattering cross-section for this process is given by \cite{2005MNRAS.358.1472L}
\be\label{sigma_R}
\sigma_{\rm R}(\nu) \approx \sigma_{{\rm T},D}\frac{81}{64}\left(\frac{\nu}{\nu_{\rm Ly\alpha}}\right)^4,
\ee
where $ \sigma_{{\rm T},D}\equiv 8\pi\alpha_D^2/(3 m_{\bf e}^2)$ is the dark Thomson cross-section and $\nu_{\rm Ly\alpha}$ is the frequency of dark Lyman-$\alpha$ photons. This expression is valid for $\nu \ll \nu_{\rm Ly\alpha}$, which is realized for the large majority of dark photons after the onset of dark recombination. The net rate of energy transfer between dark photons and DM due to Rayleigh scattering is
\be\label{R1}
\Pi_{\rm R} = 2 n_{D}(1-x_D)\int\sigma_{\rm R}(\nu)B_{\nu}(T_D)\Delta E_{\rm R}\frac{d\nu}{\nu},
\ee
where $B_{\nu}(T_D)$ is the Planck function and $\Delta E_{\rm R}$ is the average net energy gained by a dark atoms in a Rayleigh scattering event. It is given by
\be\label{Delta_E_R}
\Delta E_{\rm R} = \frac{2\pi\nu}{m_D}\left(2\pi\nu - w T_{DM}\right),
\ee
where $w=75600\zeta(9)/\pi^8$ is a constant that can be determined using detailed balance. Substituting Eq.~(\ref{Delta_E_R}) into Eq.~(\ref{R1}) and performing the integral over frequency yields the effective heating rate
\be
\Pi_{\rm R} \simeq \frac{430080\zeta(9)\alpha_D^2n_{D}(1-x_D)T_D^8(T_D-T_{DM})}{\pi^2B_D^4m_{D}m_{\bf e}^2},
\ee
where $\zeta(y)$ stands for the Riemann Zeta function. This expression is valid for $T_D\ll B_D$ which is always realized when a large population of neutral dark atoms is present.

\section{Atomic Cross Sections}\label{atomic_xs}
In this appendix, we justify the form of the cross sections used to describe collisions of DM particles.

\subsection{{\bf H-H} Cross Section}
We consider the elastic scattering of two identical hydrogen-like atoms. This problem requires one to solve for the joint wavefunction of the two atoms interacting via the singlet gerade and triplet ungerade molecular potentials \cite{PhysRevA.46.6956,0953-4075-32-14-317}. The elastic differential cross section for indistinguishable hydrogen-like atom is given by
\be
\frac{d\sigma^{\bf H-H}_{\rm el}}{d\Omega}=\frac{1}{4}\frac{d\sigma^{\bf H-H}_{\rm s}}{d\Omega}+\frac{3}{4}\frac{d\sigma^{\bf H-H}_{\rm t}}{d\Omega},
\ee
where the subscripts ``s'' and ``t'' refer to the singlet and triplet states, respectively. The singlet and triplet contributions take the form \cite{0953-4075-32-14-317}
\ba\label{sigma_HH_st}
\frac{d\sigma^{\bf H-H}_{\rm s,t}}{d\Omega}&=&\frac{1}{4}|f_{\rm s,t}(\theta)\pm f_{\rm s,t}(\pi-\theta)|^2\en
&&+\,\,\frac{3}{4}|f_{\rm s,t}(\theta)\mp f_{\rm s,t}(\pi-\theta)|^2,
\ea
where the uppermost sign is for the singlet and the lower sign for the triplet. Here, $f_{\rm s,t}(\theta)$ is the scattering amplitude for the centre-of-mass (CM) scattering angle $\theta$ and is given by
\be\label{f_HH}
f_{\rm s,t}(\theta)=\frac{1}{2ik}\sum_{l=0}^{\infty}(2l+1)(e^{2i\de_l^{\rm s,t}}-1)P_l(\cos{\theta}),
\ee
where $k$ is the wavenumber in the CM frame, $l$ is the orbital angular momentum, $P_l(\cos{\theta})$ is the $l^{\rm th}$ Legendre polynomial, and $\de_l^{\rm s,t}$ is the scattering phase shift from either the singlet or the triplet molecular potential. From Eqs.~(\ref{sigma_HH_st}) and (\ref{f_HH}), one can obtain the important result 
\ba
\sigma_{\rm mt}^{\bf  H-H}&\equiv&\int d\Omega \frac{d\sigma^{\bf H-H}_{\rm el}}{d\Omega}(1-\cos{\theta})\en
&=&\sigma_{\rm el}^{\bf  H-H},
\ea
where $\sigma_{\rm mt}^{\bf  H-H}$ and $\sigma_{\rm el}^{\bf  H-H}$ are the momentum-transfer and the total elastic cross section, respectively. This result can easily be proven using the orthogonality of the Legendre polynomial and follows from the symmetry of scattering two identical atoms. 

All the information about the atomic physics is encoded in the phase shifts $\de_l^{\rm s,t}$. These can be found by solving the Schr\"odinger equation
\be
\left[\frac{d^2}{dR^2}-\frac{l(l+1)}{R^2}+2\mu[E-V^{\rm s,t}(R)]\right]Ru_l^{\rm s,t}(R)=0,
\ee
such that the asymptotic behavior at large $R$ is
\be
u_l^{\rm s,t}(R)\simeq R^{-1}\sin{(kR-l\pi/2 + \de_l^{\rm s,t})}. 
\ee
Here, $\mu$ is the reduced mass of the colliding atoms, $V^{\rm s,t}(R)$ is the interatomic potential in the singlet or triplet state, $E$ is the relative energy of the collision and $k=\sqrt{2\mu E}$. In the semi-classical limit, the phase shift can be written as \cite{Dalgarno17041958,0953-4075-32-14-317}
\ba\label{phase_shift}
\de_l^{\rm s,t}&\approx&\int_{R_0}^{\infty}\left(k^2-2\mu V^{\rm s,t}(R)-\frac{l(l+1)}{R^2}\right)^{\frac{1}{2}}dR\en
&&\qquad-\int_{R_0'}^{\infty}\left(k^2-\frac{l(l+1)}{R^2}\right)^{\frac{1}{2}}dR,
\ea
where $R_0$ and $R_0'$ are the outermost zeros of the respective integrands. 

Now focusing on the atomic DM scenario, we would like to understand how the phase shifts are affected when the dark parameters $\alpha_D$, $B_D$ and $m_D$ are varied. These parameters enter the phase shifts through their contributions to the wave number $k$ and to the molecular potentials $V^{s,t}(R)$. While it is out of the scope of this appendix to compute the molecular potentials of dark atoms, we can nonetheless extract general properties of these potentials by studying the case of standard molecular hydrogen. To gain insights into the scaling of the molecular potential with the dark parameters, let us briefly consider the Lennard-Jones potential \cite{1924RSPSA.106..463J}
\be
V(R)=\epsilon\left[\left(\frac{R_{\rm m}}{R}\right)^{12}- 2\left(\frac{R_{\rm m}}{R}\right)^6\right],
\ee
where the first term accounts for the short-range repulsive force due to Pauli blocking while the second term describes the long-range behavior of the interaction. Here, $\epsilon$ is a constant setting the depth of the potential well and $R_{\rm m}$ is the distance where the potential admits a minimum. 

\begin{figure}[t!]
\includegraphics[width=0.5\textwidth]{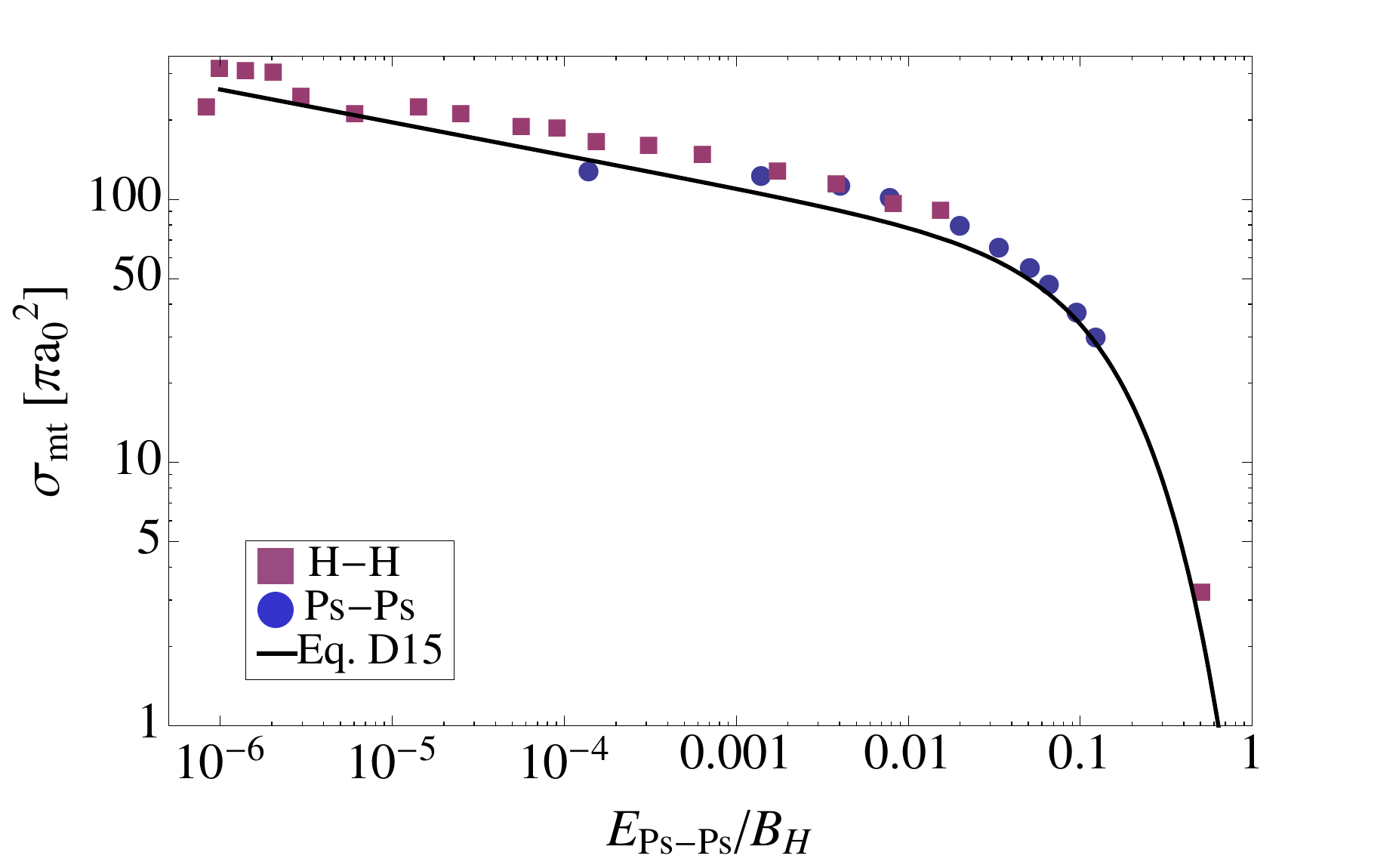}
\caption{Momentum-transfer cross section for hydrogen \cite{Krstic&Schultz}  (purple squares) and positronium (Ps) \cite{PhysRevA.65.022704}  (blue disks) self-scattering. We have rescaled the energy dependence of the H-H cross section according to Eq.~(\ref{energy_scaling}) to compare it to the positronium cross section. We also display the momentum-transfer cross section we use for dark-atom scattering (Eq.~(\ref{sig_HH})).  }
\label{HH}
\end{figure}

On purely dimensional ground, the typical energy scale of the molecular potential should be proportional to the atomic binding energy, that is, $\epsilon\propto B_D$. Also, the typical atomic separation $R_{\rm m}$ should be of the order of the dark atomic Bohr radius, $R_{\rm m}\propto a_{0,D}=2\alpha_D/B_D$. We can thus write
\be\label{approx_potential}
V_D(R/a_{0,D})\sim\frac{B_D}{B_{\rm H}}V_{\rm H}(R/a_0),
\ee
where the ``H'' subscript stands for the standard baryonic hydrogen. Defining the dimensionless distance $r\equiv R/a_{0,D}$, we can write Eq.~(\ref{phase_shift}) as
\ba\label{phase_shift2}
\de_{l,D}&\approx&\int_{r_0}^{\infty}\left(\frac{\alpha_D^2m_D}{B_D^2}[E- V_D(r)]-\frac{l(l+1)}{r^2}\right)^{\frac{1}{2}}dr\en
&&-\int_{r_0'}^{\infty}\left(\frac{\alpha_D^2m_D}{B_D^2}E-\frac{l(l+1)}{r^2}\right)^{\frac{1}{2}}dr,
\ea
where we have substituted the two-atom reduced mass $\mu=m_D/2$. We therefore observe that the phase shifts only depend on two dimensionless combinations of the dark parameters
\ba\label{phase_shiftD}
\de_{l,D}&=&\de_{l,D}\left( \frac{\alpha_D^2m_DE}{B_D^2},  \frac{\alpha_D^2m_DV_D}{B_D^2}\right)\en
&\simeq&\de_{l,D}\left( \frac{\alpha_D^2m_DE}{B_D^2},  \frac{\alpha_D^2m_DV_{\rm H}}{B_DB_{\rm H}}\right),
\ea
where we used Eq.~(\ref{approx_potential}) in the last line. Since $V_{\rm H}/B_{\rm H}$ is known, once could then use Eq.~(\ref{phase_shift2}) to obtain the phase shifts for dark atoms. However, Eq.~(\ref{phase_shiftD}) suggests that one could obtain approximate $\de_{l,D}$ directly from  the baryonic hydrogen phase shifts by rescaling the collision energy as
\be\label{energy_scaling}
E\rightarrow\left(\frac{\alpha_D}{\alpha_{\rm em}}\right)^2\frac{m_D}{m_{\rm H}}\left(\frac{B_{\rm H}}{B_D}\right)^2E.
\ee
Of course, this is only an approximation since the molecular potential would also have to be rescaled according to Eq.~(\ref{phase_shiftD}). Moreover, the molecular potential itself depends non-trivially on $\alpha_D$ and $\mu_D$ and would have to be modified accordingly. Nevertheless, studies of deuterium, tritium  \cite{Krstic&Schultz} and positronium  \cite{PhysRevA.65.022704} self-scattering support the above approximate rescaling. We therefore take our atom-atom momentum-transfer cross section to be
\ba\label{x-section_subst}
\sigma_{\rm mt}^{\bf  H-H}[E] &\approx& \left(\frac{a_{0,D}}{a_0}\right)^2\\
&&\times\,\sigma_{\rm mt}^{ H-H}\left[\left(\frac{\alpha_D}{\alpha_{\rm em}}\right)^2\frac{m_D}{m_{\rm H}}\left(\frac{B_{\rm H}}{B_D}\right)^2E\right],\nonumber 
\ea
where the leading factor accounts for the change in the geometric cross section. A numerical fit to the calculations found in Refs.~\cite{0953-4075-32-14-317,Krstic&Schultz} leads to
\be
\sigma_{\rm mt}^{ H-H}(E_{\rm CM})\approx10^2\,\pi a_0^2\left(\frac{E_{\rm CM}}{B_{\rm H}}\right)^{-\frac{1}{8}}e^{-\frac{E_{\rm CM}}{75 B_{\rm H}}},
\ee
which is valid for $10^{-3}\lesssim E_{\rm CM}/B_{\rm H}\lesssim10$. At high energy, inelastic collisions begin to dominate over the elastic channel and this explain the much steeper energy dependence in that regime. Using Eq.~(\ref{x-section_subst}), we can finally obtain an approximate momentum-transfer cross section for the neutral dark atoms
\be\label{sig_HH}
\sigma_{\rm mt}^{\bf  H-H}(E)\approx\frac{25\pi\alpha_D^2}{B_D^2}\left[\frac{\mu_{\rm H}}{\mu_D}\frac{m_D}{m_{\rm H}}\frac{E}{B_D}\right]^{-\frac{1}{8}}e^{-\frac{\mu_{\rm H}}{\mu_D}\frac{m_D}{m_{\rm H}}\frac{E}{75 B_D}},
\ee
where we have use the definition of binding energy to simplify the energy rescaling. Here, $\mu_{\rm H}$ is the reduced mass of the regular proton-electron system. In terms of the relative velocity between the two dark atoms, the cross section reads
\be\label{sig_HHv}
\sigma_{\rm mt}^{\bf  H-H}(v)\approx\frac{30\pi\alpha_D^2}{B_D^2v^{1/4}}\left[\frac{\mu_{\rm H}}{\mu_D}\frac{m_D}{B_D}\right]^{-\frac{1}{8}}e^{-\frac{\mu_{\rm H}}{\mu_D}\frac{m_D}{B_D}\frac{v^2}{300}}, 
\ee
where we used $E=m_{\rm H}v^2/4$. In the limit where $m_{\bf e}=m_{\bf p}$, Eq.~(\ref{sig_HHv}) agrees within a factor of order unity with the positronium (Ps) self-scattering cross section \cite{PhysRevA.65.022704} (see Fig.~\ref{HH}).  At very low velocities ($v\lesssim10^{-4}$), the above expression likely overestimates the momentum-transfer cross section. However, since very little energy is transfered in these low velocity collisions, we do not expect this to have a large impact on our results. 
\subsection{{\bf H-p} and {\bf H-e} Cross Section}
When a dark proton collides with a neutral dark atom such that no atomic transitions are excited, two distinct processes must be taken into account. The dark proton can either collide elastically on the dark atom or it can capture the dark electron in a symmetric charge transfer Similarly, when a dark electron encounters a neutral atom at low energy, it can either scatter elastically or eject the atomic dark electron and be captured by the atomic nucleus.  Since the final products of these two types of processes are indistinguishable, they both need to be taken into account to accurately calculate the differential elastic cross section \cite{0953-4075-32-14-317}
\ba\label{sigma_Hp_st}
\frac{d\sigma^{\bf H-p,e}_{\rm el}}{d\Omega}&=&\frac{3}{4}|f_{\rm d}(\theta)-f_{\rm ex}(\pi-\theta)|^2\en
&&+\,\,\frac{1}{4}|f_{\rm d}(\theta)+f_{\rm ex}(\pi-\theta)|^2,
\ea
where $f_{\rm d}(\theta)$ is the amplitude for the direct elastic scattering while $f_{\rm ex}(\theta)$ is the amplitude for the charge-exchange process. Quantum interference between these two types of elastic scattering generally implies that $\sigma_{\rm mt}^{\bf H-p,e}\neq\sigma_{\rm el}^{\bf H-p,e}$, in contrast to the symmetric atom-atom case. 

Much of the above discussion for the atom-atom case applies here and we can relate the dark proton-atom and electron-atom momentum-transfer cross section to that of standard proton-hydrogen and electron-hydrogen scattering as
\ba\label{x-section_subst_pH}
\sigma_{\rm mt}^{\bf  H-p,e}[E] &\approx& \left(\frac{a_{0,D}}{a_0}\right)^2\\
&&\times\,\sigma_{\rm mt}^{\rm H-p,e}\left[\left(\frac{\alpha_D}{\alpha_{\rm em}}\right)^2\frac{\mu_{D{\bf p,e}}}{\mu_{\rm Hp,e}}\left(\frac{B_{\rm H}}{B_D}\right)^2E\right],\nonumber 
\ea
where $\mu_{D{\bf p,e}}= m_Dm_{\bf p,e}/(m_D+m_{\bf p,e})$ and similarly for $\mu_{\rm Hp,e}$. From the results presented in Ref.~\cite{0953-4075-32-14-317}, we can obtain an approximate expression for the proton-hydrogen momentum-transfer cross-section (obtained through a numerical fit) 
\be
\sigma_{\rm mt}^{\rm H-p}(E_{\rm CM})\approx240\,\pi a_0^2\left(\frac{E_{\rm CM}}{B_{\rm H}}\right)^{-\frac{1}{8}}e^{-\frac{E_{\rm CM}}{100B_{\rm H}}},
\ee
again valid for $10^{-3}\lesssim E_{\rm CM}/B_{\rm H}\lesssim10$. In the case of electron-hydrogen scattering, the momentum-transfer cross section is smaller by almost two orders of magnitude
\be
\sigma_{\rm mt}^{\rm H-e}(E_{\rm CM}\ll B_{\rm H})\approx 7\pi a_0^2. 
\ee
\begin{figure}[b!]
\begin{centering}
\includegraphics[width=0.5\textwidth]{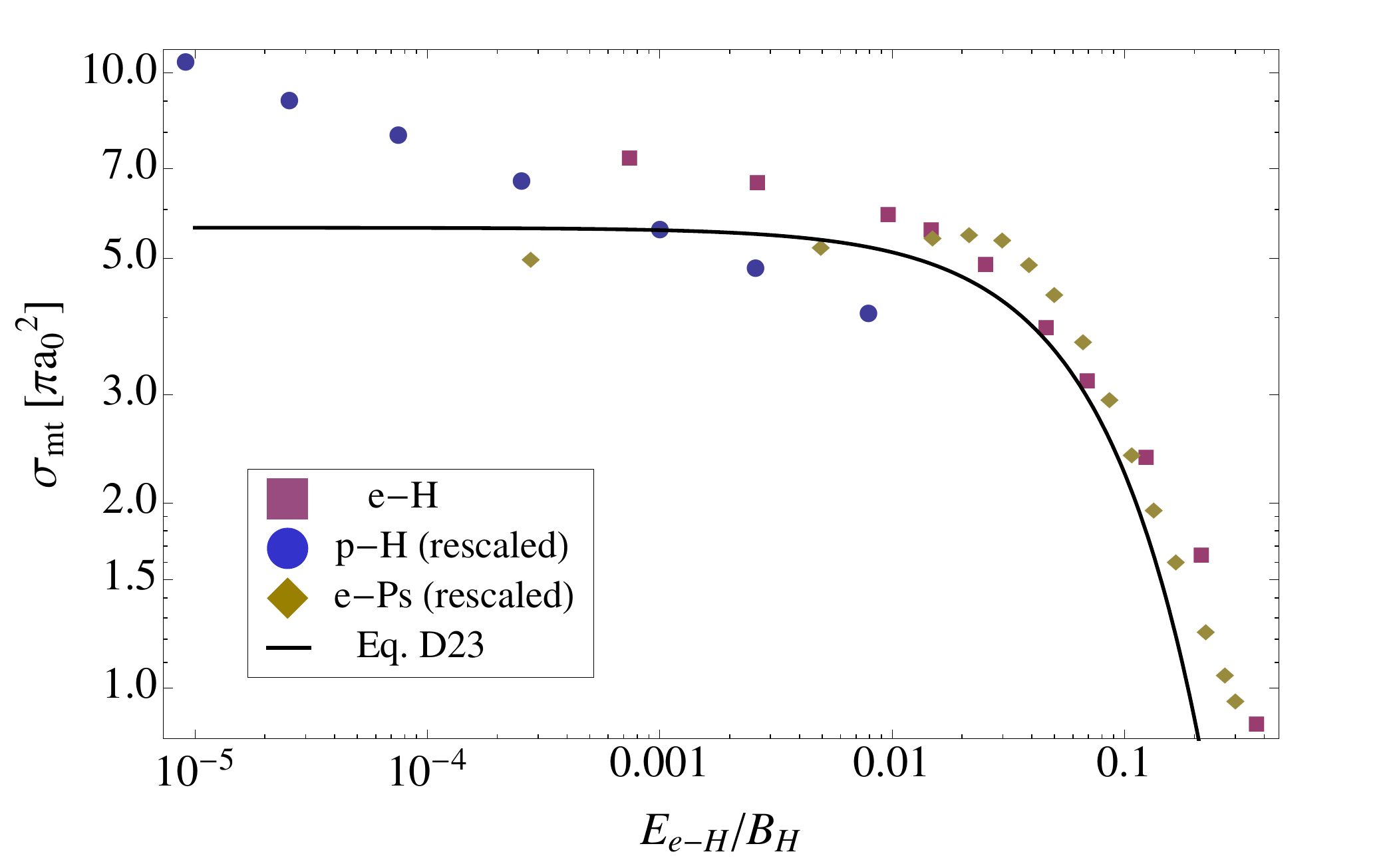}
\caption[Momentum-transfer cross section for electron-hydrogen scattering obtained from direct computation, from rescaling the proton-hydrogen cross section, and from rescaling the electron-positronium cross section.]{Momentum-transfer cross section for electron-hydrogen scattering obtained from direct computation \cite{PhysRev.125.553,Banks19661085}  (purple squares), from rescaling the proton-hydrogen cross section \cite{Krstic&Schultz}  (blue disks), and from rescaling the electron-positronium cross section \cite{0022-3700-20-1-017}  (yellow diamonds). We also display the momentum-transfer cross section we use for dark electron-atom scattering (Eq.~(\ref{sig_eH})).  }
\label{eH}
\end{centering}
\end{figure}
The main reason for this large difference is that the electron-hydrogen differential cross section is strongly peaked in the forward direction, which implies small momentum transfer. One the other hand, the proton-hydrogen differential cross section has a significant backward peak which contributes substantially to the overall momentum-transfer cross section. Phenomenologically, we can take this into account by rescaling the cross sections by the ratio of the masses of the two colliding bodies, that is
\be\label{from_p-H_get_e-H}
\sigma_{\rm mt}^{\rm H-e}(E)\approx \sqrt{\frac{m_{\rm e}}{m_{\rm H}}}\sigma_{\rm mt}^{\rm H-p}\left(\frac{m_{\rm p}}{2 m_{\rm e}}E\right),
\ee
where we have used Eq.~(\ref{energy_scaling}) to appropriately rescale the collision energy. To obtain insight on the case $m_{\bf e}\sim m_{\bf p}$, we also consider the case of an electron elastically scattering off positronium \cite{0022-3700-20-1-017}. In this case, we expect the low-energy cross section to be somewhat suppressed since the first Born approximation for the scattering amplitude exactly vanishes. Existing computations support this fact and predict a roughly constant cross section at low energy. With this in mind, we conservatively neglect the small energy dependence of the H-p cross section at low energy and write 
\be
\sigma_{\rm mt}^{\bf  H-p}(E)\approx\frac{60\pi\alpha_D^2}{B_D^2}\sqrt{\frac{m_{\bf p}}{m_D}}e^{-\frac{\mu_{\rm H}}{\mu_D}\frac{\mu_{D{\bf p}}}{\mu_{\rm Hp}}\frac{E}{100B_D}},
\ee
\be\label{sig_eH}
\sigma_{\rm mt}^{\bf  H-e}(E)\approx\frac{60\pi\alpha_D^2}{B_D^2}\sqrt{\frac{m_{\bf e}}{m_{D}}}e^{-8\frac{\mu_{\rm H}}{\mu_D}\frac{\mu_{D{\bf e}}}{\mu_{\rm He}}\frac{E}{100B_D}},
\ee
or in terms of the relative velocity between the ions and the dark atom
\be
\sigma_{\rm mt}^{\bf  H-p}(v)\approx\frac{60\pi\alpha_D^2}{B_D^2}\sqrt{\frac{m_{\bf p}}{m_D}}e^{-\frac{\mu_{\rm H}}{\mu_D}\frac{\mu_{D{\bf p}}}{B_D}\frac{v^2}{200}},
\ee
\be
\sigma_{\rm mt}^{\bf  H-e}(v)\approx\frac{60\pi\alpha_D^2}{B_D^2}\sqrt{\frac{m_{\bf e}}{m_{D}}} e^{-\frac{\mu_{\rm H}}{\mu_D}\frac{\mu_{D{\bf e}}}{B_D}\frac{v^2}{200}},
\ee
where we have used $E=\mu_{\rm He,p}v^2/2$. In Fig.~\ref{eH}, we display the H-e momentum-transfer cross section, both from direct computation and from rescaling the p-H and electron-positronium cross sections according to Eq.~(\ref{from_p-H_get_e-H}). We see that this rescaling is accurate up to a factor of order unity over the energy range of interest and thus provide the correct order of magnitude for the cross section. We also display in the figure our approximate expression for the dark electron-atom cross section Eq.~(\ref{sig_eH}). 
%

\bibliography{dao_v2}
\end{document}